# Statistical Tools and Methodologies for Ultrareliable Low-Latency Communication −A Tutorial

Onel L. A. López, Nurul H. Mahmood, Mohammad Shehab, Hirley Alves, Osmel M. Rosabal, Leatile Marata, and Matti Latva-aho

*Abstract*—Ultra-reliable low-latency communication (URLLC) constitutes a key service class of the fifth generation and beyond cellular networks. Notably, designing and supporting URLLC poses a herculean task due to the fundamental need to identify and accurately characterize the underlying statistical models in which the system operates, e.g., interference statistics, channel conditions, and the behavior of protocols. In general, multi-layer end-to-end approaches considering all the potential delay and error sources and proper statistical tools and methodologies are inevitably required for providing strong reliability and latency guarantees. This paper contributes to the body of knowledge in the latter aspect by providing a tutorial on several statistical tools and methodologies that are useful for designing and analyzing URLLC systems. Specifically, we overview the frameworks related to *i)* reliability theory, *ii)* short packet communications, *iii)* inequalities, distribution bounds, and tail approximations, *iv)* rare events simulation, *vi)* queuing theory and information freshness, and *v)* large-scale tools such as stochastic geometry, clustering, compressed sensing, and mean-field games. Moreover, we often refer to prominent data-driven algorithms within the scope of the discussed tools/methodologies. Throughout the paper, we briefly review the state-of-the-art works using the addressed tools and methodologies, and their link to URLLC systems. Moreover, we discuss novel application examples focused on physical and medium access control layers. Finally, key research challenges and directions are highlighted to elucidate how URLLC analysis/design research may evolve in the coming years.

## I. INTRODUCTION

ULTRA-reliable low-latency communication (URLLC) aims to support wireless connectivity with wired-grade reliability and latency performance and constitutes

This research was supported by the Research Council of Finland (former Academy of Finland) through the 6G Flagship programme and the academy project ReWIN-6G (Grant Number: 346208 & 357120), and the Finnish Foundation for Technology Promotion. In addition, Leatile Marata's work was partly supported by the Riitta ja Jorma J. Takanen Foundation, the Nokia Foundation, and the Botswana International University of Science and Technology.

Authors are with Centre for Wireless Communications (CWC), University of Oulu, FIN-90014 Oulu, Finland. (e-mail: {onel.alcarazlopez, nurul-huda.mahmood, mohammad.shehab, hirley.alves, osmel.martinezrosabal, leatile.marata, matti.latva-aho}@oulu.fi).

a key service class of the fifth-generation (5G) cellular networks. Moreover, URLLC will undoubtedly remain relevant for the sixth-generation (6G) wireless systems and beyond according to the latest report on future technology trends in the field by the International Telecommunication Union in [1]. Indeed, URLLC will expand in several directions in the coming years. Specifically, i) reliability and latency targets will be more stringent [2], ii) massive and/or broadband connectivity will be supported to some extent [3], and iii) new metrics such as jitter, synchronization accuracy, and dependability measures may be considered. The potential applications of URLLC in 6G are indeed numerous and include smart grid, professional audio, intelligent transport systems, industrial and process automation, augmented reality, holographic communication, digital twinning, entertainment industry, e-health, tactile and haptic Internet, as well as yet many unforeseen use cases [4]–[10].

In URLLC, reliability is often defined as the probability that a data unit is successfully transferred within a certain time period[1]. Noteworthy, although reliability and latency are inherently tied by definition, the technological enablers of URLLC can often be classified as delay-reducing or reliability-promoting [12]. For instance, latency figures can be mainly improved by leveraging short codes and transmission time intervals [7], [13], [14], edge caching/computing/slicing [5], [8], [12], limited-overhead protocols [5], [6], [15]–[17], and non-orthogonal multiple-access (NOMA) [5], [7], [18]. Meanwhile, high-reliability levels can be attained by exploiting diversity techniques as in multi-connectivity [6], [7], [19]–[21], data replication [16], [22], automatic repeat request protocols [7], [23], and multiple-input multiple-output (MIMO) systems [6], [7], [17], [24], as well as efficient network coding and relaying [12], [25]–[27] and space-time block codes [12], [13], [24]. A robust URLLC design may inevitably exploit several of the above technologies so that latency and reliability metrics perform at the desired levels.

Remarkably, designing and supporting URLLC poses a herculean task due to the fundamental need to identify and accurately characterize the underlying statistical models in which the URLLC system operates, e.g., interference statistics, channel conditions, and the behavior of protocols, as this is ultimately required for providing strong quality-of-service (QoS) guarantees [4], [6], [28], [29]. This calls for multi-layer end-to-end approaches considering all the potential delay and error sources [6]–[10][2], and proper statistical tools and methodologies [12], [14], [28]–[32]. The latter may rely on model-driven and/or data-driven approaches and constitute the scope of this work.

---

[1]The data unit (and also reliability/latency figures) in question can be application-dependent or pre-specified. A well-known example of the latter is the layer 2 protocol data unit of 32 bytes, as considered by the third generation partnership project (3GPP) [11]. Still, one can find alternative proposals as well. See Section II-A for a more detailed definition of "reliability", especially within the context of reliability theory.

## A. Model- vs Data-driven Approaches

Model-driven approaches exploit mathematical models that describe the system's behavior based on a priori knowledge of its physical and operational characteristics. In contrast, data-driven approaches leverage empirical data to extract/infer patterns and relationships without necessarily making assumptions about the underlying statistical processes. Usually, model-driven approaches are simpler but may have limited accuracy/flexibility, especially in complex and dynamic environments, due to inherent underlying assumptions. Meanwhile, data-driven approaches are particularly useful when the underlying statistical models are difficult to characterize or when the system behavior is highly nonlinear or unpredictable, as they may successfully capture complex system behaviors and interactions. However, they face serious challenges related to overfitting, lack of interpretability, high data requirements, and/or sensitivity to data quality [33].

Machine learning (ML) is a prevalent data-driven approach. There are three main types of ML: supervised learning (learns from labeled data), unsupervised learning (learns from unlabeled data), and reinforcement learning (RL) (learns through interaction with the environment and feedback from rewards/penalties). In the field of wireless communications, ML has been applied, for instance, in traffic and interference prediction [34], [35], physical layer design [36], heterogeneous network traffic control [37], and multi-access edge communications [38]. Notably, it is also extended to URLLC scenarios in recent years, e.g., for fast uplink scheduling relying on traffic prediction [39], [40] and communication-control co-design [41].

Although the imminent benefits from ML, it may not always be suitable for handling strict URLLC requirements due to challenges such as difficulty in exploiting prior contextual information, limited availability of large training datasets, and high computation complexity and incurred delay. Moreover, the complexity and black-box nature of some ML algorithms may make it difficult to understand their decision-making processes, and thus establish confidence levels. Furthermore, the performance of ML algorithms can also be affected by biases in the data or algorithm design, which may lead to unfair or discriminatory outcomes. All these limitations are being tackled by the research community, which has made significant progress recently in areas such as

- explainable ML [42], [43], which facilitates understanding the ML decision-making processes, allowing for increased transparency and trust in the algorithms;
- ensemble learning [44], [45], which involves combining multiple ML models to improve the accuracy and robustness of the predictions;
- transfer learning [32], [46], federated learning (FL) [38], and meta-learning [47], which are specially efficient in low-data scenarios.

Finally, to combine the strengths of model- and data-driven approaches, She *et al.* [32] advocate integrating domain knowledge with learning, where only the unknown parameters of a model that is known to describe the system behavior are learned. Meta-probability notions may be exploited here to support strong QoS guarantees.

All in all, a fundamental understanding of URLLC and the established statistical framework for ensuring and explaining reliability and low-latency communication is essential and may help to decide and design eventual model-driven or data-driven approaches, including ML, to be used. This paper contributes to the body of knowledge here by providing a tutorial on several statistical tools and methodologies that are useful for designing and analyzing URLLC systems. We specifically focus on physical and medium access control layer perspectives.

## B. State-of-the-Art Tutorials

There are already some surveys, tutorials, and overview papers on tools and methodologies for designing and analyzing URLLC systems in the literature. Table I collects those that cover at least two tools/methodologies[3], and summarizes their main contributions. Next, we discuss the scope and contributions of such works in more detail.

In 2016, Durisi *et al.* [14] overviewed the main theoretical principles for the design and performance analysis of short packet communications, which are key components in URLLC. Specifically, they delved into the structure of a data packet, finite block length (FBL) error characterization, and proper communication protocols. They discussed three exemplary scenarios, i.e., *i)* the two-way channel, *ii)* the downlink broadcast channel, and *iii)* the uplink random access channel, and illustrated how the transmission of control information can be optimized for short packets. The main conveyed message was that it is crucial to consider all the communication resources that are invested in the transmission of metadata when data messages are short.

In 2018, Benis *et al.* [12] discussed definitions of latency and reliability, examined various URLLC enablers and their inherent tradeoffs, and overviewed a wide variety of techniques and methodologies for URLLC. The latter included risk analysis and optimization tools, extreme value theory (EVT), effective bandwidth, stochastic network calculus (SNC), meta-distribution, and mean-field (MF) game theory. Moreover, via four selected use cases related to *i)* ultra-reliable millimeter-wave communication, *ii)* virtual reality, *iii)* mobile edge computing, and *iv)* multiconnectivity for ultra-dense networks[4], the authors showed how such tools provide a principled and clean-slate framework for modeling and optimizing URLLC-centric problems at the network level. The authors agreed that URLLC mandates a departure from expected utility-based approaches relying on average quantities to approaches taming risk and distribution tails.

---

[2]Among sources of delay, we can mention link establishment, signal propagation, waiting times between transmissions, network queues, backhauling, random backoff times, inter/intra-layer processing and computing tasks, whereas relatively low signal-to-interference-plus-noise ratio (SINR), energy shortage, hardware component failure, and multi-user access collision classify as error sources.

[3]Note that for a thorough understanding of a specific tool or methodology, the best approach might be to consult a book or a comprehensive tutorial paper on the matter.

TABLE I
REPRESENTATIVE SURVEY/OVERVIEW/TUTORIAL PAPERS RELATED TO TOOLS AND METHODOLOGIES FOR DESIGNING/ANALYZING URLLC WIRELESS SYSTEMS

| Ref. | Main Contributions | Tools/Methodologies |
|---|---|---|
| [14] | Overview of FBL theoretical and protocol principles | i) FBL Coding Theory, ii) Short Packet Protocols |
| [12] | Discussion about URLLC enablers and overview of techniques and methodologies for the design of URLLC systems | i) Risk Analysis/Optimization Tools, ii) EVT, iii) Effective Bandwidth, iv) SNC, v) Meta-Distribution, vi) MF Game Theory |
| [30] | Revision of the security aspects of compressed sensing and applications to secure wireless communications | i) Compressed Sensing, ii) Security Tools |
| [31] | Discussion of requirements and challenges in URLLC vehicular networks, and overview of frameworks to enable them | i) FBL Coding Theory, ii) Effective Bandwidth, iii) SNC, iv) Lyapunov Optimization, v) Markov Decision Process |
| [32] | Review promising network architectures and DL frameworks for 6G URLLC | i) DL Frameworks, ii) FBL Coding Theory, iii) Queuing Theory Tools, iv) Stochastic Geometry for Delay Analysis |
| this work | Comprehensive overview and exemplification of tools and methodologies for the analysis and design of URLLC systems | i) Reliability Theory, ii) Inequalities and Distribution Bounds, iii) EVT, iv) Rare Events Simulation Tools, v) FBL Coding Theory, vi) Short Packet Protocols, vii) Effective Capacity, viii) SNC, ix) AoI, x) Meta-Distribution, xi) Clustering, xii) Compressed Sensing, xiii) MF Game Theory |

In 2019, Zhang *et al.* [30] reviewed the state-of-the-art of compressed sensing-based secure wireless communications. For this, they first studied the security aspects of compressed sensing according to different types of random measurement matrices such as the Gaussian matrix, circulant matrix, and other special random matrices. Meanwhile, applications such as the wireless wiretap channel, wireless sensor network, Internet of Things, crowdsensing, smart grid, and wireless body area networks were thoroughly discussed immediately after.

In 2020, Yang *et al.* [31] investigated the challenges and solutions for enabling URLLC in vehicular networks. Specifically, the underlying application scenarios exploiting vehicle-to-vehicle/infrastructure/network/pedestrian communications were overviewed, together with the corresponding URLLC requirements and potential challenges. Moreover, some frameworks for optimizing latency and/or reliability and a few case studies were discussed. The optimization frameworks in question were related to FBL coding theory at the physical layer, and effective bandwidth/capacity, SNC, Lyapunov optimization, and Markov decision process at the medium access control layer.

In 2021, She *et al.* [32] illustrated how to integrate domain knowledge (models, analytical tools, and optimiza-tion frameworks) of communications and networking into deep learning (DL) algorithms for URLLCs. For this, they first overviewed URLLCs together with promising architectures and DL frameworks for 6G. Then, they revisited model-based analytical tools and cross-layer optimization frameworks, including FBL coding theory, queuing theory, and stochastic geometry for delay analysis, and hinted at how to improve learning algorithms with such knowledge. Finally, they validated the effectiveness of several learning algorithms via simulation and experimental results.

### C. Our Approach and Contributions

It is worth noting that the state-of-the-art tutorials covering tools and methodologies for the analysis and design of URLLC systems have mostly focused on a few approaches only. In some cases, the main contributions are not even around them, e.g., [30], [31]. Therefore, our approach here is more exhaustive and fully focused on tools and methodologies that seem appealing for the analysis and design of URLLC systems.

The last row of Table I collects the 13 specific tools and methodologies that we cover in this tutorial. Observe that some of them have been already discussed in prior tutorials, e.g., FBL coding theory in [14], [31], [32], short packet protocols in [14], EVT in [12], meta-distribution in [12], [32], compressed sensing in [30], effective capacity/bandwidth and SNC in [12], [31], and MF game theory in [12]. For these, we provide fresh and intuitive discussions, in many cases accompanied by novel examples to propitiate a better understanding of the approaches and their, not always evident, link to URLLC design/analysis. Meanwhile, the state-of-the-art tutorials have not covered some other tools and methodologies we discuss here. That is the case with reliability theory, inequalities and distribution bounds, rare events simulation tools, clustering, age of information (AoI), and time series prediction tools. Therefore, our contributions in this regard are even stronger and more evident, and comprise an overview of the tool/methodology, its potential for URLLC design/analysis, and novel examples. In several cases, we refer to prominent data-driven algorithms within the scope of the discussed tools/methodologies, including pointers to relevant literature. We also identify convenient links among the different approaches, i.e., by highlighting how a certain tool/methodology can be leveraged by another one to (more efficiently) meet the desired goals. Additionally, throughout the paper, we briefly review the state-of-the-art works using the addressed tools/methodologies, especially when they link them to URLLC systems, and identify key challenges. Finally, key research directions are highlighted, which might hint at how URLLC analysis/design research may evolve in the coming years.

---

[4]Refer to [48] for a survey on ultra-dense networks, including definitions and distinctions compared to traditional networks.

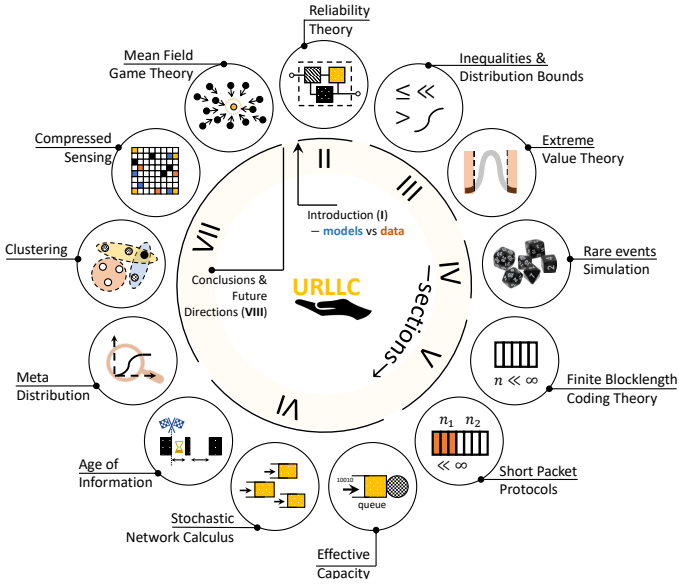

Fig. 1. Tools and methodologies are overviewed in this work for the design and analysis of URLLC systems. Here, they appear in the same order as addressed in the paper, and the corresponding section number is also indicated.

### D. Organization

In the following, Sections II-VII overview the specific tools and methodologies for the design and/or analysis of URLLC systems as illustrated in Fig. 1, while providing examples and fresh views. Section VIII concludes the paper and discusses interesting challenges and research directions. The acronyms and main notations used throughout this article are listed in Table II and Table III, respectively.

## II. RELIABILITY THEORY

Reliability theory comprises a set of mathematical methods for analyzing the life cycles and failures of technical systems [49]. The subsystems and components of a system, hereinafter referred to as "items", can be classified as repairable or non-repairable. In the case of the latter, it does not matter what happens after the first failure occurs, i.e., the item might be discarded or even repaired, while all the operational and failed cycles matter in the case of repairable systems. Notice that a wireless communication system can be composed of several items, e.g., transmitters, receivers, and links/channels, all of which might constitute sources of failure. Next, we overview key aspects of the reliability theory framework and exemplify their potential application to URLLC analysis/design.

### A. Key Dependability Quantities

Here, we overview key dependability quantities within the reliability theory framework[5]. Such quantities can be estimated from data or inferred from models and must respectively capture the uncertainty due to limited data or modeling assumptions to robustly support URLLC design/analysis. This calls for meta-probability-based approaches.[6] In the following, refer to "up" and "down" as operational

---

[5]Notice that the term *dependability* is a broader concept that fits better here as it encompasses availability, reliability, maintainability, and in some cases, other characteristics such as durability, safety, and security [49].

[6]Refer to Section VII-A for more information about meta-probability, especially within the stochastic geometry framework.

| Notation | Description |
|---|---|
| $f_X(x)$, $F_X(x)$ | PDF/CDF of RV $X$ |
| $f(x;y)$ | joint PDF of RVs $X$ and $Y$ |
| $\bar{F}_X(x)$ | complementary CDF of RV $X$ |
| $\Pr[A]$ | probability of event $A$ occurrence |
| $A|B$ | RV $A$ conditioned on the realization of RV $B$ |
| $\mathbb{E}_X[\cdot]$, $\mathbb{V}_X[\cdot]$ | expected value/variance of the argument w.r.t RV $X$ |
| $\binom{\cdot}{\cdot}$ | binomial coefficient |
| $\Gamma(\cdot)$, $\Gamma(\cdot,\cdot)$ | complete/upper incomplete gamma function |
| $\text{li}(\cdot)$ | logarithmic integral function |
| $\Upsilon$ | Euler–Mascheroni constant ($0.577215664901532\cdots$) |
| $Q(\cdot)$ | Gaussian $Q$ function |
| $I_\nu(\cdot)$, $K_\nu(\cdot)$ | modified Bessel function of first/second kind & order $\nu$ |
| $\text{sgn}(\cdot)$ | sign/signum function |
| $\oplus$ | exclusive OR operator |
| $\sim$ | distributed as |
| $\text{Exp}(\lambda)$ | exponential distribution with mean $1/\lambda$ |
| $\mathbb{G}(\alpha,\beta)$ | gamma distribution with shape $\alpha$ and scale $\beta$ |
| $\ll$, $\gg$ | much less/greater than |
| $\cdot^T$, $\cdot^H$ | transpose, Hermitian transpose operation |
| $\mathbb{C}^{N \times M}$ | domain of $N \times M$ complex matrices |
| $|\cdot|$ | absolute value |
| $\|\cdot\|_\#$ | $\ell_\#$–norm |
| $\inf \cdot$ | infimum operation |
| $a \to b$ | $a$ approaches (converges to) $b$, domain-image mapping |
| $\mathbb{R}$, $\mathbb{R}^+$ | real, and non-negative real domain |
| $\mathbb{I}\{\cdot\}$ | indicator function: 1 (0) if the input is true (false) |

and failure (i.e., in repair if repairable) state, respectively, and assume that the considered item is "up" at time $t = 0$.

*1) Availability:* The instantaneous availability or point availability of an item is defined as [49]

$$A(t) = \Pr[\text{"item is up at time } t\text{"}]. \tag{1}$$

Meanwhile, steady-state availability is defined as

$$A = \lim_{t \to \infty} A(t), \tag{2}$$

and allows characterizing the long-term probability that an item is available. Indeed, (2) converges to the fraction of the mission time in which the item remains in the "up" state.

*2) Reliability:* Reliability refers to the failure-free operation probability of the item during the interval $[0, t]$, i.e.,

$$R(t) = \Pr[\text{"item is up throughout the interval } [0, t]\text{"}]. \tag{3}$$

$R(t)$ is also referred to as the survival function. Meanwhile, the amount of time elapsed before a failure occurs can be measured/characterized by the time to failure random variable (RV) $T$, and notice that $R(t) = 1 - F_T(t), t \geq 0$.

Reliability is usually the measure of interest for non-repairable systems, where failures are permanent for the remainder of the mission, while *availability* is preferred for repairable systems. In general, $A(t) \geq R(t)$, while $\lim_{t \to \infty} R(t) = 0$ and $\lim_{t \to \infty} A(t) = A \geq 0$. In both cases, the equality holds only in the special case of no repairs.

*3) Mean downtime/uptime:* The mean downtime (MDT) and mean uptime (MUT) constitute the mean duration of a system failure and mean system operational time, respectively. Notice that $A = \text{MUT}/(\text{MUT} + \text{MDT})$.

*4) Mean time to first failure (MTTFF):* MTTFF characterizes the average duration of an item in "up" state before the failure occurs, thus, it can be written as [49]

$$\text{MTTFF} = \int_0^\infty R(t)\mathrm{d}t. \tag{4}$$

This metric is equivalent to the mean time to failure and MUT in non-repairable systems, and also to the mean time between failures in memoryless repairable systems.

*5) Failure Rate:* Failure rate, also known as Hazard function, refers to the conditional probability that an item that operates failure-free until time $t$, i.e., $R(t) = 1$, also survives an additional infinitesimal time period. Mathematically, the failure rate is given by

$$\lambda(t) = \lim_{\Delta t \to 0} \frac{1}{\Delta t} \frac{F_T(t + \Delta t) - F_T(t)}{1 - F_T(t)} = \frac{f_T(t)}{R(t)}. \tag{5}$$

Moreover, as shown in [50], the probability that the item fails within the (short) time interval $\Delta t$ after being operational until time $t$ approximates to $\lambda(t)\Delta t$.

Several practical systems exhibit a Weibull-distributed time to failure, such that $\lambda(t) = K t^m$ with $m > -1$ [49]. Note that the exponential distribution (a Weibull distribution with $m = 0$) is the only continuous distribution with a constant failure rate, thus, memoryless, i.e., the distribution of the additional system lifetime does not depend on its age.

*6) Maintainability:* Maintainability refers to the ability to repair a system within a given period of time. Notice that if the repair time follows an exponential distribution, it is possible to define the repair probability density function (PDF) and cumulative density function (CDF) based on a single parameter, the repair rate, $\mu$.

*7) Importance measures:* Importance measures, including the Birnbaum importance, criticality importance measure, risk achievement worth, risk reduction worth, Fussell-Vesely, and differential importance, are not dependability quantities per se, but rather they evaluate the contribution of the elements (e.g., failure modes, items, or events) to a dependability quantity of interest [49]. Assessing their potential for URLLC design/optimization/analysis is an interesting research direction.

## B. Structure Functions

Assume a system is composed of $n$ items, and let the vector $\mathbf{x} = [x_1, x_2, \cdots, x_n]^T$ indicate which of the items are functioning and which have failed. Mathematically,

$$x_i = \begin{cases} 1, & \text{if the } i-\text{th item is "up"} \\ 0, & \text{if the } i-\text{th item is "down"} \end{cases}. \tag{6}$$

The structure function of the system, denoted as $\phi(\mathbf{x})$, captures the operability of the whole system given a state vector $\mathbf{x}$ as

$$\phi(\mathbf{x}) = \begin{cases} 1, & \text{if the system is "up" given } \mathbf{x} \\ 0, & \text{if the system is "down" given } \mathbf{x} \end{cases}. \tag{7}$$

Meanwhile, the probability of the system being "up" or available can be written as $A = \Pr[\phi(\mathbf{x}) = 1] = \mathbb{E}_{\mathbf{x}}[\phi(\mathbf{x})]$. Notice that $A$ is solely a function of $\mathbf{p} = [p_1, p_2, \ldots, p_n]^T$, where $p_i \triangleq \Pr[x_i = 1]$, when all the items' states are independent. In such case, $A(\mathbf{p})$ is an increasing function of $\mathbf{p}$, and $A(\mathbf{p}^\alpha) \geq A(\mathbf{p})^\alpha$ with $\alpha \in [0,1]$ [51, Ch. 9].

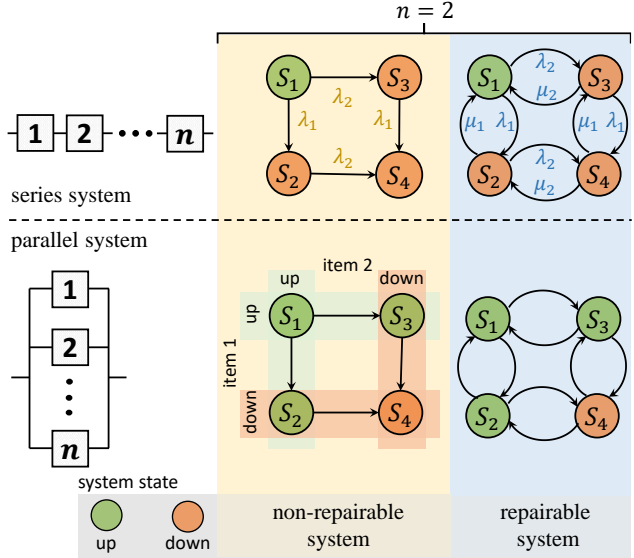

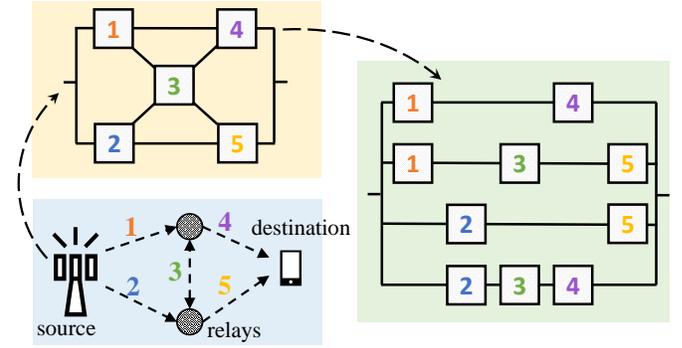

Fig. 3. a) Example of a wireless relaying scenario (bottom-left), and its b) structure representation (top-left) and c) series-parallel equivalent representation (right).

Fig. 2. RBD for the two special systems, and Markov modeling for non-repairable and repairable systems with two items. Here, $\lambda_i$ and $\mu_i$ respectively denote the failure and repair rate of item $i \in \{1,2\}$.

A system, as a composition of several items, can be assessed via a fault tree analysis or a reliability block diagram (RBD). The former is a graphical tool used mainly for system-level risk assessment and mitigation purposes, while the latter illustrates, in a simple manner, how component reliability contributes to the success or failure of a system. Both can be used to model and analyze similar types of logical configurations required for system reliability, although RBDs are usually more mathematically abstracted and can be handled more straightforwardly. Here, we focus only on RBDs, while the interested reader can refer to [52] for further details on fault tree analysis.

The way in which the items are interconnected determines the redundancy level of the whole system. Mathematically, a $k$-out-of-$n$ structure models a system that is functioning if and only if at least $k$ of the $n$ items are "up" [49], i.e.,

$$\phi_{k,n}(\mathbf{x}) = \begin{cases} 1, & \text{if } \mathbf{1}^T\mathbf{x} \geq k \\ 0, & \text{otherwise} \end{cases}. \qquad (8)$$

There are two special systems: *i)* parallel or full redundancy system, which corresponds to a 1-out-of-$n$ structure; and *ii)* series or non-redundancy system, which corresponds to an $n$-out-of-$n$ structure. Their RBDs are illustrated in Fig. 2 and their structure functions are given by

$$\phi_{1,n}(\mathbf{x}) = \max\{x_1, x_2, \ldots, x_n\} = 1 - \prod_{i=1}^{n}(1 - x_i), \qquad (9)$$

$$\phi_{n,n}(\mathbf{x}) = \min\{x_1, x_2, \cdots, x_n\} = \prod_{i=1}^{n} x_i. \qquad (10)$$

Moreover, $A_{1,n}(\mathbf{p}) = 1 - \prod_{i=1}^{n}(1 - p_i)$ and $A_{n,n}(\mathbf{p}) = \prod_{i=1}^{n} p_i$ in case of independent items' states. Notice that any general structure (and corresponding structure function) can be scrutinized as a combination of series and parallel structures, thus, they may be seen as canonical structures.

EXAMPLE 1 (RBD ANALYSIS OF A RELAYING SYSTEM).
*Consider that the wireless link between a source and a destination node is obstructed, thus, a potential direct communication between these nodes is discarded. Instead, the source sends a data packet to the destination via a wireless network composed of two relays as illustrated in Fig. 3a. There are five active wireless links, which constitute the system items here. Assume that each link $i \in \{1,2,3,4,5\}$ remains in state $x_i$ during a transmission round, which starts with the packet being transmitted by the source and ends with the packet being received by the destination or dropped due to errors. Note that the realizations of $\{x_i\}$ are commonly affected in practice by the fading, shadowing, path-loss, and interference phenomena, to name a few.*

*Based on Fig. 3a, one can straightforwardly draw the structure representation of the system, followed by its series-parallel equivalent representation, as shown respectively in Fig. 3b and Fig. 3c. This allows one to easily obtain the structure function and availability of the system, for which one can exploit $x_i^2 = x_i$.*

Previous example is quite simple and abstract, and the RBD analysis fits nicely here. Indeed, the reliability theory framework was originally conceived for industrial life cycle analysis, where each item's design/performance can be analyzed separately without significantly influencing the other items, but this does not completely hold in wireless communications systems, where transmitters, receivers, antennas, signal processing algorithms, and network protocols, must work tightly together. Nevertheless, adapting industrial life cycle analysis to wireless communications setups may be still feasible by adopting, e.g., [49]

- System-level analysis, e.g., considering/simulating items' interactions and interdependencies instead of analyzing them in isolation.
- Iterative design, such that items' design is iteratively refined considering the impact on other items. Iterations and feedback loops may support system optimization.

- Sensitivity analysis to reveal system sensitivity to changes in individual items. Critical factors impacting the system performance can be identified by systematically varying items' parameter/characteristics.

Moreover, techniques like RBDs and fault tree analysis may not be useful due to the highly complex and hard-to-capture interactions and dependencies between items in wireless communications. Still, one may simplify and abstract them to a manageable level, e.g., using hierarchical splitting, breaking down the analysis into specific subsystems, and/or identifying the most impactful interactions and representing them in a simplified manner. By focusing on the key dependencies and interactions or reducing their scope, the analysis can still provide valuable insights and identify critical failure paths.

A crucial limitation of RBDs and fault tree analysis is that they are Boolean tools, i.e., with only "up" and "down" states. This is overcome by Markov models, which can include several component states capturing key aspects of the temporal and sequential dependence among events [53]. Next, they are overviewed within the reliability framework, while the interested reader can refer to [54] for detailed analysis and mathematical characterization.

### C. Markov Models

To formulate a Markov model, the system behavior is abstracted into a set of mutually exclusive system states, and a set of equations, describing the probabilistic transitions among states and initial state probability distributions, is set. Moreover, the transition from one state to another depends only on the current state, thus, the way in which the entire past history affects the future of the process is completely summarized in the current state of the process. A Markov model can be defined over discrete (finite or countable infinite) or continuous state spaces. The former case is commonly known as Markov chain, whereas the latter is defined as a continuous-space Markov model. Regarding the time granularity of the state transitions, Markov models are classified as continuous-time or discrete-time if they allow transitions at any time or at fixed intervals, respectively. Next, we discuss and exemplify key principles and methodological aspects of Markov modeling within the reliability theory framework.

In Markov models for repairable systems, the repairs are represented by cycles depicting the loss and the restoration of the functionality of an item. Meanwhile, the Markov models for non-repairable systems are acyclic. As an example, the last two blocks in Fig. 2 illustrate the Markov models for the series and parallel repairable and non-repairable systems with two items.

EXAMPLE 2 (MARKOV-ENABLED RELIABILITY ANALYSIS).
*Consider that a Rayleigh-faded receive signal can be successfully decoded if the instantaneous signal-to-noise ratio (SNR) at time $t \in \mathbb{R}$, $\gamma(t)$, is above a certain threshold, $\gamma_{th}$. Thus, there are two states, i.e.,*

$$x(t) = \begin{cases} 0, & \text{if } \gamma(t) < \gamma_{th} \quad \rightarrow \text{``down''} \\ 1, & \text{if } \gamma(t) \geq \gamma_{th} \quad \rightarrow \text{``up''} \end{cases}. \qquad (11)$$

*According to the Gilbert-Elliot model [55], one can interpret the wireless channel as a repairable item and model it as a continuous-time Markov chain of two states. The corresponding failure rate and repair rate are respectively given by [56]*

$$\lambda = \sqrt{\frac{2\pi\gamma_{th}}{\bar{\gamma}}} f_D, \qquad \mu = \frac{\lambda}{\exp(\gamma_{th}/\bar{\gamma}) - 1}, \qquad (12)$$

*and are obtained based on a level-crossing analysis. In (12), $\bar{\gamma}$ is the average SNR over time, and $f_D$ is the maximum Doppler frequency. Notice that the rates $\lambda$ and $\mu$ are constants because $\gamma(t)$ is exponentially distributed.*

*Meanwhile, the steady-state probabilities are given by*

$$P_0 = \Pr[x(t) = 0] = \frac{\lambda}{\lambda + \mu}, \qquad P_1 = \Pr[x(t) = 1] = \frac{\mu}{\lambda + \mu}. \qquad (13)$$

*They are obtained by solving $\{[P_0, P_1]\mathbf{M} = \mathbf{0}, P_1 + P_2 = 1\}$, where $\mathbf{M}$ is the system transition matrix, i.e.,*

$$\mathbf{M} = \begin{bmatrix} -\mu & \mu \\ \lambda & -\lambda \end{bmatrix}. \qquad (14)$$

*With all this at hand, one obtains*

$$A(t) = A = P_1 = \frac{\mu}{\lambda + \mu}, \qquad (15a)$$

$$R(t) = \exp(-\lambda t), \qquad (15b)$$

$$\text{MTTFF} = \text{MUT} = \frac{1}{\lambda}, \qquad (15c)$$

$$\text{MDT} = \frac{1}{\mu}. \qquad (15d)$$

*Notice that $1 - A = \frac{\lambda}{\lambda + \mu} = 1 - \exp(-\gamma_{th}/\bar{\gamma})$, which, as expected, matches the so-called outage probability or packet loss rate in wireless communications.*

Some comments are in order regarding the previous example:

- In URLLC systems with short packet transmissions, there is no such threshold $\gamma_{th}$ that fully guarantees a failure-free system operation. The reasons are discussed in Section V. Instead, an appealing approach that is leveraged in [21] relies on defining the system "up" state as the state in which the failure-free operation is guaranteed with a certain (potentially high) confidence level. By setting a target confidence level, the value of $\gamma_{th}$ can be computed using (50).
- There is no system redundancy. The interested reader can refer to [57], [58], where the authors assume there are $n \geq 1$ links and the system is operational if at least one of the links is "up", i.e., redundancy via selection combining. In the case of [57], this leads to a birth-death continuous-time Markov chain, while the Markov chain is more evolved in [58] since the authors assume Rician fading and incorporate another cause of failure, interference. The analysis in [58] is facilitated by a hybrid series-parallel structure representation of the system (see Section II-B).
- One may wonder how the dependability metrics in (15d) are really relevant in a URLLC context. In the

case of the steady-state availability, $A$, there is not much more to say as this quantity (or its complement, the outage probability $1 - A$) has been the most commonly adopted for assessing the performance of URLLC systems at the physical layer in the literature. In the case of the instantaneous availability and the reliability quantities, the main attraction lies in their time-dependency feature, especially in systems with a non-constant failure rate. Finally, although MTTFF, MUT and MDT are quantities comprising averaging over time, which is not per se appealing in a URLLC context [12], they provide an overarching view of the system in terms of average times in "up", "down" and transition states. Perhaps most importantly, they can facilitate more rigorous reliability analysis as we illustrate in the following example.

**Example 3 (Minimum Duration Outage).**
*Consider the same system setup as in Example 2. Moreover, assume that such a wireless system can tolerate error bursts of $u$ symbols by implementing an appropriate channel coding mechanism. The symbol duration is denoted by $T_s$. Herein, we are interested in determining the SNR margin, defined as $F = \bar{\gamma}/\gamma_{th}$, such that the packet loss rate does not exceed the value of $\xi \ll 1$.*

*Obviously, $uT_s$ must be greater than* MDT, *but by how much? The notion of minimum duration outage (MDO), introduced in [59], addresses this kind of problem. Specifically, an MDO occurs only when the signal level remains below a pre-specified threshold for a certain minimum duration. Mathematically, the probability of MDOs is given by*

$$P_{\text{MDO}} = \Pr[\tau > uT_s | x(t) = 0]\Pr[x(t) = 0], \quad (16)$$

*where $\tau$ is an RV defined as the duration of a fade. Observe that this formulation neglects the possibility that a given packet experiences more than one deep fade event. Nevertheless, this would rarely occur for $\xi \ll 1$, thus, we can safely adopt it. Then, one must configure $F$ such that $P_{\text{MDO}} \le \xi$.*

*Since $\Pr[x(t) = 0]$ can be easily computed as $1 - A = \lambda/(\lambda + \mu)$, the only problem in evaluating (16) is finding the distribution of $\tau$. Herein, we adopt a much simpler approach that exploits the already available MDT quantity. Specifically, based on the simplest form of Markov's inequality (see Section III-A), one obtains*

$$\Pr[\tau > uT_s | x(t) = 0] \le \frac{\mathbb{E}[\tau]}{uT_s} = \frac{\text{MDT}}{uT_s}. \quad (17)$$

*By substituting (17) into (16) and leveraging (15d), we have*

$$P_{\text{MDO}} \le \frac{\text{MDT}}{uT_s}(1 - A)$$
$$= \frac{\lambda}{uT_s\mu(\lambda + \mu)} = \sqrt{\frac{2F}{\pi}}\frac{\cosh(1/F) - 1}{uT_s f_D}. \quad (18)$$

*As expected, the upper bound of $P_{\text{MDO}}$ is a decreasing function of $F$, and $\lim_{F\to\infty}P_{\text{MDO}}(F) = 0$. This means that there is a unique solution $F^{\star}$ for the equation*

$$\sqrt{\frac{2F}{\pi}}\frac{\cosh(1/F) - 1}{uT_s f_D} = \xi, \quad (19)$$

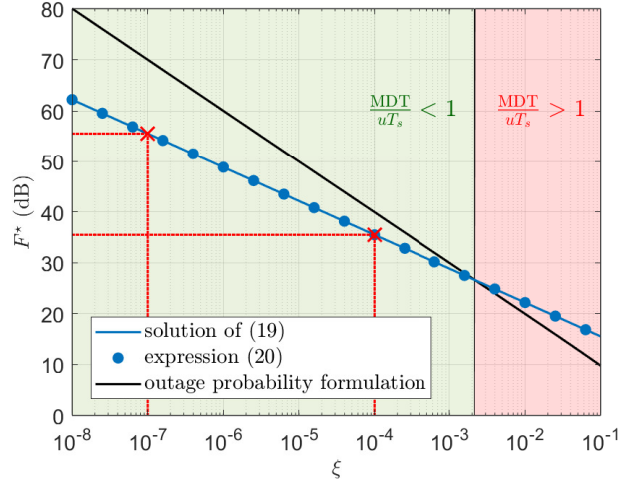

Fig. 4. The SNR margin, $F^{\star}$, as a function of a target MDO probability constraint, $\xi$. We set $uT_s = 0.2$ ms and $f_D = 93.33$ Hz (corresponding to an operating frequency of 28 GHz and a relative velocity of 1 m/s between transmitter and receiver). A curve corresponding to the SNR margin required to only satisfy an outage probability constraint, $\xi$, is also illustrated as a benchmark.

*which, in fact, constitutes our solution to the original problem. Finally, we can obtain a closed-form approximation for (19) for the high-SNR regime. Specifically, by using $\cosh(1/F) \approx 1 + 1/(2F^2)$, which comes from the Taylor series expansion, one obtains*

$$F^{\star} \approx \frac{1}{\sqrt[3]{2\pi\xi^2 u^2 T_s^2 f_D^2}}. \quad (20)$$

*Fig. 4 shows $F^{\star}$ as a function of $\xi$ for a certain system configuration, while confirming the high accuracy of (20) driven by the relatively large values of $F^{\star}$. Notice that the adopted upper-bound in (18) is only relevant when $\text{MDT}/(uT_s) < 1$ since $P_{\text{MDO}} < 1 - A$ must hold. Therefore, as shown in the figure, it can be only used when $\xi < 2 \times 10^{-3}$. According to the figure, $F = 35.5$ dB and $F = 55.5$ dB can respectively guarantee an MDO probability below $10^{-4}$ and $10^{-7}$. Meanwhile, in terms of outage probability, which is agnostic of the channel coding correcting error potentialities, the required SNR margin values would be $F = 40$ dB and $F = 70$ dB, respectively.*

Table IV summarizes the main advantages and disadvantages of Markov modeling within the reliability theory framework [60]. In general, Markov modeling should not be used when: *i)* the system can be well-modeled with simpler combinatorial methods, e.g., RBDs and fault tree analysis, *ii)* modeling the system requires a very large number of states[7], and *iii)* system behavior is too detailed or complex to be expressed in a Markov/semi-Markov model. In the case of the latter two, one may rely on complex event processing (CEP) techniques.

---

[7] Observe that the state space grows exponentially with the number of items. Specifically, given $n$ items and $s$ states per item, the state space would be composed of $s^n$ states. Considering a system with 10 items and 2 states per item, there are $2^{10} = 1024$ system states, while that number increases to $3^{10} = 59049$ if an additional state is considered for each item.

TABLE IV
Advantages and Disadvantages of Markov Modeling

| Advantages | Disadvantages |
|---|---|
| • Repairs (individually or in groups) can be modeled in a natural way, including sequential and partial repairs<br>• Inherent ability to probabilistically model system failures<br>• Feasibility to model imperfect fault coverage, which arises when a dynamic reconfiguration process triggered by an item failure may lead to the system failure<br>• Feasibility to model standby spares (namely, items that replace those in "down" state) and dependencies on sequences of events | • System modeling may require many states, making the whole modeling complex and often intractable<br>• Some models might be difficult to construct and validate<br>• The Markov assumptions/properties and the specific item failure distributions that are required may not be suitable for proper system modeling |

### D. Complex Event Processing

CEP comprises techniques to analyze and process large data sets in real-time systems, and thus support fault detection, fault tolerance, and the implementation of proactive measures. CEP techniques can be mainly divided into time-series modeling/prediction and pattern recognition, which are briefly discussed in the following.

Time-series modeling/prediction is useful for analyzing a system's data over time and predicting future behavior. This requires first collecting the data on the items of interest over a period of time, including its conversion into a time-series format, and then adopting appropriate statistical processing methods. The latter include autoregressive moving average (and its variants), Markov models (refer to Section II-C)), Bayesian network [61], Kalman filter [62], and ML. The ML methods suitable for time-series modeling/prediction include deep belief networks [63], convolutional neural networks [64], recurrent neural networks [65], long short-term memory networks [35], and the novel "transformer" architecture [66] which may provide unprecedented support to tackle sophisticated and high-dimensional problems.

Meanwhile, pattern recognition leverages ML tools such as support vector machines, neural networks, and clustering (refer to Section VII-B) to identify and interpret patterns within data. The main phases are: *i)* data representation, where raw data is converted into feature vectors that capture the relevant characteristics or attributes of the patterns; *ii)* feature extraction, where features discriminating between different patterns are selected/derived from the data; *iii)* training, where a pattern recognition model is constructed using a labeled dataset; and *iv)* classification/recognition, where unseen patterns are classified/identified according to the trained model. Two relevant pattern recognition frameworks are [67]: *i)* association rule mining, e.g., Eclat [68], frequent pattern growth [69], and Apriori [70] algorithms, which can be used to identify co-occurring events and generate rules based on their proba-bilities or support; and *ii)* anomaly detection, which is used to identify unusual/outlier events exploiting methods like z-score, Gaussian mixture models, and isolation forests. In general, the success of pattern recognition systems depends on the quality of the features, the classification algorithm choice, and the availability of a diverse and representative training dataset. For this, domain expertise, feature engineering, and iterative refinement are needed.

Notably, ML tools play a crucial role in modern CEP. One of the main challenges is the learning uncertainty as most ML tools cannot specify it. To output a prediction together with a confidence level, supervised ML can exploit Bayesian inference techniques, including Gaussian processes [71] and neural processes [72]. Specifically, Gaussian processes assign a probability distribution to the space of functions that can describe the data. Establishing such a prior distribution over functions, they capture the uncertainty about the true underlying function before observing any data. For this, the process uses a mean function and a covariance function (or kernel), which encode the beliefs about the average behavior and the similarity between function values at different input points, respectively. Meanwhile, neural processes aim to provide a flexible and scalable framework for modeling complex, structured data. Instead of relying on explicit covariance functions or kernels, they use neural networks to learn the mapping between inputs and outputs. In any case, providing high levels of confidence may be still a critical challenge for modern ML algorithms. Please, refer to Section I-A, where we highlight relevant challenges related to high-reliability and low-latency support and corresponding research areas.

### III. Taming distribution tails

URLLC is about taming the occurrence of extreme and rare events, specifically the tail distribution of latency and/or reliability system performance. Next, we overview key bounds and approximations for taming distributions' tails. The interested reader can also refer to risk assessment tools and metrics such as value-at-risk, expected shortfall, and risk-sensitive learning [73], and their application to URLLC design/analysis problems [2], [74], [75].

### A. Bounds & Limiting Forms

Assume $X$ is an RV, and let us define $Y = g(X)$ where $g$ is a real monotonic function. Then,

$$F_X(x) = \Pr[X < x] = \begin{cases} \Pr[g(X) < g(x)] = F_Y(g(x)), \text{ or} \\ \Pr[g(X) > g(x)] = 1 - F_Y(g(x)), \end{cases} \quad (21)$$

if $g$ is strictly increasing or decreasing, respectively. In addition, if $g$ is a function such that $X \leq Y$ is guaranteed, then $F_X(x) = \Pr[X < x] \geq \Pr[Y < x] = F_Y(x)$; and similarly $F_X(x) \leq F_Y(x)$ if $X \geq Y$. All in all, the underlying message is that one may tractably bound the distribution of $X$ by properly designing $g(\cdot)$. Alternatively, bounds to the distribution of $X$, if known, can be directly established, e.g., by applying fundamental algebra.

In practice, bounding the entire distribution of $X$ is not often necessary, but only one of its tails according to the specific URLLC problem at hand. Therefore, one may resort to limiting forms, which are often obtained by taking only the asymptotically dominant terms of a series (e.g., Taylor, Laurent, and Puiseux) expansion or similar (e.g., Poincaré expansion and Padé approximant). Table V lists simple limiting forms of popular functions, which can be used to construct the limiting forms of more evolved functions.

TABLE V
LIMITING FORMS OF SOME POPULAR FUNCTIONS

| Function | $x \to 0$ | $x \to \infty$ |
|---|---|---|
| $\exp(x)$ | $1+x$ | – |
| $\sin x, \ \tanh x,$ | $x$ | – |
| $\ln(1+x), \tan x$ | | |
| $\cos x$ | $1 - x^2/2$ | – |
| $\sinh x$ | $x$ | $\exp(x)/2$ |
| $\cosh x$ | $1 + x^2/2$ | $\exp(x)/2$ |
| $\binom{2x}{x}$ | – | $4^x/\sqrt{\pi x}$ |
| $\Pi_{i=1}^N (1+x_i)$ | $\left(1 + \sum_{i=1}^N x_i/N\right)^N$ | – |
| $\Gamma(x)$ | $1/x$ | $\sqrt{2\pi/x}(x/e)^x$ |
| $\Gamma(a,x)$ | $-x^a/a \ \ \forall a \in \mathbb{R}^+$ | $x^{a-1}\exp(-x)$ |
| $\mathrm{li}(1+x)$ | $\Upsilon + \ln x$ | $\frac{1+x}{\ln(1+x)}$ |
| $Q(x)$ | $\frac{1}{2} - \frac{x}{\sqrt{2\pi}}$ | $\frac{1}{\sqrt{2\pi}x}\exp(-x^2/2)$ |
| $I_\nu(x)$ | $\frac{x^\nu}{2^\nu\Gamma(\nu+1)} \ \ \forall \nu \neq -1,-2,\cdots$ | $\exp(x)/\sqrt{2\pi x}, \ x \in \mathbb{R}^+$ |
| $K_\nu(x)$ | $-\ln x$ for $\nu = 0$, and | $\sqrt{\pi/(2x)}\exp(-x),$ |
| | $2^{\nu-1}x^{-\nu}\Gamma(-\nu) \ \forall \nu \in \mathbb{R}^+$ | $\forall x \in \mathbb{R}^+$ |

Notice that if the distribution of $X$ is known, the limiting forms may help to simplify it in the tail region (see Example 4 below). Otherwise, if RV $X$ is, or can be represented, as a transformation of other RV(s), one may leverage their limiting forms to bound/approximate the tail distribution of $X$. For instance, assume $X = \exp(Y-1) + \cosh(1/Z) - 2$, where $Y, Z \geq 1$ are some arbitrary RVs, and one is interested in the left tail of the distribution of $X$, i.e., $F_X(x)$ as $x \to 0$. In this case, finding the whole distribution of $X$ may be cumbersome even in the case that the distributions of $Y$ and $Z$ are simple. Instead, one may realize that the small realizations of $X$ are obtained when $Y$ and $Z$ approach 1 and $\infty$, respectively, and consequently may leverage the limiting forms in the rows one and four of Table V. As a result, one obtains that the distribution of $Y + 1/(2Z^2) - 1$ converges to the distribution of $X$ in the left tail. Notably, obtaining the distribution of such an expression seems much more analytically friendly than directly obtaining the distribution of $X$.

EXAMPLE 4 (RELIABLE INTERFERENCE NETWORKS [76], [77]). *Consider a point-to-point wireless communication link subject to the interference of $K$ neighboring nodes. Assume single-antenna nodes and that all channels are subject to Rayleigh fading. Then, the receive SINR can be written as*

$$\mathrm{SINR} = \frac{\bar{\gamma}_0 h_0}{\sum_{k=1}^K \bar{\gamma}_k h_k + 1}, \tag{22}$$

*where $\bar{\gamma}_0$ and $h_0 \sim \mathrm{Exp}(1)$ denote the average SNR of the received signal from the intended transmitter and the experienced normalized channel power gain, respectively. Similarly, $\bar{\gamma}_k$ and $h_k \sim \mathrm{Exp}(1)$ denote the average SNR of the* receive interfering signal from node $k$ and the corresponding normalized channel power gain, respectively.

*The CDF of the SINR, i.e., outage probability for a certain SINR threshold $\gamma_{\mathrm{th}}$, can be computed as follows*

$$F_{\mathrm{SINR}}(\gamma_{\mathrm{th}}) = 1 - \Pr[\mathrm{SINR} > \gamma_{\mathrm{th}}]$$

$$\overset{(a)}{=} 1 - \Pr\left[h_0 > \gamma_{\mathrm{th}}\Big(\sum_{k=1}^K \bar{\gamma}_k h_k + 1\Big)/\bar{\gamma}_0\right]$$

$$\overset{(b)}{=} 1 - \mathbb{E}\left[\exp\Big(-\gamma_{\mathrm{th}}\Big(\sum_{k=1}^K \bar{\gamma}_k h_k + 1\Big)/\bar{\gamma}_0\Big)\right]$$

$$\overset{(c)}{=} 1 - \exp(-\gamma_{\mathrm{th}}/\bar{\gamma}_0)\prod_{k=1}^K \mathbb{E}\left[\exp(-\gamma_{\mathrm{th}}h_k\bar{\gamma}_k/\bar{\gamma}_0)\right]$$

$$\overset{(d)}{=} 1 - \exp(-\gamma_{\mathrm{th}}/\bar{\gamma}_0)\prod_{k=1}^K \frac{1}{1 + \frac{\bar{\gamma}_k}{\bar{\gamma}_0}\gamma_{\mathrm{th}}}, \tag{23}$$

*where $(a)$ comes from using (22), $(b)$ from using the complementary CDF of $h_0$, $(c)$ from exploiting the property $\exp(\sum_i a_i) = \prod_i \exp(a_i)$, and $(d)$ after taking the expectation over every $h_k$. Observe that although (23) is in closed form, it is of little, if any, relevance for designing practical resource allocation mechanisms. For instance, let us focus on a power control mechanism and try to shed some light on the following problem: what is the average SNR $\bar{\gamma}_0$ required for satisfying $F_{\mathrm{SINR}}(\gamma_{\mathrm{th}}) \leq \xi$, where $\xi \ll 1$? It is important to note that neither the specific values of the average SNR of the interfering signals $\{\bar{\gamma}_k\}$ nor the number of interferers are generally known in practice. Nevertheless, observe that $\gamma_{\mathrm{th}}/\bar{\gamma}_0$ must be small to guarantee $\xi \ll 1$, thus, one may leverage the limiting forms in the seventh and first rows[8] of Table V to state*

$$\prod_{k=1}^K \left(1 + \frac{\bar{\gamma}_k}{\bar{\gamma}_0}\gamma_{\mathrm{th}}\right) \underset{\gamma_{\mathrm{th}}/\bar{\gamma}_0 \to 0}{=} \left[1 + \frac{\gamma_{\mathrm{th}}}{K\bar{\gamma}_0}\sum_{k=1}^K \bar{\gamma}_k\right]^K$$

$$= \left[1 + \frac{\gamma_{\mathrm{th}}}{K\bar{\gamma}_0}\bar{\gamma}\right]^K$$

$$\underset{\gamma_{\mathrm{th}}/\bar{\gamma}_0 \to 0}{=} \exp(\gamma_{\mathrm{th}}\bar{\gamma}/\bar{\gamma}_0), \tag{24}$$

*where $\bar{\gamma} = \sum_{k=1}^K \bar{\gamma}_k$. Substituting (24) into (23) yields*

$$F_{\mathrm{SINR}}(\gamma_{\mathrm{th}}) \underset{\gamma_{\mathrm{th}}/\bar{\gamma}_0 \to 0}{=} 1 - \exp\left(-\gamma_{\mathrm{th}}/\Upsilon\right), \tag{25}$$

*where $\Upsilon = \bar{\gamma}_0/(1+\bar{\gamma})$ is the average-signal-to-average-interference-plus-noise ratio.[9]*

*Note that the complexity can be further reduced, e.g., for diversity order analysis, by exploiting again the limiting form in the first row of Table V such that (25) converges to $\gamma_{\mathrm{th}}/\Upsilon$ as $\gamma_{\mathrm{th}}/\Upsilon \to 0$. Fig. 5 illustrates the incredible tightness of (25) with respect to the left tail of (23). This evinces that the outage probability in the URC regime for the*

---

[8]The limiting form in the seventh row comes from the inequality relating the geometric and arithmetic mean. Indeed, both means are equivalent when the terms are equal, which is what asymptotically happens when the terms are $\{1+x_i\}$ and each $x_i$ tends to zero. Finally, the limiting form in the first row is applied in the reverse direction for convenience.

[9]Strictly speaking, the right-hand term of (25) constitutes an upper bound for $F_{\mathrm{SINR}}(\gamma_{\mathrm{th}})$ because the arithmetic mean is always greater or equal than the geometric mean and $1 + x \leq \exp(x), \ \forall x \geq 0$, which are respectively used in the first and third line of (24).

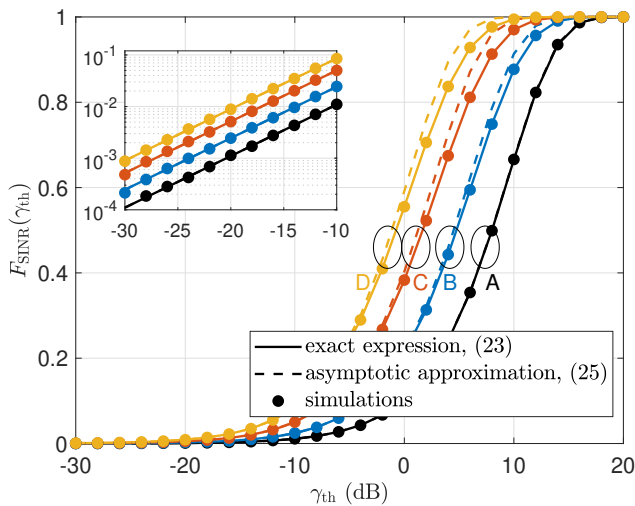

Fig. 5. CDF of the SINR for the system model of Example 4. We set $\bar{\gamma}_0 = 10$ dB, and consider four scenarios: A) $\bar{\gamma}_k = -10$ (dB) with $K = 1$; B) $\bar{\gamma}_k = -20 + 5k$ (dB) with $K = 4$; C) $\bar{\gamma}_k = -20 + 4k$ (dB) with $K = 6$; and D) $\bar{\gamma}_k = -21 + 3k$ (dB) with $K = 9$. Y-axis is illustrated in linear scale to be able to visualize the divergence of the bound (25) with respect to the exact (23) and simulation curves as the value of $\gamma_{\text{th}}$ increases.

considered scenario depends merely on the $\Upsilon$, whose value may be acquired in practice. Back to the original power control problem, $F_{\text{SINR}}(\gamma_{\text{th}}) \leq \xi$ with $\xi \ll 1$ is satisfied for $\bar{\gamma}_0 \geq -(1 + \bar{\gamma})\gamma_{\text{th}}/\ln(1-\xi)$ or simply $\bar{\gamma}_0 \geq (1 + \bar{\gamma})\gamma_{\text{th}}/\xi$.

When some information on the statistics of $X$ is known beforehand, one may exploit the following result.

**Theorem 1** (Markov's inequality [78]). *Let $X \in \mathbb{R}^+$ be a RV and $g : \mathbb{R}^+ \mapsto \mathbb{R}^+$ a non-decreasing function such that $\mathbb{E}[g(X)] < \infty$ exists, then*

$$\bar{F}_X(x) \leq \frac{\mathbb{E}[g(X)]}{g(x)}. \qquad (26)$$

Observe that (26) includes an infinite variety of bounds since there are infinite choices for $g$. The most popular bounds derived from (26) are illustrated in Table VI.

**Example 5** (Channel History-Based Precoding [79]). *Assume that a critical data message arriving at the data queue of a multi-antenna node A demands an urgent URLLC transmission to a single-antenna node B. Because of the stringent latency constraints, the transmission needs to be carried out immediately in one shot, thus instantaneous channel state information (CSI) cannot be acquired at A. However, in the past, A had kept a record of L CSI entries of the communication channel with B, which it may now exploit for the current URLLC transmission. The set of CSI entries is given by $\mathcal{L} = \{\mathbf{h}_1, \mathbf{h}_2, \ldots, \mathbf{h}_L\}$, where $\mathbf{h}_l \in \mathbb{C}^{M \times 1}$ is the channel vector coefficient corresponding to the $l-$th CSI entry, and M is the number of transmit antennas.*

*Assume an interference-free scenario, and denote by $\mathbf{h} \in \mathbb{C}^{M \times 1}$ the unknown channel coefficient for the URLLC transmission. Then, the corresponding SNR is given by*

$$\text{SNR} = \frac{|\mathbf{w}^H \mathbf{h}|^2}{N}, \qquad (27)$$

TABLE VI
Corollaries of Theorem 1

| Corollary | Transformation of (26) | Resulting Bound |
|---|---|---|
| Simple form of Markov's inequality | $g(t) = t$ | $\bar{F}_X(x) \leq \frac{\mathbb{E}[X]}{x}$ |
| Chebychev's bound | $g(t) = t$ & $X \leftrightarrow X - \mathbb{E}[X]$ | $\bar{F}_{|X-\mathbb{E}[X]|}(x) \leq \frac{\sqrt{\mathbb{V}[X]}}{x^2}$ |
| Chernoff's bound | $g(t) = e^{\theta t},\ \theta \geq 0$ | $\bar{F}_X(x) \leq C_1(x)$ |
| | | $= \inf_{\theta \geq 0} e^{-\theta x} \mathbb{E}_X[e^{\theta X}]$ |
| Moment bound | $g(t) = t^\theta,\ \theta \geq 0$ | $\bar{F}_X(x) \leq C_2(x)$ |
| | | $= \inf_{\theta \geq 0} \frac{1}{x^\theta} \int_0^\infty t^\theta f_X(t) \mathrm{d}t$ |

where $N$ is the receive noise power, and $\mathbf{w} \in \mathbb{C}^{M \times 1}$ is the transmit precoder. Here, we aim to attain the lowest-power precoder to guarantee a data reception's SNR of at least $\gamma_{\text{th}}$ with probability $1 - \xi$, where $\xi \ll 1$, thus $\min_{\mathbf{w}}\{\|\mathbf{w}\|^2 : \Pr[\text{SNR} < \gamma_{\text{th}}] \leq \xi]\}$. Observe that for a sufficiently large record of CSI entries, i.e., $L \to \infty$, the distribution of $\mathbf{h}$ (thus, the distribution of the SNR for a fixed $\mathbf{w}$) can be acquired, making the design of $\mathbf{w}$ relatively easy to optimize. However, in a practical setup, the number of channel samples $L$ may be rather limited due to limited training, memory and/or channel coherence time. Herein, we employ the simple form of Markov's inequality to state

$$\Pr[\text{SNR} < \gamma_{\text{th}}] = \Pr[\text{SNR}^{-1} > \gamma_{\text{th}}^{-1}] \leq \frac{\mathbb{E}[|\mathbf{w}^H \mathbf{h}|^{-2}]}{(N\gamma_{\text{th}})^{-1}} \leq \xi. \qquad (28)$$

Now, the best estimator for $\mathbb{E}[|\mathbf{w}^H \mathbf{h}|^{-2}]$ is the sample mean $\frac{1}{L} \sum_{l=1}^L |\mathbf{w}^H \mathbf{h}_l|^{-2}$. Then, the problem translates to $\min_{\mathbf{w}}\left\{\|\mathbf{w}\|^2 : \frac{N\gamma_{\text{th}}}{L} \sum_{l=1}^L |\mathbf{w}^H \mathbf{h}_l|^{-2} \leq \xi\right\}$, which is solved numerically to illustrate some performance results in the following.

Fig. 6 shows the truly attainable information outage as a function of the number $L$ of channel history samples for $\xi = 10^{-3}$ and quasi-static Rician fading channels with line-of-sight (LOS) factor of $-10$ dB. Herein, both the average and 50% confidence intervals for the information outage realizations considering randomly generated history sets are illustrated. Observe that the outage constraint may not be met for a relatively small number of samples. This is because the sample mean converges to the population mean only as $L \to \infty$. Thus, substituting the sample mean into (28) does not fully guarantee that the bound still holds, but only for large $L$. The situation may worsen if tighter bounds such as Chernoff's or moment bounds are used. In such cases, one might need to limit the possible range of values for parameter $\theta$ (see Table VI) to prevent excessive skewness of the distribution of $g(\text{SNR})$ that may significantly slow down the convergence of the sample mean to the population mean. Alternatively, and/or in addition, one might leverage a probabilistic upper bound of the sample mean to ensure (28) is met with a certain confidence level. Such bound can be designed exploiting the central limit theorem, as in [79]. In general, the confidence intervals narrow as $L$ increases.

Fig. 7 leverages $10^4$ randomly generated sets of $L = 32$ channel history samples to illustrate the empirical distribution of the outage probability (Fig. 7a) and transmit power

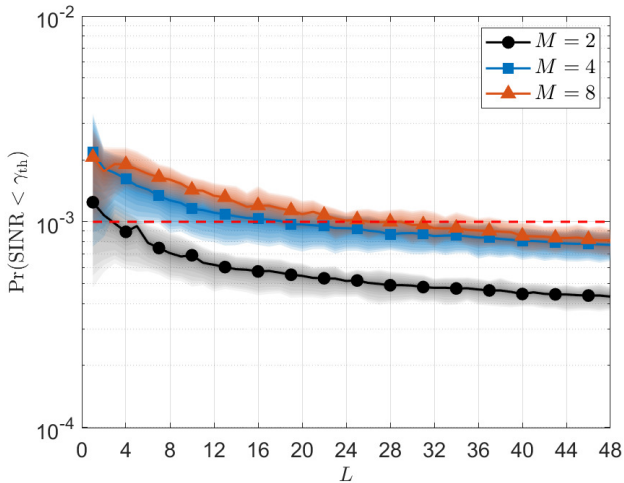

Fig. 6. Outage probability as a function of the number of history channel samples. The channel is subject to quasi-static Rician fading with LOS factor of $-10$ dB. We set $\gamma_{\text{th}} = 1$, $\xi = 10^{-3}$, and average channel pathloss to noise power ratio equal to 10 dB. Lines with markers represent the average over different realizations of the set $\mathscr{L}$ for each $L$, while the colored region corresponds to the 50% confidence interval.

*(Fig. 7b) for quasi-static Rician fading with different LOS factors. Results evince that the target outage probability $\xi = 10^{-3}$ is met regardless of the channel features, although less tightly as the LOS factor increases since the channel becomes more deterministic. Moreover, the transmit power requirements tend to decrease with the LOS factor.*

Theorem 1 has been extensively utilized in the literature for the analysis and design of wireless communication systems, e.g., [16], [27], [80], including URLLC. However, a fixed transformation function $g$ is always assumed, thus intrinsically discarding an optimization over $g$ for potentially tightening the bounds. For instance, let's say we have a function space of positive non-decreasing real functions $\mathscr{G}$, then (26) can be tightened as

$$\bar{F}_X(x) \leq \inf_{g \in \mathscr{G}} \frac{\mathbb{E}[g(X)]}{g(x)}. \tag{29}$$

In practice, the main challenge lies in having accurate estimates of $\mathbb{E}[g(X)]$ as these statistics converge at a different pace for different functions $g$ given limited data samples.

Finally, a selection of other less popular, but valuable, distribution bounds is compiled in Table VII.

### B. Extreme Value Theory

Estimating the extreme values of an underlying process, which are scarce by definition, inevitably requires extrapolation from available observations. This is facilitated by the EVT [81], which revolves around the statistical characterization of the RV

$$M_n = \max\{X_1, X_2, \cdots, X_n\} \text{ as } n \to \infty, \tag{30}$$

where $X_1, X_2, \cdots, X_n$ are i.i.d. RVs with unknown CDF $F_X(x)$.

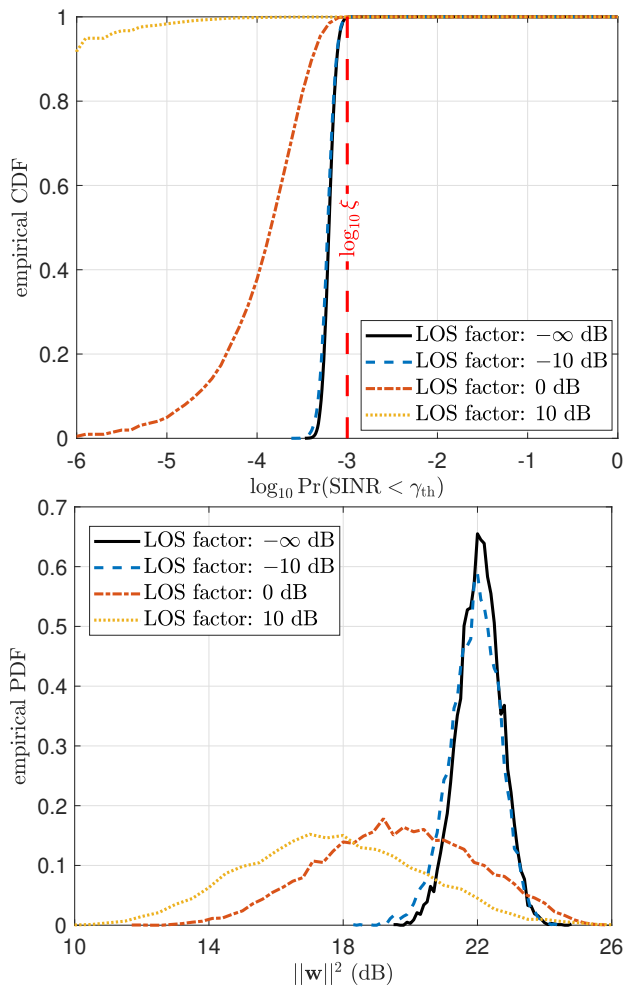

Fig. 7. Empirical distribution of the a) outage probability (top) and b) transmit power (bottom). The channel is subject to quasi-static Rician fading. We set $M = 4$, $L = 32$, $\gamma_{\text{th}} = 1$, $\xi = 10^{-3}$, and average channel pathloss to noise power ratio equal to 10 dB. We consider $10^4$ random sets of the channel history.

THEOREM 2 (*Extreme Value Distribution [81, Th.3.1.1]*). *If there is a sequence of constants $\{a_n > 0\}$ and $\{b_n\}$ such that*

$$\Pr\left[(M_n - b_n)/a_n \leq z\right] \to F_Z(z) \text{ as } n \to \infty \tag{31}$$

*for a non-degenerate distribution function $F_Z(z)$, then*

$$F_Z(z) = \exp\left(-\left[1 + \kappa\left(\frac{z - \mu}{\sigma}\right)\right]^{-1/\kappa}\right), \tag{32}$$

*defined on $\{z : 1 + \kappa(z - \mu)/\sigma > 0\}$, where $-\infty < \mu < \infty$ is the location parameter, $\sigma > 0$ is the scale parameter, and $-\infty < \kappa < \infty$ is the shape parameter.*

Theorem 2 states that the distribution convergence of suitably normalized maxima of RVs implies the convergence to $F_Z(z)$ in (32) for some $\mu, \sigma$, and $\kappa$. The apparent difficulty that $\{a_n\}$ and $\{b_n\}$ are unknown in practice is easily resolved as follows. Assume that $\Pr\left[(M_n - b_n)/a_n \leq z\right] \approx F_Z(z)$ for large enough $n$. Then, $\Pr\left[M_n \leq z\right] \approx F_Z((z - b_n)/a_n) = F_Z^*(z)$, which has also the form (32) but with potentially different parameters $\mu$ and $\sigma$. However, this is irrelevant since the parameters of the distribution must be estimated anyway.

| | Assumptions | Bound |
|---|---|---|
| Paley-Zygmund [82] | $X$ is non-negative, $0 \leq \theta \leq 1$ | $\tilde{F}_X(x) \geq \frac{(1-\theta)^2 x^2}{\theta^2 \mathbb{E}[X^2]}$ |
| Hoeffding [83] | $X_i \in [a_i, b_i], \forall i$ are i.i.d. RVs | $\tilde{F}_{\sum_{i=1}^{n} X_i}(x) \leq \exp\left(-\frac{2(x+\sum_{i=1}^{n} \mathbb{E}[X_i])^2}{\sum_{i=1}^{n} (b_i - a_i)^2}\right)$ |
| Vysochanskij–Petunin [84] | $X$ is unimodal | $\tilde{F}_{|X-\mathbb{E}[X]|}(x) \leq \frac{4V|X|}{9x^2}, \forall x \geq \sqrt{8V|X|/3}$ |
| Cantelli [85] | $X$ is real-valued | $\tilde{F}_X(x) \leq \frac{V|X|}{V|X|+(x-\mathbb{E}[X])^2}, \forall x > \mathbb{E}[X]$ |
| (tightened) Gauss (Dharmadhikari & Joag-Dev [86]) | $X$ is unimodal, $r > 0$, $s > r + 1$ and $s(s - r - 1) = r^r$ | $\tilde{F}_{|X|}(x) \leq \max\left\{\frac{r}{(r+1)x}\right)^r \mathbb{E}[|X|^r], \frac{s\mathbb{E}[|X|^r]}{(s-1)x^r} - \frac{1}{s-1}\right\}$ |
| Dvoretzky–Kiefer–Wolfowitz [87] | $X$ is non-negative | $\tilde{F}_{\sup_{y \in \mathbb{R}} (F_X^{(n)}(y) - F_X(y))}(x) \leq 2e^{-2nx^2}, \tilde{F}_{\sup_{y \in \mathbb{R}} (F_X^{(n)}(y) - F_X(y))}(x) \leq e^{-2nx^2} \ \forall x \geq \sqrt{\frac{\ln 2}{2n}}$ |

Finally, notice that Theorem 2 can be easily transformed to characterize the asymptotic distribution of the minima of RVs [81, Th.3.3], and the asymptotic distribution of the $k$−th ordered statistics [81, Th.3.5].

**THEOREM 3** (*Limiting distribution of excesses [81, Th.4.1]*). *For large enough $u$, the distribution of $X - u$ given $X > u$ and $X \in \{X_1, X_2, \cdots, X_n\}$, with* $\Pr[\max_{i=1,\dots,n} X_i < z] \approx F_z(z)$ *in* (32), *follows approximately a generalized Pareto distribution (GPD) such as*

$$F_{X-u|X>u}(y) \approx 1 - \left(1 + \frac{\kappa y}{\tilde{\sigma}}\right)^{-1/\kappa}, \tag{33}$$

*where $\tilde{\sigma} = \sigma + \kappa(u - \mu)$. Such distribution is defined on $\{y : y > 0, (1 + \kappa y/\tilde{\sigma}) > 0\}$.*

Although Theorem 2 and Theorem 3 assume an underlying process consisting of a sequence of independent RVs, their claims are still usually valid when such an assumption does not hold, as in the case of time-correlated data [81]. Meanwhile, in the case of Theorem 3, $u$ must be first selected to then proceed to fit $\kappa$ and $\tilde{\sigma}$, which is not straightforward. Indeed, extreme events may not be captured accurately if $u$ is too small, while, if too large, there might not be enough data for efficiently estimating the distribution parameters. A reasonable approach is to set $u$ as a certain quantile of the empirical distribution of the data, as in Example 6, but this may not always be appropriate depending on the data and the specific application. Some alternative selection methods have been developed, such as the mean residual life plot and the parameter stability method, which are applied respectively before and after the GPD fitting [81]. These methods are leveraged, for instance, in [88] to accurately model the behavior of extreme events in a URLLC wireless channel.

**EXAMPLE 6** (CHANNEL HISTORY -BASED PRECODING (II)).
*Let's exploit the EVT framework to solve the problem described in Example 5, i.e.,*

$$\min_{\mathbf{w}} \{\||\mathbf{w}\||_2^2, \Pr[\text{SNR} < \gamma_{\text{th}}] \leq \xi\}, \tag{34}$$

*where $\xi \ll 1$. Here, the difficulty lies in addressing the URLLC constraint, thus, we proceed as follows*

$$
\begin{aligned}
\Pr[\text{SNR} < \gamma_{\text{th}}] &= \Pr\left[N|\mathbf{w}^H \mathbf{h}|^{-2} > 1/\gamma_{\text{th}}\right] \overset{(a)}{=} \Pr\left[\Theta > f(1/\gamma_{\text{th}})\right] \\
&\overset{(b)}{=} \Pr\left[\Theta - u > f(1/\gamma_{\text{th}}) - u \mid \Theta > u\right] \Pr[\Theta > u] \\
&\overset{(c)}{\approx} \frac{1}{L}\left(1 - F_{\Theta-u|\Theta>u}\left(f(1/\gamma_{\text{th}}) - u\right)\right) \sum_{l=1}^{L} \mathbb{1}\{\Theta_l > u\} \\
&\overset{(d)}{\approx} \frac{1}{L}\left(1 + \frac{\kappa(\mathbf{w}, \mathscr{L})}{\tilde{\sigma}(\mathbf{w}, \mathscr{L})}\left(f\left(\frac{1}{\gamma_{\text{th}}}\right) - u(\mathbf{w}, \mathscr{L})\right)\right)^{-\frac{1}{\kappa(\mathbf{w}, \mathscr{L})}} \\
&\qquad \times \sum_{l=1}^{L} \mathbb{1}\{\Theta_l > u(\mathbf{w}, \mathscr{L})\},
\end{aligned}
\tag{35}
$$

*where $(a)$ comes from introducing the increasing function $f$ and defining $\Theta \triangleq f(N|\mathbf{w}^H \mathbf{h}|^{-2})$. Note that by adopting a concave $f$ one can mitigate the impact of potentially dispersed samples $\Theta_l \triangleq f(N|\mathbf{w}^H \mathbf{h}_l|^{-2})$ as a more compact sample set $\{f(\Theta_l)\}$ is obtained, thus, facilitating a better GPD fitting in the next steps. Thereafter, $(b)$ comes from exploiting the probability conditioned on exceeding a threshold $u$, which must be set relatively large but smaller than $1/\gamma_{\text{th}}$. Then, the empirical distribution of $\Theta$ (obtained from $\Theta_l$) is used in $(c)$ to estimate $\Pr[\Theta > u]$. Finally, Theorem 3 result is used to obtain $(d)$, where we highlight the dependency of $\kappa$, $\tilde{\sigma}$, and $\mu$ on the adopted beamformer $\mathbf{w}$ and channel sample set $\mathscr{L}$.*

*Then, one may use the following procedure to determine if a certain beamformer $\mathbf{w}$ satisfies the URLLC constraint:*

1) *Set $u(\mathbf{w}, \mathscr{L})$ as the quantile of $\{\Theta_l\}$ for the cumulative probability $\rho \in (0.5, 1)$. This leads to $\frac{1}{L}\sum_{l=1}^{L} \mathbb{1}\{\Theta_l > u(\mathbf{w}, \mathscr{L})\} = \rho$.*

2) *Fit the GPD* (33) *to the samples $\{\Theta \mid \Theta > u\}$ and obtain a $100 \times \alpha\%$ confidence intervals $[\kappa_{lb}, \kappa_{ub}], [\tilde{\sigma}_{lb}, \tilde{\sigma}_{ub}]$ respectively for $\kappa(\mathbf{w}, \mathscr{L}), \tilde{\sigma}(\mathbf{w}, \mathscr{L})$.*

3) *By exploiting* (35), *$\mathbf{w}$ surely satisfies the URLLC constraint if $\left(1 + \frac{\kappa(\mathbf{w}, \mathscr{L})}{\tilde{\sigma}(\mathbf{w}, \mathscr{L})}\left(\frac{1}{N\gamma_{\text{th}}} - u(\mathbf{w}, \mathscr{L})\right)\right)^{-1/\kappa(\mathbf{w}, \mathscr{L})} \leq \xi/\rho, \forall \kappa(\mathbf{w}, \mathscr{L}) \in [\kappa_{lb}, \kappa_{ub}], \tilde{\sigma}(\mathbf{w}, \mathscr{L}) \in [\tilde{\sigma}_{lb}, \tilde{\sigma}_{ub}]$.*

*Observe that $\rho$ and $\alpha$ are pre-defined heuristic parameters. The closer $\rho$ is to 1, the more extreme and right-tailed the data samples $\{\Theta|\Theta > u(\mathbf{w}, \mathscr{L})\}$, thus, the better the suitability of a GPD fitting. However, fewer samples are employed for the*

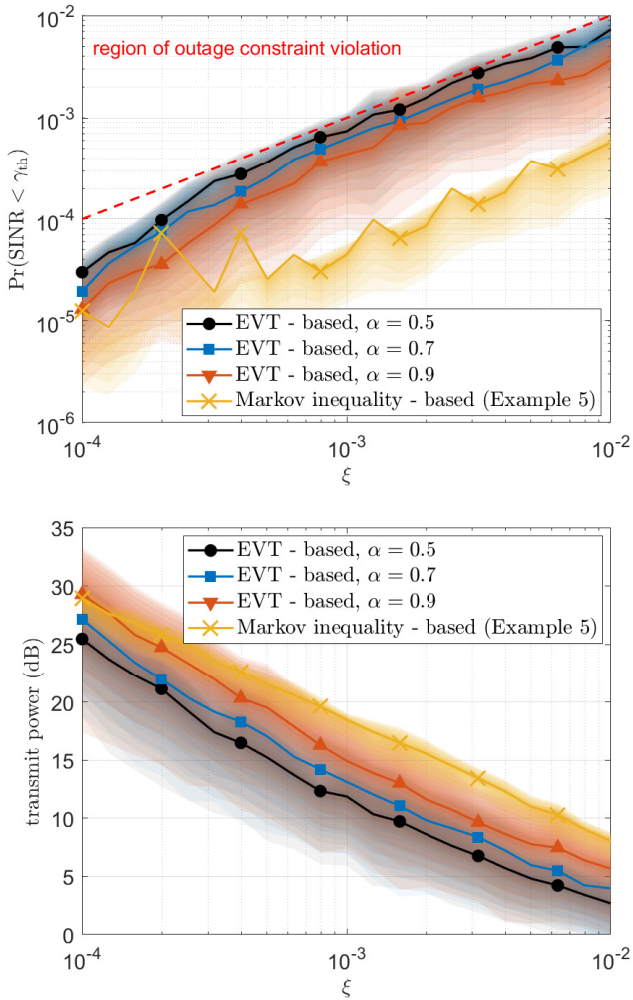

Fig. 8. a) Outage probability (top) and b) required transmit power (bottom) as a function of outage constraint $\xi$ for $f(x) = x$. Lines with markers represent the average over different realizations of the set $\mathcal{L}$ for each $L$, while the colored region corresponds to the $100 \times \alpha\%$ confidence interval. In the case of Markov inequality-based results, the colored region corresponds to the 50% confidence interval. Since satisfying increasingly stringent reliability requirements requires more extreme data samples in the EVT framework, here $L$ and $\rho$ are varied according to $\xi$ as $L = \lceil 1/\xi \rceil$ and $\rho = 1 - \sqrt{\xi}$. The channel is subject to quasi-static Rician fading with LOS factor of 0 dB, and we set $M = 8$, $\gamma_{\text{th}} = 1$, and average channel pathloss to noise power ratio equal to 10 dB.

*fitting, reducing accuracy. Meanwhile, the closer $\alpha$ is to one, the more flexible the GPD fitting becomes, but the chances of misdetecting an appropriate beamformer increase.*

*We resort to a brute force numerical optimization where the lowest-power beamformer satisfying the URLLC constraint, out of $10^5$ randomly generated beamformers, is selected. The performance resulting from employing the EVT framework is illustrated in Fig. 8 and compared to that of the Markov inequality -based solving framework used in Example 5. Observe that EVT-based solutions may be more energy efficient, especially by reducing GPD estimator confidence intervals (smaller $\alpha$), although this leads to greater chances of violating the reliability constraints.*

Another useful EVT result is the Pickands-Balkema-de Haan theorem [89, Th. 2.1.1], which states that

$$F_X(x) \to \alpha x^{\beta} \quad \text{as} \quad x \to 0, \quad \text{for some } \alpha, \beta > 0, \quad (36)$$

(or alternatively, $F_X(x) \to 1 - \nu x^{-\beta}$ as $x \to \infty$ when the interest is in the right tail) holds for a large class of distributions. This has been exploited in [4] for estimating channel tails followed by transmit rate adaptation with ultra-reliability guarantees. Notably, estimators of the tail index, i.e., $1/\beta$, include Pickands estimator [90], the moment estimator [91], and Hill's estimator [92]. Note that the asymptotic convergence of (36) tends to be slower compared to the Pareto approximation, thus although simpler, its accuracy is more limited, especially when exploiting limited data sets.

It is worth noting that the fitting confidence of the EVT parameters, e.g., $\kappa, \tilde{\sigma}$ in Theorem 3 and $\nu, \beta$ in (36), may also effectively measure the confidence in satisfying the URLLC performance requirements. Therefore, EVT allows quantifying the uncertainty associated with the estimation of rare event probabilities, thus supporting informed decisions on the performance/operation of URLLC systems.

Finally, two critical issues regarding the exploitation of EVT in URLLC problems are related to data availability/characteristics and fitting complexity. First, EVT-based solutions require the input data to capture some rare events, which might be challenging under stationary conditions[10]. Therefore, in the case of limited data sets, Markov-inequality -based solutions, which are based on averages, may be preferable. Second, EVT analysis is usually computationally intensive, which may hinder the support of real-time URLLC applications. The complexity comes mainly from the need for numerical methods for parameter fitting, as in the case of setting $u$ and estimating $\kappa, \tilde{\sigma}$ when exploiting (33) [81], especially when using data involving multiple sources of variability, i.e., high-dimensional data. This issue tilts the scale for sub-optimal low-complexity and/or ML optimization approaches.

## IV. RARE EVENT SIMULATION

In wireless communications, we are often confronted with noisy observations from which we would like to estimate some unknown distribution and/or its parameters, or the unknown parameters. Among the many solutions, constructing posterior distributions is normally adopted as it facilitates the incorporation of prior information into the observed data. However, closed-form solutions for such problems are often infeasible, leading to the development of approximate inference techniques, such as Monte Carlo (MC) methods [93], [94].

MC methods comprise a large class of simulation-based algorithms optimized to perform random drawings from a target probability distribution.[11] These methods can be used *i)* to generate random samples from a probabilistic model, or *ii)* to perform numerical integration, or *iii)* in optimization. Notably, the focus in URLLC is on accurately estimating/simulating the occurrence of rare events.

---

[10]EVT results are built on the assumption of stationary data, however, non-stationarity can arise in URLLC applications due to changes in the communication environment, such as interference or congestion, especially if data sets are relatively large.

In the remainder of this section, we overview classical MC methods and assess their limitations in terms of computation time and accuracy. We then overview alternative solutions dedicated to rare-event sampling and provide examples of their efficiency in terms of computation time, complexity, and accuracy.

### A. Conventional MC

To introduce the MC method, let us define

$$p = \Pr[S(x_i) \geq x_{\text{th}}] = \int_{\mathbf{X}} f_{\mathbf{X}}(x) \mathbb{I}\{S(x_i) \geq x_{\text{th}}\}\, dx$$
$$= \mathbb{E}_{\mathbf{X}} [\mathbb{I}\{S(x_i) \geq x_{\text{th}}\}], \tag{37}$$

where $\mathbf{X}$ is a $d \times N$ random matrix, being $d$ the dimensions and $N$ the number of samples, while $S(\mathbf{x}) : \mathbb{R}^d \to \mathbb{R}$ is a real-valued function. Moreover, $x_{\text{th}}$ is a threshold parameter and determines the rarity of the event. In other words, this parameter indicates how far into the tail of the distribution we are attempting to estimate the probability $p$.

The conventional MC (CMC) method (also known as naive or crude MC) is a convenient way to estimate $p$, i.e., $\hat{p} = \mathbb{E}_X [\mathbb{I}\{S(x_i) \geq x_{\text{th}}\}]$. The idea behind CMC is to estimate the probability $\hat{p}$ using many independently generated samples of $X = [X_1, \cdots, X_N]^T$ per dimension. In this case, the unbiased estimator of choice is the sample mean, i.e., $\hat{p} = \frac{1}{N}\sum_{i=1}^{N} \mathbb{I}\{S(x_i) \geq x_{\text{th}}\}$; in other words, the ratio between the samples that belong to the event $\{S(x_i) \geq x_{\text{th}}\}$ and the total number of samples. Therefore, CMC relies on the law of large numbers and the central limit theorem to increase the accuracy of the estimation, as the number of samples grows to infinity, the sample mean converges to the expectation, i.e.,

$$\mathbb{E}[\hat{p}] = \lim_{N \to \infty} \frac{1}{N}\sum_{i=1}^{N} \mathbb{I}\{S(x_i) \geq x_{\text{th}}\}$$
$$= \frac{1}{N}\sum_{i=1}^{N} \mathbb{E}_X [\mathbb{I}\{S(x_i) \geq x_{\text{th}}\}] = p. \tag{38}$$

We resort to the coefficient of variation, $\delta(\hat{p})$ as a measure of the accuracy of CMC, and therefore as $\delta(\hat{p}) \to 0$ the more accurate is the estimation. We define this measure as $\delta(\hat{p}) = \frac{\sqrt{\mathbb{V}[\hat{p}]}}{\mathbb{E}[\hat{p}]}$, which is the ratio of the standard deviation and expected value of the estimate $\hat{p}$. Following the same steps as in (38) we calculate the estimate's variance

$$\mathbb{V}[\hat{p}] = \mathbb{V}_X \left[ \frac{1}{N}\sum_{i=1}^{N} \mathbb{I}\{S(x_i) \geq x_{\text{th}}\} \right]$$
$$\overset{(a)}{=} \frac{1}{N^2}\sum_{i=1}^{N} \left( \mathbb{E}_X [\mathbb{I}\{S(x_i) \geq x_{\text{th}}\}^2] - p^2 \right)$$
$$\overset{(b)}{=} \frac{1}{N^2}\sum_{i=1}^{N} \left( p - p^2 \right) = \frac{p(1-p)}{N}, \tag{39}$$

where (*a*) expands the definition of the variance, and (*b*) uses the identity $\mathbb{I}\{\cdot\}^2 = \mathbb{I}\{\cdot\}$ and (38). Now, we can calculate the coefficient of variance as

$$\delta(\hat{p}) = \sqrt{\frac{1-p}{Np}} \approx \frac{1}{\sqrt{Np}}, \tag{40}$$

where the last step holds accurate for $p \ll 1$. The latter corresponds to the probability of a rare event, which is of particular interest in URLLC setups.

One advantage of CMC is that it does not depend on the dimension of the input state for the accuracy measure. Note that the accuracy varies with the number of samples and the target probability; therefore, the lower the desired estimate, the higher the number of samples required. Let us exemplify this point with the 3GPP target probability for URLLC of $p = 10^{-5}$. We expect a low variability, therefore, a coefficient of variation of 1% is appropriate. Then, replacing these values into (40), the required number of samples for such an estimate is $N = 10^9$, which is costly with conventional simulation tools. In addition, for every sample, there are additional computations required, e.g., $\mathbb{I}\{\cdot\}$ and $S(\cdot)$, which may increase the computational cost and render the URLLC analysis infeasible with CMC.

Notably, CMC is particularly useful when the target probability is not far from the mean of the underlying distribution, otherwise, it may be very inefficient. The latter is the case of URLLC scenarios, where the target probability is at the distribution's tail (refer to Section III). Therefore, robust MC methods are needed to evaluate URLLC with limited costs (e.g., computation and time). Indeed, we are interested in computationally affordable methods that allow evaluating reliability levels in the order of five or more nines, as in (extreme) URLLC use-cases [2], [97], by effectively sampling the region of the distribution's domain where such rare events occur. Fortunately, several methods can help in this task.

### B. Importance Sampling

Importance sampling (IS) is perhaps one of the most common MC methods for rare event simulation, which is a variance reduction technique in the estimation of (37). In general, for a fixed degree of relative error, IS achieves sufficient variance reduction to decrease total computational effort by several orders of magnitude compared to that of CMC. This is achieved because IS draws samples from another distribution where the original rare events happen more often. Precisely, assume a PDF $g_Y$ such that if $g_Y(x) = 0$ then $\mathbb{I}\{S(x_i) \geq x_{\text{th}}\} f(x) = 0 \; \forall x$; thus, we can re-write the expectation in (37) as

$$\mathbb{E}_X\left[\mathbb{I}\{S(x_i) \geq x_{\text{th}}\}\right] = \int \mathbb{I}\{S(x_i) \geq x_{\text{th}}\} \frac{f_X(x)}{g_Y(x)} g_Y(x) dx$$

$$= \mathbb{E}_Y\left[\mathbb{I}\{S(x_i) \geq x_{\text{th}}\} \frac{f_X(X)}{g_Y(X)}\right]. \quad (41)$$

---

[11] Notice that there are alternative sampling solutions to MC methods. For instance, the class of variational inference, particularly variational Bayesian, methods emerges as a computationally efficient alternative to sampling methods. We refer the reader to [95], [96] for a tutorial introduction to variational inference.

Notice that one must carefully select the PDF such that the variance of $X$ is reduced. For this, the PDF must have *i)* finite variance, i.e.,

$$\mathbb{E}_X\left[\mathbb{I}\{S(x_i) \geq x_{\text{th}}\}^2 \frac{f_X^2(X)}{g_Y^2(X)}\right] = \mathbb{E}_Y\left[\mathbb{I}\{S(x_i) \geq x_{\text{th}}\}^2 \frac{f_X^2(X)}{g_Y^2(X)}\right] < \infty,$$

and *ii)* heavier tails than $f_X$ and the bounded likelihood ratio $\frac{f_X}{g_Y}$. Notably, a poor selection of the PDF compromises the quality of the estimate [98].

Observe that the optimum PDF is the one that minimizes the variance of the estimate, i.e.,

$$\arg\min_{g_Y} \mathbb{V}_{g_Y}\left[\mathbb{I}\{S(x_i) \geq x_{\text{th}}\} \frac{f(X)}{g(X)}\right], \quad (42)$$

whose solution is the zero-variance IS density [98]

$$\pi(x) = \frac{\mathbb{I}\{S(x_i) \geq x_{\text{th}}\} f(x)}{l}. \quad (43)$$

However, (43) depends on the unknown estimate $l$, thus it cannot be directly used. Instead, we approximate the $\pi$ and the IS density $g_Y$, e.g., by resorting to differential entropy and by allowing $g_Y$ to be in product form $g_Y(x) = \prod_{i=1}^{d} \pi_i(x_i)$, which usually simplifies the analysis.

Though relevant, IS is particularly useful for light-tailed distributions, e.g., where the rare-event probability decays exponentially. Moreover, the selection of the desired density can be optimized, e.g., with the cross-entropy method [98], which minimizes the divergence between the density $g_Y$ and a parametric family of probability densities instead of a single function. This approach allows for the iterative update of the target density, facilitating effective sampling.[12]

## C. Markov-chain MC Algorithms

Current MC methods are based on the initial works of Stanislaw Ulam, John von Neumann, and others, during the Manhattan Project in the 1940s. One of the key contributions was the Metropolis algorithm, which relies on a Markov chain for the generation of a desired distribution based on symmetric proposal densities. After improvements to accept asymmetric proposal densities by Hastings, the algorithm is now known as Metropolis-Hastings (MH) [93]. The Markov-chain MC algorithms rely on a basic idea: run a sufficiently long Markov chain such that its stationary distribution approximates an arbitrary distribution, known as target distribution, from which we can generate samples. In other words, suppose our target distribution is $f_X(x)$, as in (37), then construct a Markov chain with states $\{X_t, t = 0, 1, 2, \cdots\}$, where $X_t$ is a given state at time $t$ and the stationary probability distribution is $f_X(x)$. The MH's

---

**Algorithm 1:** MH Algorithm [98]

**Input:** $X_0$, $N$, $f_X(x)$, $q_Y(y|x)$
1  Initialization: $t = 0$
2  **repeat**
3      Draw $Y \sim q_Y(y|X_t)$
4      Calculate $\alpha(X_t, Y)$ in (44)
5      Draw $U \sim \mathcal{U}(0, 1)$ (uniform distribution)
6      $X_{t+1} = \begin{cases} Y, & \text{if } U \leq \alpha(X_t, Y) \\ X_t, & \text{otherwise} \end{cases}$
7      Increase $t$
8  **until** *N samples were generated, i.e. $t = N$;*
**Output:** $X_1, \cdots, X_N$

---

core idea lies in calculating the transition from one state to the next. To do so, we first draw a proposal state, e.g., $Y$ from a transition density $q_Y(\cdot|X_t)$. Then, the proposal state is accepted based on the acceptance probability

$$\alpha(x, y) = \min\left\{\frac{f_X(y) q_Y(x|y)}{f_X(x) q_Y(y|x)}, 1\right\}. \quad (44)$$

Otherwise, the proposed state is rejected and the chain remains in the current state. This can be observed in Algorithm 1 lines 4-6 when the proposal is accepted or rejected compared to a draw form a uniform distribution. Notice that the transition density satisfies the balanced detailed equations, i.e., $f_X(x) q_Y(y|x) = f_X(y) q_Y(y|y)$. Therefore, $f_X(x)$ is the stationary distribution of the chain. Notably, if the event $\{X_{t+1} = X_t\}$ has a positive probability, i.e., $\Pr[\alpha(X_t, Y) < 1 | X_t] > 0$ and $q_Y(x|y) > 0 \, \forall x, y$ in the domain, then the Markov chain is ergodic and its limiting distribution is $f_X(x)$ [98].

Note that computing the acceptance probability $\alpha(x, y)$ requires knowing the distribution $f_X(x)$ only up to a normalizing constant, i.e., $f_X(x) = c\tilde{f}(x)$ where $\tilde{f}(x)$ is a known function while $c > 0$ is an unknown constant. Moreover, the efficiency of MH heavily depends on the choice of the transition density, which ideally should be close to the target function $f_X(y)$ regardless of $x$. In such circumstances, the proposal function can be independent of $x$, thus $q_Y(y|x) = g_Y(y)$ for some distribution $g_Y(y)$. Consequently, (44) reduces to $\alpha(x, y) = \min\left\{\frac{f_X(y) g_Y(x)}{f_X(x) g_Y(y)}, 1\right\}$, and this method is known as an independence sampler. However, notice that the generated samples are dependent and the proposal function should be close to the target function. Moreover, if $f_X(x) \leq c \, g_Y(x) \forall x$, then, the acceptance rate in Algorithm 1 Step 6 is at least $1/c$. A simple alternative is to consider a symmetric proposal function, i.e., $q_Y(y|x) = q_Y(x|y)$, hence $\alpha(x, y) = \min\left\{\frac{\tilde{f}(y)}{\tilde{f}(x)}, 1\right\}$.

A particular case of MH is the Gibbs sampler, commonly used for $n$-dimensional random vectors. In this case, the conditional distribution governs the construction of the Markov chain, thus, Gibbs sampler is helpful when sampling this distribution is less computationally expensive than the joint distribution. In addition, the Gibbs sampler distribution, $f_X$, is the stationary distribution of the Markov chain and has a geometrically fast convergence [98].

---

[12]Please, see [98] for a detailed discussion on alternative algorithms.

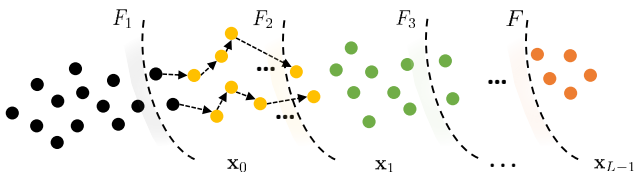

Fig. 9. Schematic of the SS Algorithm. First, a small number of samples are generated from a target distribution (black circles). These samples are sorted and after selecting the samples closer to the desired domain (boundaries between subsets in gray) Markov-chain MC is applied using those samples as seeds. For example, see region $F_2$, notice that the black circles are seeds, and the arrows indicate the Markov-chain MC sampling; yellow circles are new samples from the conditional distribution. Then, the process repeats until the last domain $L$.

---

**Algorithm 2:** SS Algorithm [99]

**Input:** $N$, $d$, $S(\cdot)$, $q$

1  Initialization: $l = 0$
2  Draw the first subset:
   $x_0 = x_0^1, \cdots, x_0^n \approx \pi(\cdot), n < N, N = n^d$
3  **repeat**
4     Calculate the response $y_l = S(x_l)$
5     Order $x_l$ based on the response $y_l = S(x_l)$
6     Define the intermediate failure domain
      $F_l = \{x : S(x_i) > y_l^*\}$
7     Draw a sample from $\pi(\cdot|F_l)$ using Markov-chain
      MC and $nq$ samples from $x_l$ as seeds
8     Increment $l$
9  **until** $\frac{n(l+1)}{n} < p$;
   **Output:** $\Pr[F] \approx q^L \frac{n(L)}{n}$

---

## D. Subset Simulation

With origins in structural reliability theory, subset simulation (SS) is a powerful and efficient method for the simulation of rare events and the estimation of tail probabilities. Moreover, SS decomposes the rare event into a sequence of nested events such that the most frequent events are supersets of the less frequent ones. [99]. In other words, let the rare event $F$ be decomposed into a sequence of $L$ progressively rarer events from $F_1$ to $F_L$, i.e., $F = F_L \subset F_{L-1} \subset \cdots \subset F_1$, where $F_1$ is a frequent event. Therefore, the probability of the rare event $\Pr[F]$ follows the chain rule as

$$\Pr[F] = \Pr[F_L] = \Pr[F_1]\Pr[F_2|F_1]\cdots\Pr[F_L|F_{L-1}]. \quad (45)$$

However, in many applications, it is cumbersome to decompose the events and obtain closed-form or tractable expressions for the conditional probabilities. Thus, to overcome this issue, it relies on Markov-chain MC methods to adaptively obtain the conditional probabilities, and consequently the intermediate events.

In a nutshell, the SS algorithm generates a small number of samples to explore the input space, thus, generating a rough approximation of the failure domain $F$. This process is repeated iteratively such that the domain approximations converge to the domain $F$, and sufficient samples are generated in $F$ to estimate the rare event, as illustrated in Fig. 9. Notice that some samples serve as seeds for the Markov-chain MC algorithm at the upper level $L$, thus, drawing from $\pi(\cdot|F_L)$. These steps are summarized in Algorithm 2. Notice that any Markov-chain MC algorithm can be used to sample from the conditional distribution. The particular choice depends on the scenario, e.g., the difficulty of sampling from the conditional distribution.

Based on [99], we describe the key points of the SS algorithm. For the first domain, SS generates the initial sample set comprising the realizations $x_0 = \left[ x_0^1, \cdots, x_0^n \right]^T \sim \pi(x)$, $n < N$, $N = n^d$, as shown in line 2 of Algorithm 2. Then, the SS algorithm computes the response $y_0 = S(x_0)$ in line 4. As the sample size is small and we want to estimate a rare event, it is likely that the samples $x_0$ do not belong to $F$. However, $x_0$ contains relevant information about $F$ since by ordering $y_0$ we can identify which are the samples that closest to the target set. As in (37), we can define the first intermediate failure domain, i.e., $F_1 = x : S(x_i) > y_1^*$, which is the set of samples that lead toward the rare event and are ordered in line 5. By construction, $x_0^1, \cdots, x_0^{nq}$ belong to $F_1$ and $x_0^{np+1}, \cdots, x_0^n$ do not, which is obtained by the intermediate failure domain $y_1^* = (y_0^{nq} + y_0^{nq+1})/2$ (line 6), where $q \in (0, 1)$ is an arbitrary input parameter such that $nq$ is a natural number. Consequently, the probability estimate of $F_1$ is $\Pr[F_1] = q$, which can be estimated via CMC since $F_1$ is by design not rare and the number of samples, $n$, is small. Moreover, we can calculate the rare-event probability as in (45), which for the first domain reduces to $q\Pr[F|F_1]$. Therefore, $\Pr[F|F_1]$ using a Markov-chain MC method (line 7), e.g., a modified MH or Gibbs sampling [98], [99]. Once we generate the samples based on $\pi(x|F_1)$, we apply the same steps and determine the next domain. This process is repeated iteratively until the stopping criterion is met (see [99] for details on the stopping criteria and an alternative implementation of the algorithm). By design, the SS algorithm controls the probability of the intermediate steps, therefore, (45) approximates to $q^L\Pr[F|F_l]$ where $l \leq L$ indicates the number of domains, $\Pr[F|F_l] = n(L)/n$, and $n(L)$ is the number of samples in the last domain. Thus, the output of Algorithm 2 line 9 is the rare-event probability, which is given as

$$\Pr[F] \approx q^L \frac{n(L)}{n}. \quad (46)$$

A key advantage of the SS algorithm is that by design one can control input variable $q$, which influences the number of chains, the number of samples in each chain, and the stopping criteria. Therefore, with proper selection of $q$, the SS algorithm efficiently samples the conditional distributions to (approximately) match this probability ensuring that each factor in (45) is larger than $q$. Such an iterative process yields an accurate estimate of the rare event even for a small sample size.

**Example 7.** *Multi-connectivity is a promising technique for URLLC as it allows the devices to connect to two or more*

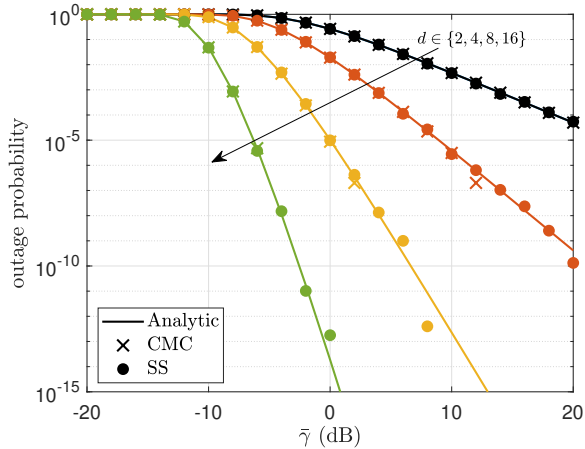

Fig. 10. Comparison between CMC and SS algorithms in terms of outage probability against the average SNR $\bar{\gamma}$. The closed-form result is shown as the baseline. We set $r = 1$ bpcu and consider different numbers of connected base stations, i.e., $d \in \{2, 4, 8, 16\}$.

base stations.[13] Herein, we assume a simple system model where a device connects to $d \in \{2, 4, 8, 16\}$ base stations. The channel coefficients follow a Rayleigh distribution and we are interested in coherently combining these signals via maximal ratio combining. Then, the SNR of each link $X$ follows an exponential distribution[14], i.e., $X \sim \text{Exp}(1)$ and the SNR of the combined received signal can be described as $Y = \sum_{i=1}^{d} X_i$. For this particular case, the distribution of the sum is known in closed-form, i.e., $Y \sim \text{G}(d, 1)$, which helps the comparison between CMC and SS algorithms here. However, in many cases, we are unable to express the target distribution in closed form due to the complexity of the model, or intractability of the functions that lead to intricate integrals that require numerical analysis. This is one more example of the usefulness of MC methods in practice.

We are interested in the outage probability, i.e.,

$$P_{\text{out}} = \Pr\left[\log_2(1 + \bar{\gamma}Y) < r\right] = \Pr\left[Y < (2^r - 1)/\bar{\gamma}\right],$$

where $r$ represents a target rate and $\bar{\gamma}$ the average SNR. In Fig. 10, we compare the analytical expressions to the CMC and SS algorithms as a function of $\bar{\gamma}$ for different numbers of base stations (dimensions) and $r = 1$ bits per channel use (bpcu). For CMC, we draw $N = 10^7$ samples for each dimension $d$, while for the SS algorithm, we draw $n = 10^4$ samples and we assume $q = 0.1$. We implement Algorithm 2 using the modified MH algorithm, as in Algorithm 1.

As discussed above, CMC is inefficient as the $\bar{\gamma}$ grows, thus, as the events become rare, and therefore, it can only reliably estimate the probabilities as $P_{\text{out}} > 10^{-6}$. Increasing the sample size, $N \gg 10^7$ becomes computationally expensive. On the other hand, the SS algorithm uses a small sample size for each domain yielding a faster simulation

---

[13]See for instance [19] for a comprehensive overview and [6], [21], [100] for related works and applications.

[14]For simplicity of presentation, we assume all links have the same mean, which in practice presumes that the nodes are equidistant from the base station, or the devices employ some path-loss compensation.

---

compared to CMC. The number of samples grows with the increase in $\bar{\gamma}$. For instance, for $d = 4$, $\bar{\gamma} = 20$ dB, SS requires approximately 91000 samples to estimate an event with probability around $4 \times 10^{-10}$ (see Fig. 10). Moreover, it can estimate the probabilities for regions below $10^{-10}$. This is particularly relevant for the analysis of performance metrics in extreme URLLC cases. Notice that the SS algorithm can be optimized to reduce the variance for extreme events, e.g. by selecting a different target distribution or different input parameters such as the number of samples or $q$.

### E. Complexity of the MC Methods and Availability

There are many other MC methods in the literature beyond those discussed so far. Currently, the most efficient general-purpose samplers are based on the Hamiltonian MC (HMC) method [101]. The core idea lies in exploiting Hamiltonian dynamics to define a dynamical system that simulates the potential energy landscape of a target distribution. Hamiltonian dynamics preserve the energy and volume of the system by using a suitable symplectic solver, which is an integrator designed to solve Hamilton's equations [101]. By sampling from the simulated system, HMC can explore and generate samples from complex probability distributions.

HMC introduces an intricate proposal distribution to MH, which allows the efficient simulation of very long trajectories with a high acceptance probability. HMC performs well for unimodal distributions and requires tuning the integrator's accuracy and number of steps. Notably, the larger the number of steps, the longer the paths for the Markov-chain MC chain. However, large steps may cause the trajectories to turn back and eventually return to start, which leads to a long simulation time. Dynamic HMC algorithms emerged to solve this issue.

The No-U-Turn Sampler (NUTS) [102] is the first and most popular dynamic HMC algorithm. NUTS avoids revisiting local spaces, significantly reducing simulation time compared to HMC due to efficient exploration of the sampling space. Moreover, NUTS iteratively extends (forward and backward) a simulated trajectory until it observes a U-turn, i.e., the path turns back, which guarantees reversibility. It also samples the next point among a set of candidates generated along the path rather than a single accept/reject choice. The NUTS algorithm is highly efficient and effective, especially for high-dimensional problems. It often outperforms other Markov-chain MC algorithms.

Table VIII compares the different MC and Markov-chain MC algorithms in terms of complexity and availability in Matlab and Python. We assume that $N$ is the total number of samples (though the total number of samples for each method may differ), and $L$ is the total number of levels for the SS algorithm. We highlight that many of these algorithms are natively implemented in Matlab [103], [104] and/or Python[15] [107] or via a Matlab dedicated

---

[15]Note that modern ML packages such as PyTorch and Tensorflow have dedicated probabilistic programming libraries comprising Markov-chain MC algorithms, e.g., NumPyro [105] and TensorFlow Probability [106].

| Algorithm | Complexity | Matlab | Python |
|---|---|---|---|
| CMC | $\mathcal{O}(N)$ | ✔ | ✔ |
| IS | $\mathcal{O}(N)$ | ✔ | ✔ |
| MH | $\mathcal{O}(N)$ | ✔ mhsample [103] | ✔ pymc [107] |
| SS | $\mathcal{O}(LN)$ | ✔ SS Toolbox [108] | – |
| HMC | $\mathcal{O}(N^{\frac{5}{4}})$ | ✔ hmcSampler [104] | ✔ pymc [107] |
| NUTS | $\mathcal{O}(N^{\frac{5}{4}})$ | ✔ | ✔ pymc [107] |

✔(✔) refers to native (application-specific) implementations in Matlab/Phyton.

toolbox [108]. Notably, all these libraries are extensively documented and carefully maintained. Meanwhile, others, although not native, can be implemented in Matlab or Python but are often related to a specific application.

The complexity of MH is $\mathcal{O}(N)$ due to evaluating the proposal distribution at each iteration. Likewise, the complexity of IS is $\mathcal{O}(N)$, as it draws $N$ samples from an importance distribution and evaluates the proposal distribution for each sample. The complexity of SS is $\mathcal{O}(LN)$ because the algorithm requires evaluating the posterior distribution for each level. In most cases, each evaluation has a complexity $\mathcal{O}(N)$. However, SS can be much more efficient than other algorithms when the posterior distribution is sparse, meaning it has a few dominant modes [99], [109].

For the HMC, the leapfrog algorithm is often used to simulate Hamiltonian dynamics as it takes small steps through phase space. The number of these steps depends on the curvature of the posterior distribution. As the number of dimensions in the target distribution increases, the curvature increases, and more steps are required for accurate simulation [101]. In general, the leapfrog algorithm has complexity $\mathcal{O}(N^{\frac{1}{4}})$, which leads to a total complexity $\mathcal{O}(N^{\frac{5}{4}})$. Furthermore, NUTS explores the target distribution more efficiently than HMC by using a tree-depth parameter to dynamically adjust the number of leapfrog steps. In addition to the leapfrog algorithm, each step requires the evaluation of the gradient of the log posterior distribution and the computation of the Hamiltonian function, each of which has time complexity $\mathcal{O}(N)$. Therefore, the total complexity is the same as the HMC, $\mathcal{O}(N^{\frac{5}{4}})$.

## V. FINITE BLOCK LENGTH CODING

Errors during transmission in a wireless system can occur due to a number of reasons, for example, deep fades or outages due to strong interference. One way to recover such errors is to use error correction codes, where redundant information is added to the transmitted message to facilitate error detection and/or correction. This requires additional decoding algorithms at the receiver end to recover the original message from the transmitted encoded information. The first error-correcting code was the Hamming (7,4) code invented in 1950 [110]. Error-correcting codes are usually classified into convolutional codes and block codes. Convolutional codes work on a bit-by-bit basis as a sliding window of the bits. A $(k_c, k_d)$ block code acts on a block of $k_d$ information bits of input data to produce $k_c(k_c > k_d)$ bits of coded output data, also known as a codeword symbol. The ratio $r_c \triangleq k_d / k_c$ is known as the coding rate.

### A. Finite Block Length Theory

The capacity of a wireless channel dictates the maximum data rate that can be transmitted with asymptotically small error rates in the absence of any complexity or delay constraints in the coding and encoding process. For example, the well-known Shannon capacity of additive white Gaussian noise (AWGN) channels assumes that the channel's mutual information is maximized over all possible input distributions and the block lengths are infinite. Notably, encoding over a large number of bits increases the diversity order of the code and hence provides better error resilience at the expense of added complexity and delay. Moreover, the payload in the novel URLLC and massive MTC service classes introduced in 5G are usually small, while the low-latency constraint in URLLC implies that messages have to be transmitted as soon as they arrive and cannot afford to be buffered for long in a queue. Alongside this, the massive contention to access the channel and/or strict energy limitations that are common in massive MTC impose further constraints. Altogether, many URLLC and massive MTC use cases do not allow coding over a large block length and hence there is a need to evaluate the performance of a wireless system transmitting FBL data.

We next present Polyanskiy, Poor, and Verdú's *FBL theory framework* [111] to obtain tight bounds on the maximum transmission rate $r(N, \epsilon)$ as a function of the block length $N$ and the error probability $\epsilon$.

THEOREM 4 (MAXIMUM FBL TRANSMISSION RATE [111]).
*For a given channel with capacity $C$, the difference between the channel capacity and the maximum transmission rate in the FBL regime is a function of the block length $N$, the error probability $\epsilon$, and the channel dispersion $V$. The latter is a measure of the stochastic variability of the channel relative to a deterministic channel with the same capacity. In concrete terms,*

$$r(N, \epsilon) = C - \sqrt{\frac{V}{N}} Q^{-1}(\epsilon) + \mathcal{O}\left(\frac{\log N}{N}\right), \qquad (47)$$

*where $Q^{-1}(\cdot)$ is the inverse of the Gaussian Q function.*

From (47), one can obtain the asymptotic result $\lim_{N \to \infty} r(N, \epsilon) = C$. Meanwhile, notice that for an AWGN channel with SNR $\gamma$, the capacity matches the Shannon rate given by $C(\gamma) = \log_2(1 + \gamma)$ in bpcu, and the dispersion is given by [111]

$$V(\gamma) = \gamma \frac{2 + \gamma}{(1 + \gamma)^2} \log_2^2 e. \qquad (48)$$

Moreover, it has been shown in [14], [111] that $\frac{1}{2N} \log_2 N$ can be used to capture the residual terms on the right-hand side of (47) with sufficient accuracy.

Fig. 11 leverages (47) and (48) to illustrate the maximum transmission rate achievable in AWGN channels as a function of the error probability $\epsilon$ for SNR values $\gamma \in \{10, 20\}$ dB, and block lengths $N \in \{100, 1000\}$. The transmission rate is normalized by the corresponding AWGN capacity in the infinite block length regime. We can observe that the penalty incurred with FBL transmissions is higher for shorter block lengths, lower SNRs, and tighter error probability targets, as typical in the case of URLLC transmissions.

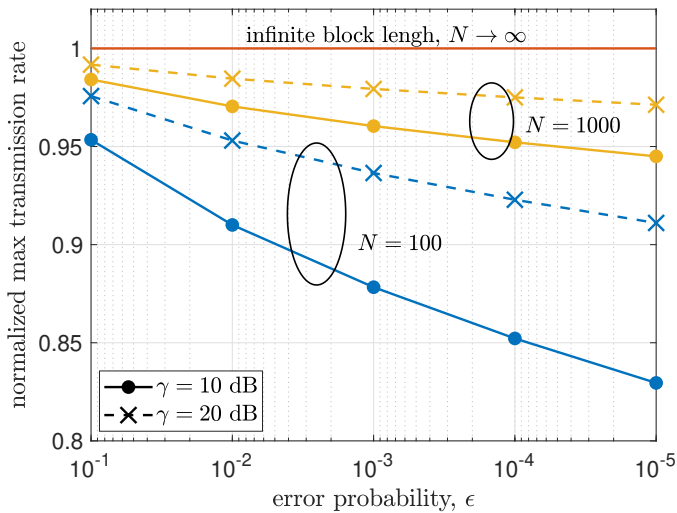

Fig. 11. The normalized maximum transmission rate with FBL transmission for different values of SNR $\gamma$, block length $N$, and target error probability $\epsilon$. The contribution of the residual terms on the right-hand side of (47) is modeled as $\frac{1}{2N}\log_2 N$ [14], [111].

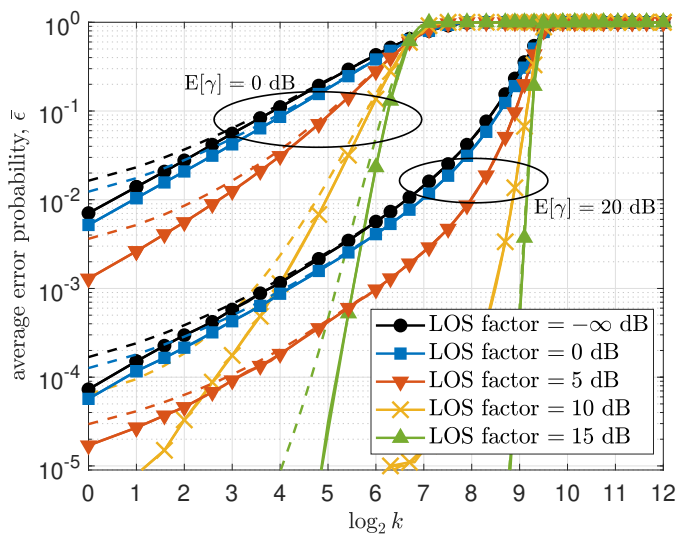

Fig. 12. Average error probability $\bar{\epsilon}$ as a function of $\log_2 k$, where $k$ is the number of transmit bits, for $N = 100$ channel uses and Rician block fading channels with different LOS factors. The dashed lines represent the asymptotic outage probability.

Meanwhile, one can obtain the achievable error probability given a fixed transmission rate $r = k/N$ from (47) as

$$\epsilon(\gamma) \approx Q\left(\frac{C(\gamma) - k/N}{\sqrt{V(\gamma)/N}}\right), \tag{49}$$

for which we have ignored the residual terms of order $\frac{1}{N}\log N$ in (47) for simplicity. Notice that the FBL error probability is an RV in fading scenarios, for which authors in [75] provide accurate PDF analytical approximations.

The average error probability for block fading channels is given by $\bar{\epsilon} = \mathbb{E}_\gamma[\epsilon(\gamma)]$.[16] Interestingly, it has been shown in [14], [112], [113] that the effect of the fading on (49) vanishes the FBL impact in two cases, when *i)* $k$ is not extremely small, and *ii)* there is not a strong LOS component (e.g., as in Rayleigh fading). In such cases, the asymptotic outage probability is a good approximation for $\bar{\epsilon}$, i.e., $\bar{\epsilon} = \Pr[\gamma < 2^{k/N} - 1] = F_\gamma(2^{k/N} - 1)$, as shown in Fig. 12. Unfortunately, this is not yet completely understood by the scientific community, which often adopts $\mathbb{E}_\gamma[\epsilon(\gamma)]$ in situations where $F_\gamma(2^{k/N} - 1)$ is very accurate and much easier to evaluate.

The basic result for the AWGN channel is extended to MIMO quasi-static Rayleigh fading channels in [114]. FBL coding results for MAC are derived in [115], while second-order asymptotic results are derived in [116] for the Gaussian MAC with degraded message sets. Other multiuser channels such as the asymmetric broadcast channel are studied in terms of dispersion expressions in [117]. FBL results for other fading channel models are studied in [118].

Next, we present an application example of the above complete framework and some follow-up discussions.

**Example 8** (Error-Constrained FBL Power Control).
*Assume a node A wants to transmit a data message of $k$ bits through $N$ channel uses to a node B. For this, A has already acquired the CSI of the communication link. Specifically, A perfectly knows the value of $\gamma'$, which is defined as the receive SNR at B given a normalized transmit power. What is the minimum transmission power that A must set to ensure a decoding error probability $\epsilon$ at B?*

*In this case, both the transmission rate $r = k/N$ and the error probability $\epsilon$ are fixed. Also, given a transmit power $p$, the receive SNR at B becomes $\gamma = p\gamma'$. Therefore, the problem reduces to finding the required SNR $\gamma$ such that (49) (or alternatively (47)) holds. It has been shown in [119] that the required $\gamma$ is the fixed point solution of*

$$\gamma^{(t+1)} = 2^{k/N + \sqrt{V(\gamma^{(t)})}Q^{-1}(\epsilon)/\sqrt{N}} - 1, \tag{50}$$

*where $t$ is the iteration index. Here, one may use $V(\gamma^{(0)}) = \log_2^2 e$ for initialization, which comes from letting $\gamma^{(0)} \to \infty$. The corresponding iterative procedure was shown in [119] to converge fast, i.e., in no more than five iterations for $N \le 1000$ with an accuracy superior to 99.9%. Finally, after solving (50), the required transmit power is given by $p = \gamma/\gamma'$.*

In the previous example, the solution of (50) becomes $\gamma = 2^{k/N} - 1$ when $N \to \infty$ (and also $k \to \infty$ such that $k/N = r$ is fixed). This agrees with Shannon's framework in which the error probability becomes arbitrarily small when the received SNR exceeds $2^{k/N} - 1$. Here, an interesting question is: *how much greater must the SNR at an FBL $N$ be in comparison with the asymptotic result?* The answer comes by defining and simplifying the quotient between

---

[16]As shown in [75], statistically different risks may be experienced even on services with similar performance in terms of average reliability.

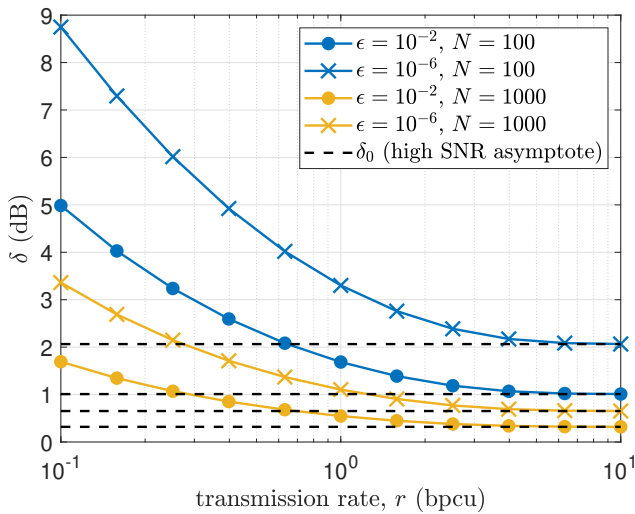

Fig. 13. $\delta$ and $\delta_0$ as a function of $r$ for $N \in \{100, 1000\}$ channel uses and $\epsilon \in \{10^{-2}, 10^{-6}\}$.

the required SNR at FBL and infinite block length as [119]

$$\delta \triangleq \frac{2^{r+\sqrt{V(\gamma^{(\infty)})}Q^{-1}(\epsilon)/\sqrt{N}} - 1}{2^r - 1}$$

$$= 2^{\sqrt{V(\gamma^{(\infty)})}Q^{-1}(\epsilon)/\sqrt{N}} + \frac{2^{\sqrt{V(\gamma^{(\infty)})}Q^{-1}(\epsilon)/\sqrt{N}} - 1}{2^r - 1}. \quad (51)$$

Interestingly $\delta$ is a decreasing function of $r$ [119], thus

$$\delta \geq \delta_0 = \lim_{r \to \infty} \delta = e^{Q^{-1}(\epsilon)/\sqrt{N}}, \quad (52)$$

which exploits the fact that $r \to \infty$ implies $\gamma \to \infty$, for which $V(\gamma) \to \log_2^2 e$. Fig. 13 corroborates this behavior and already evinces the convergence of $\delta$ to $\delta_0$ for $r \approx 4$ bpcu. Notably, the main remark from (52) and Fig. 13 is that the SNR has to be at least $e^{Q^{-1}(\epsilon)/\sqrt{N}}$ times greater than the asymptotic threshold of $2^r - 1$ to reach an error probability no lower than $\epsilon$ while transmitting the information through $N$ channel uses. Obviously, this criterion is also applied to the transmit power.

### B. Encoding of data and metadata in URLLC transmissions

Most wireless systems group the information bits to be transmitted into transmission blocks whose duration is sufficiently short compared to the channel coherence time. Thus, all bits in the block experience the same channel condition. The bits in a transmission block are of two types, metadata, i.e., the control information needed to decode the packet, and the intended data message.

In conventional mobile broadband (MBB) transmissions, the modulation and coding scheme of the data message within each transmitted block is selected to meet a target block error rate (BLER). It has been shown that for a large range of system parameters, a 10% BLER leads to near-optimal performance [23]. In contrast, the metadata is transmitted with a much lower BLER target, which means that the contribution of the metadata in the final error probability calculations is negligible.

| Transmission type | metadata considered | | metadata neglected | |
|---|---|---|---|---|
| | $n = 0$ | $n = 1$ | $n = 0$ | $n = 1$ |
| MBB | 0.9 | 0.99 | 0.899 | 0.989 |
| URLLC | 0.999 | 0.999999 | 0.998 | 0.999 |

This is no longer the case for URLLC, where both metadata and the data have to be transmitted with ultra-low BLER targets. Instead, the final success probability is the product of the probabilities of successfully decoding both metadata and the data, i.e.,

$$P_s = (1 - P_{e,m})(1 - P_{e,d}), \quad (53)$$

where $P_{e,m}$ and $P_{e,d}$ are the error probabilities for the metadata and the data, respectively. The success probability can be further improved via retransmissions in the event of a failure [23]. In this case, the error probability of the feedback link should not be ignored. The success probability $P_{s,n}$ after the $n$−th ($n > 1$) retransmission is given by

$$P_{s,n} = P_{s,n-1} + \underbrace{(1 - P_{e,m})P_{e,d}}_{\substack{\text{metadata decoded,} \\ \text{data failed}}} \underbrace{(1 - P_{e,f})}_{\text{feedback success}}$$
$$\times \underbrace{(1 - P_{e,m})(1 - P_{e,d})}_{\text{success after retransmission}}, \quad (54)$$

where $P_{e,f}$ is the error probability of the feedback link. Eq. (54) indicates that feedback is only possible if the metadata is successfully decoded, allowing the receiver to establish the transmitter and the packet identity. It is worth mentioning here that this only holds under the assumption of independence between subsequent transmission slots. Such assumptions may not hold under very stringent latency constraints where the retransmission occurs within the channel coherence time [120].

Let us illustrate the importance of considering the metadata error probability in the success probability calculation through an example.

EXAMPLE 9 (IMPACT OF THE METADATA ERROR PROBABILITY). *Consider two services: i) MBB transmissions characterized by $\{P_{e,m} = P_{e,f} = 10^{-3}, P_{e,d} = 0.1\}$, and ii) URLLC transmissions characterized by $\{P_{e,m} = P_{e,f} = P_{e,d} = 10^{-3}\}$. The success probabilities up to the first retransmission ($n = 1$) are shown in Table IX for the case when the error probability of the metadata is considered (the middle column) and compared against the case when the metadata is ignored (the right column). We can observe that the difference between the two cases (considering vs. neglecting the metadata error probability) is almost negligible for conventional MBB transmissions, whereas it is several orders of magnitude for URLLC transmissions. This highlights the importance of considering the metadata error probability when analyzing URLLC transmissions.*

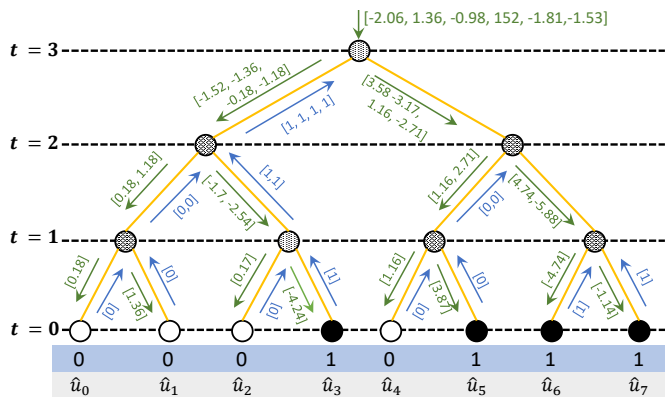

Fig. 14. Successive cancellation decoding of an (8,4) polar code over a binary phase shift keying (BPSK) modulated transmission; white and black circles represent frozen and information bits, respectively.

## C. Polar codes for FBL transmissions

We have seen in Section V-A that using FBL in URLLC transmissions reduces the maximum achievable rate. One of the reasons for this is the reduced coding gain due to the reduced block length. Hence there is a need to study channel codes that perform well in the FBL regime. The channel code selected by 3GPP for MBB data transmissions, namely low-density parity-check codes, are not particularly suitable for FBL scenarios [13]. Recently, 3GPP has adopted polar codes for short blocks of metadata in MBB transmissions. Current investigations also demonstrated that polar codes outperform low-density parity-check codes in short block lengths and low code rates, therefore they might be appealing for URLLC use cases [13].

Polar codes date back to 2009 with Arikan's work in [121] and are the first codes with an explicit construction to achieve the channel capacity using low-complexity encoding and decoding. Specifically, low-complexity successive cancellation and belief propagation decoding are used at the receiver side. A polar code $\mathscr{P}(k_c, k_d)$ is a code sequence containing a codeword symbol of $k_c$ bits carrying $k_d$ information bits. Channel polarization is proposed in [121] to construct code sequences achieving the symmetric channel capacity of the binary-input discrete memoryless channel.

Channel polarization refers to the fact that it is possible to synthesize the channel into a set of $k_c$ binary-input channels such that, as $k_c$ becomes large, a fraction of the channels becomes reliable (i.e., the symmetric capacity approaches 1) while the remaining fraction becomes unreliable (i.e., the symmetric capacity approaches 0). Thus, the information bits can be transmitted at rate 1 through the $k_d$ most reliable channels, while "dummy" frozen bits (usually zero) are transmitted through the unreliable channels. Reliable transmission can be achieved if $k_d$ is less than equal to the number of reliable polarized channels.

### 1) Encoding of polar: codes

The transformation matrix for a polar code of length $k_c$ is constructed through a $\log_2 k_c$– Kronecker product of the polarization kernel

$$\mathscr{G} = \begin{bmatrix} 1 & 0 \\ 1 & 1 \end{bmatrix}.$$

As $k_c \to \infty$, this construction creates channels that are either perfectly noiseless or completely noisy. For smaller values of $k_c$, the synthetic channels polarization may be incomplete, generating intermediary channels that are only partially noisy.

### 2) Decoding of polar: codes

The successive cancellation algorithm used to decode polar codes can be represented as a depth-first binary tree search with priority to the left branch, which has a decoding complexity of $\mathscr{O}(k_c \log_2 k_c)$ [121], [122]. The leaf nodes are the $k_c$ bits to be estimated. The soft information on the received code bits is input at the root node. Fig. 14 shows the decoding tree of a (8,4) polar code where bits numbered 1,2,3,5 are frozen. The black leaf nodes represent information bits and the white ones are the frozen bits.

Each node at stage $t_p$ uses the $2^{t_p}$–length soft input vector $\alpha_{t_p}$, received from its parent node, to calculate the soft output $\alpha_{t_p-1}^l$ and forward it to its left child. The $i$–th element of $\alpha_{t_p-1}^l$ is given by $\alpha_{t_p-1,i}^l = f(\alpha_{t_p,i}, \alpha_{t_p,i+2^{t_p-1}})$, with $f(a,b) = \text{sgn}(a)\text{sgn}(b)\min(|a|,|b|)$. The soft output $\alpha_{t_p-1}^r$ to the right child is calculated by combining $\alpha_{t_p}$ with the $2^{t_p-1}$–length hard decision $\beta_{t_p-1}^l$ received from its left child, where $\alpha_{t_p-1,i}^r = g(\alpha_{t_p,i}, \alpha_{t_p,i+2^{t_p-1}}, \beta_{t_p-1,i}^l)$ with $g(a,b,c) = b+(1-2c)a$. The final hard decision vector is then obtained as $\beta_{t_p} = [\beta_{t_p-1}^l \oplus \beta_{t_p-1}^r, \beta_{t_p-1}^r]$. Lastly, the hard decision at the $i$–th leaf node is given by $\hat{u}_i = \beta_{0,i} = (1 - \text{sgn}(\alpha_{0,i}))/2$; while frozen bits are always decoded as zero.

The BLER of polar codes for different codeword sizes ($k_c$), data size ($k_d$), and different coding rates ($k_d/k_c$) are illustrated in Fig. 15. The codeword and the data sizes are chosen to cover a range of coding rates, from $\sim 0.24$ to a value close to one. The BLER performance results indicate that polar codes can deliver very low BLERs in the FBL regime with block lengths in the order of $100-200$ bits. The performance is compared against the ideal FBL error rates obtained by Eq. (49). We observe that the performance gap is small with robust coding whereas a larger gap is seen for coding rates close to one. It is worth highlighting that even lower BLERs can be achieved through retransmissions (cf. Section V-B), albeit at the cost of higher latency.

The recursive nature of the successive cancellation decoding algorithm may impose a large latency and low error-correction performance in the FBL regime [122]. Different approaches have therefore been proposed to enhance the coding efficiency and reduce the latency for URLLC applications. For example, the binary tree can be pruned to decrease the number of calculations, thereby reducing the decoding latency and complexity [123]. Altering the polar code construction is another way to further reduce the latency of polar codes, though this induces a trade-off with the error-correction performance [124]. This can be addressed by taking advantage of the high degree of parallelism to provide a favorable tradeoff between throughput and energy efficiency at short to medium block length [125], or through a more efficient memory utilization [126].

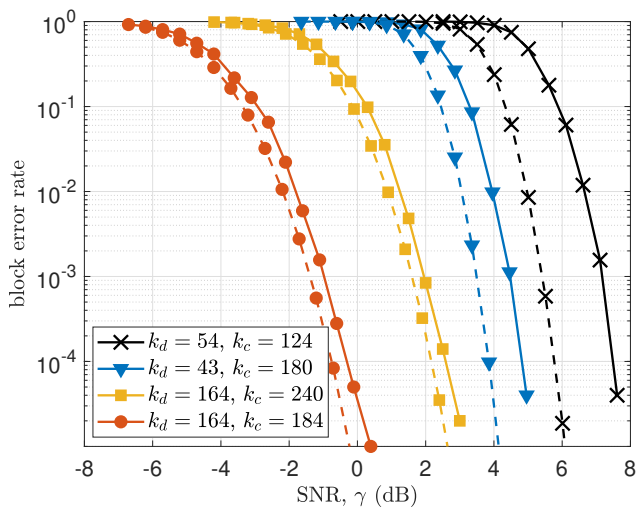

Fig. 15. BLER performance of Polar codes in the downlink for different code rates and message lengths and using quadrature phase shift keying modulation. The dashed lines show the error rate with FBL theory, as given by Eq. (49).

### D. GRAND: a generic maximum likelihood decoder

Recently, [127] and [128] proposed a universal maximum likelihood decoding algorithm for discrete channels with random codebooks, short code-lengths, and *high code-rates* called guessing random additive noise decoding or GRAND for short. We briefly introduce it here as it may meet fundamental URLLC decoding requirements such as low latency, outstanding error-rate performance, and low decoding complexity [129]. The first two features facilitate the efficient decoding of short blocklength transmissions, while the third feature enables the support of URLLC services on low-capability devices.

The basic principle of the GRAND algorithm is as follows. Let $X^N, Y^N$, and $Z^N$ be RVs representing the coded transmitted message, received message, and random noise of block length $N$ over a discrete channel. The arbitrary codebook $\mathscr{C}_N$ used to generate the encoded message $X^N$ is commonly known to the transmitter and the receiver. Assume that the input-output relation of the channel is reversible such that noise can be subtracted from the received message to recover the transmitted message, i.e.,

$$Y^N = X^N + Z^N \quad \rightarrow \quad X^N = Y^N - Z^N. \quad (55)$$

For a given received message $y^N$, the receiver orders the possible noise sequences from the most likely to the least likely and then queries one by one whether the sequence $y^N - z_p^N$ is an element of the codebook $\mathscr{C}_N$. Here $z_p^N$ is the $p$−th ordered noise sequence. For the channel structure described above, irrespective of how the codebook is constructed, the first instance such that $x^N = y^N - z_p^N \in \mathscr{C}_N$ corresponds to the maximum likelihood decoding.

GRAND operates by attempting to identify the random noise that has corrupted the transmitted message and hence is independent of the code structure. It is inherently highly parallelizable, resulting in the low latency desired in URLLC and other similar applications [130]. GRAND's accu-racy depends on the ability to query the noise sequences, whose probability distribution depends on the channel statistics. For instance, in a binary symmetric channel, the relationship among the binary output symbol $y^N$, the binary input codeword $x^N$, and the independent binary additive channel noise $z^N$ is given in (55).

## VI. Queuing Theory & Information Freshness

URLLC is about taming the tail distributions of reliability and latency (see Section III) while considering the inherent trade-offs associated with energy consumption, throughput, and data freshness. All this makes the design/analysis of URLLC systems extremely challenging in practice and motivates the exploitation of queuing theory metrics/tools such as effective capacity, AoI, and SNC, which can natively capture several of these trade-offs and constitute the scope of our discussions here. But before delving further into the section, it is worth highlighting that URLLC messages are inherently small (see Section V) and are mostly transmitted promptly without waiting for a long queue, which must be considered when exploiting the queuing theory metrics/tools and performing related analysis.

### A. Effective capacity

Metrics such as effective capacity and effective bandwidth capture tail statistical delay requirements in parallel with transmission throughput. Unlike Shannon capacity which considers only the transmission rate, effective capacity and effective bandwidth account for the queuing aspect which results in packet delivery delay. Specifically, the effective capacity is defined as the highest arrival rate that can be served by the network under a particular delay constraint [131]. Conversely, its dual twin, the effective bandwidth, characterizes the minimum service rate required to support the arrival of data in a certain network subject to a QoS constraint [132]. Here, we focus on the foundations of effective capacity, which can be easily extended to effective bandwidth as well.

To begin with, the effective capacity, denoted as $C_e$, is a metric that captures the physical (comprising coding, modulation, and transmission) and link layers characteristics in terms of specific delay QoS guarantees, thus, allowing a further investigation of the latency-reliability trade-off. For low-latency communication, $C_e$ is a powerful metric that characterizes the relation between the communication rate and the tail distribution of the packet delay violation probability [12].

A statistical delay violation model implies that for a relatively large delay, an outage occurs when a packet delay $\delta$ exceeds a maximum delay bound $\delta_{\max}$, and its probability is defined as [133]

$$P_{\text{out}} = \Pr[\delta \geq \delta_{\max}] \approx e^{-\theta C_e \delta_{\max}}. \quad (56)$$

Let us define $Q$ as the stationary queue length, then $\theta$ in (56), known as delay exponent or QoS exponent, is the decay rate of the tail distribution of $Q$, i.e.,

$$\theta = -\lim_{q \to \infty} \frac{\log \Pr[Q \geq q]}{q}, \quad (57)$$

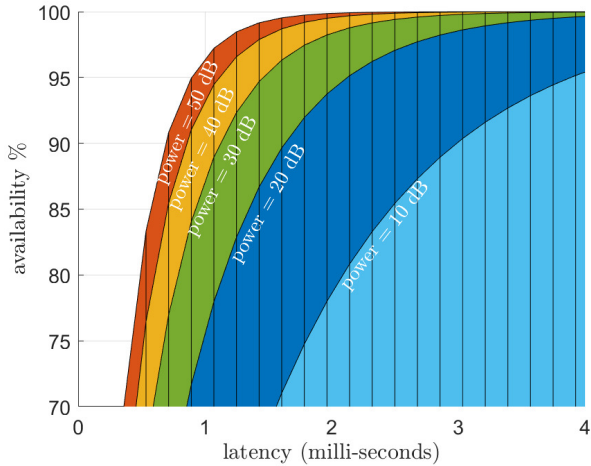

Fig. 16. Trade-off between availability, latency, and power consumption for 5G New Radio numerology 1 in Rayleigh fading. The latency is computed as $\delta_{\max} \times 35.7 \ \mu s$.

and quantifies the system's tolerance to long delays. The system tolerates large delays for small values of $\theta$ (i.e., $\theta \to 0$) while it becomes more delay-sensitive for large values of $\theta$. Therefore, $\theta$ must be intrinsically small in URLLC systems, and thus one may need to leverage the methods discussed in Section III for estimating/bounding it from queue information data over time in practical systems.

Notably, the packet delay is lower bounded by the packet transmission time. This translates to $\delta \geq N$ considering the time expressed in channel uses. Notably, the effective capacity and effective bandwidth are adequate metrics/tools for scenarios where packets span several, e.g., at least 100, symbols [133], [134]. The following examples facilitate a practical understanding of the effective capacity, the parameter $\theta$, and the corresponding trade-off between latency and availability in a real communication scenario. Notice that the system's availability (or reliability in case of no repairs) is given by $A = 1 - P_{\text{out}}$.

EXAMPLE 10. *Consider 5G New Radio numerology 1 transmission with $C_e = 1$ bpcu, for which the symbol period is 35.7 μs. For a delay outage probability of $P_{\text{out}} = 10^{-5}$ (i.e., 99.999% availability), the network can tolerate, on average, a maximum delay of $\delta = 1151$ symbol periods ($\approx 41$ ms) for $\theta = 0.01$, and $\delta = 115$ symbol periods ($\approx 4.1$ ms) when $\theta = 0.1$.*

EXAMPLE 11. *Consider a Rayleigh block fading channel with block length $N$. Then, for a transmission rate $r$ in bpcu and channel fading coefficient $|h|^2$, we can write*

$$C_e(r, \theta) = -\frac{1}{\theta N} \ln \left( \mathbb{E}_{|h|^2} \left[ e^{-\theta N r} \right] \right) \tag{58}$$

*given in bpcu. In the case of fixed-rate transmission with no CSI availability at the transmitter, a transmission outage may occur if the transmission rate is higher than the channel capacity. For the Rayleigh fading channel, the mean outage*

probability for a transmission SNR $\rho$ is given by[17]

$$\epsilon = \Pr[\log_2(1 + \rho z) < r] = 1 - e^{-\frac{2^r - 1}{\rho}}. \tag{59}$$

*Herein, the transmission might be successful with probability $1 - \epsilon$. In case of transmission failure, the transmission rate is effectively zero and the average effective capacity is given by*

$$C_e(r, \theta) = -\frac{1}{\theta N} \ln \left( \epsilon + (1 - \epsilon) e^{-\theta N r} \right). \tag{60}$$

*Fig. 16 depicts the trade-off between availability, latency, and power consumption from the effective capacity perspective for 5G New Radio with a symbol period of 35.7 μs (numerology 1) and transmission rate of 0.5 bpcu. The maximum arrival rate is set to $C_e = 0.2$ bpcu. The figure shows a contour plot for the achievable availability-latency regions for different transmission power ranges. This reveals the contradicting achievability of URLLC transmission where achieving both high availability and low latency together requires high power consumption. We also observe that there is a dead zone of extremely low latency (below 1.5 ms) and high availability (>99%) that can not be achieved at the same time for this setup except at extremely high transmit power. For more details about how Fig. 16 was generated, one can refer to Algorithm 3.*

Notice that the role of the communication bandwidth is not explored in the previous example. Nevertheless, it is shown in [6] that this is often a relevant degree of freedom for jointly improving reliability and latency figures.

### B. Stochastic Network Calculus

SNC is a framework that enables the end-to-end analysis of networks by means of non-asymptotic performance bounds. SNC builds upon the theory of network calculus, which analyzes performance guarantees of queuing systems via convolutional forms based on (min, +) dioid algebra. Network calculus enables the derivation of work-case performance bounds such as backlog and delay. In the deterministic case, network calculus models the arrival and service process as envelope functions, and thus, it does not capture the stochasticity of the arrival and service processes. SNC addresses this issue by relaxing the deterministic assumptions introducing envelope violation probability [135], [136]. Compared to conventional queuing theory, SNC is broader and includes several stochastic processes such as long-range dependent, self-similar, and heavy-tailed traffic. In addition, it can incorporate the variability of wireless fading channels.

Fidler and Rizk provide a comprehensive guide to SNC in [135]. Commonly, SNC resorts to envelope functions of the MGF of the arrival and service processes. Nonetheless, the framework is robust and supports general envelope models and models that provide strong guarantees. For instance,

---

[17]Recall from our discussions in Section V-A, and specifically around Fig. 12, that the asymptotic outage probability approaches accurately the average error probability for Rayleigh block fading channels as long as $r$ is not significantly small. For better accuracy at relatively small $r$, one may need to substitute (59) by $\mathbb{E}_z[\epsilon(\rho z)]$, where $\epsilon(\gamma)$ is given in (49) and $k = rN$.

---

**Algorithm 3:** Solving the statistical interplay between availability, latency, and power consumption

**Input:** $C_e$, $N$, and $r$.

1 **for** *every $\rho$ and $\delta_{max}$* **do**
2      Compute $\epsilon$ using (59)
3      Solve (60) for $\theta^*$
4      Compute $A = 1 - P_{\text{out}}$ using (56)
5 **end**

**Output:** $\{A, \delta_{max}, \rho\}$ triplets

---

statistical delay analysis in fading channels is introduced [136] using a new representation based on the Mellin transform, and a (min, ×) calculus, leading to tractable closed-form results that incorporate the channel variability to the model, yielding non-asymptotic performance bounds of the fading and arrival processes [136]. Due to its properties, SNC, and its simplified variants, effective bandwidth and effective capacity, are suitable for modeling end-to-end performance in URLLC (e.g., see [137]–[141]).

Next, we focus on the fundamental components of SNC, namely the arrival process (cumulative number of bits arriving) between times $\tau$ and $t$, $t \geq \tau \geq 0$, $A^{\circ}(\tau, t)$, the service process (e.g., dynamic server), $S^{\circ}(\tau, t)$, and the departure process, $D^{\circ}(\tau, t)$. We can relate these quantities as

$$D^{\circ}(t) \geq \min_{\tau \in [0, t]} \left\{ A^{\circ}(\tau) + S^{\circ}(\tau, t) \right\}, \tag{61}$$

where $A^{\circ}(0) = A^{\circ}(t, t) = 0$ and $S^{\circ}(0) = S^{\circ}(t, t) = 0$, $\forall t \geq 0$, and both arrival and service functions are non-negative. In addition, assuming a first-come first-served approach, the delay bound at time $t \geq 0$ is

$$W(t) \leq \min \left\{ w \geq 0 : \max_{\tau \in [0, t]} \left\{ A^{\circ}(\tau) - S^{\circ}(\tau, t + w) \right\} \leq 0 \right\}. \tag{62}$$

For other arrival-service approaches see [135].

To calculate the actual delay bounds, we resort to *i)* affine traffic envelope functions, introduced in deterministic network calculus, and denoted as $\varrho(t-\tau) + b$, where $\varrho$ is a rate parameter related to the traffic and $b$ is the burst parameter; and *ii)* MGFs, from effective bandwidth theory, since it determines the distribution of a random process, and the sum of two or more random processes can be calculated by the product of their MGFs. We write the MGF of the arrival process as

$$M_A(\theta, t - \tau) = \mathbb{E} \left[ \exp \left( \theta A^{\circ}(t, \tau) \right) \right] \leq \exp \left( \theta (\varrho(t-\theta) + b \right), \tag{63}$$

where $\theta \geq 0$ is a free parameter. Similarly, we can write the MGF of the service process as

$$M_S(\theta, t - \tau) = \mathbb{E} \left[ \exp \left( -\theta S^{\circ}(t, \tau) \right) \right] \leq \exp \left( -\theta (\varrho(t-\theta) + b \right). \tag{64}$$

Then, we calculate the delay bound as in (62) and upper bound the delay violation probability, i.e., $P_{out} = \Pr[W(t) > w]$ as in (71). Note that when we introduce fading to the model, the rate expressions include logarithmic terms, which makes it cumbersome to find closed-form expressions for MGFs and the metrics. To alleviate this issue, [136] introduces a transformation facilities analysis under fading conditions.

EXAMPLE 12. *The log-normalized MGFs of the arrival and service process are known as effective bandwidth [135] and effective capacity [133], which are respectively given as*

$$B_e(\theta) = \frac{1}{\theta t} \ln M_A(\theta, t), \tag{65}$$

$$C_e(\theta) = -\frac{1}{\theta t} \ln M_S(-\theta, t). \tag{66}$$

*Then, we can derive $C_e$ as in* (58) *in Example 11.*

Let us consider a simple arrival model known as the two-state (ON-OFF) Markovian model. In this model, the data arrival process is described as a two-state discrete-time Markov chain, $r$ bits arrive during ON state with an arrival rate of $r$ bits/block, while no arrivals occur during the OFF state. Such system has a transition probability matrix $\mathbf{J} = \begin{pmatrix} p_{11} & p_{12} \\ p_{21} & p_{22} \end{pmatrix}$, where $p_{11} \in [0, 1]$ denotes the probability of staying in the off state, while $p_{22} \in [0, 1]$ denotes the probability of staying on the ON state, while the transition probabilities are $p_{21} = 1 - p_{22}$ and $p_{12} = 1 - p_{11}$. At the steady state, the probability of ON state is $p_{\text{ON}} = \frac{1 - p_{11}}{2 - p_{11} - p_{22}}$ [142]. Then, working out the MGF expressions for the arrival process, we obtain a closed-form expression for the effective bandwidth as

$$\begin{aligned} B_e(\theta) &= \frac{1}{\theta} \log \Big( \tfrac{1}{2} \left( p_{11} + p_{22} e^{r\theta} \right) \\ &\qquad + \tfrac{1}{2} \sqrt{\left( p_{11} + p_{22} e^{r\theta} \right)^2 - 4(p_{11} + p_{22} - 1) e^{r\theta}} \Big) \\ &\overset{(a)}{=} \frac{1}{\theta} \log \left( 1 - s + s e^{r\theta} \right), \end{aligned} \tag{67}$$

*where (a) comes from a simplified version of the source with $p_{11} = 1 - s$ and $p_{22} = s$, hence $p_{\text{ON}} = s$. Thus, the parameter $s$ can be interpreted as a measure of the burstiness, which is relevant to modeling different traffic generated by a device. Then, the maximum average arrival rate is $\bar{r}_{\max} = r \, p_{\text{ON}}$.*

For the particular case when $B_e(\theta) = C_e(\theta)$, we can calculate the maximum average arrival rate of the discrete-time Markov source supported by the wireless channel as

$$\bar{r}_{\max} = \frac{s}{\theta} \log \left( \frac{e^{\theta E_e(\theta)} - (1 - s)}{s} \right). \tag{68}$$

Note that the SNC is general and accommodates the cases when $B_e(\theta) \neq C_e(\theta)$ given queuing stability constraints [135], [136]. Moreover, this framework can be extended to other types of sources. The maximum average arrival rate enables us to characterize the impact of the variability of the arrival process as a function of the service process. Therefore, when the effective capacity is optimized, as in Example 11, we assess the impact of the arrivals so that fluctuations are accommodated, reducing delay violation and latency. All in all, SNC constitutes a valuable framework for the statistical QoS delay and end-to-end performance analysis of URLLC networks, whose traffic diverges from conventional light-tailed models as discussed in [6], [12].

TABLE X
Peak AoI Formulae for some popular queuing models

| Queue model | Characteristics | Peak AoI formula |
| --- | --- | --- |
| M/M/1 | single server, exponential inter-arrival and service times | $\frac{1}{\mu}\left(1 + \frac{1}{\rho_o} + \frac{\rho_o}{1-\rho_o}\right)$ |
| M/G/1 | single server, exponential inter-arrival times | $\frac{1}{\lambda} + \frac{1}{\mu} + \frac{\lambda}{2\mu^2(1-\rho_o)}$ |
| M/G/1/1 | M/G/1 with a finite buffer capacity of 1 | $\frac{1}{\mu} + \rho_o\left(1 + \frac{1}{\lambda}\right)$ |
| M/M/1/1 | M/M/1 with a finite buffer capacity of 1 | $\frac{1}{\lambda} + \frac{2}{\mu}$ |
| M/M/1/2 | M/M/1 with a finite buffer capacity of 2 | $\frac{3}{\mu} + \frac{1}{\lambda} + \frac{2}{(\lambda+\mu)}$ |

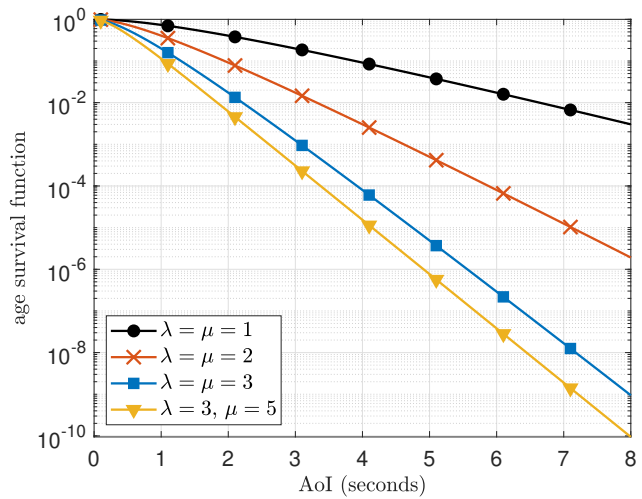

Fig. 17. Age survival function for an M/M/1 last-come first-serve transmission with preemption.

### C. Age of Information

AoI was introduced in [143] as a tool to measure the degree of freshness of the received data corresponding to a certain process. It is defined as the time elapsed since the latest received packet about a process under observation was generated. Hence, the AoI at time $t$ can be written as

$$\Delta(t) = t - t_g, \tag{69}$$

where $t_g < t$ is the generation time (i.e, timing anchor) of the most recently received packet [144]. Herein, in order to maintain fresh information, AoI should be minimized.

Note that the AoI minimization is a non-trivial problem of great interest. In fact, the system utilization can be maximized by making the source send more and faster updates, which would lead to packet congestion and higher delay. The delay could be reduced by decreasing the updating rate but this can lead to the destination having outdated information. Hence, timely updating a destination about a remote process is neither the same as maximizing the system utilization nor reducing the delay. However, increasing the effective service rate is beneficial for reducing the AoI as it allows emptying faster the queue.

The design of URLLC focuses more on peak and tail age metrics rather than the average AoI. In what follows, we discuss two important metrics of information freshness in time-sensitive applications, e.g., remote factory automation, where decisions should be based on fresh information.

*1) Peak AoI:* a measure of the maximum AoI that could occur within a network. Thus, the peak AoI measures the worst-case scenario with respect to reliability. Table X lists formulae for the average peak AoI of some popular queuing models according to [145]. Notice that for an average constant arrival rate $\lambda$, and average service rate $\mu$, the server utilization is given by $\rho_o = \frac{\lambda}{\mu}$.

Let us discuss briefly the just-in-time sequence parsing policy. In this case, a packet is generated at time $t_i$ and received at $t_{i+1} = t_i + d_{i+1}$, where $d_{i+1}$ is the service time of packet $i$. A new packet is instantaneously generated by the source node and starts its service time right after the current update packet in service is received at the destination node.

The peak AoI for the just-in-time policy is given by [146]

$$\hat{\Delta}_{JiT} = \frac{1}{N-1}\sum_{i=1}^{N-1} d_i + d_{i+1}, \tag{70}$$

where $N$ is the total number of packets. In this system, there is no queueing of updates and the server is always busy. Since each delivered status update is as fresh as possible, the performance of this system is a lower bound for the performance of any queue in which updates are generated as a stochastic process independent of the current state of the queue [146]. Notably, the peak AoI here depends only on the average service time.

*2) Age survival function:* complimentary CDF of the AoI, i.e., $\bar{F}_{AoI}(t)$. It allows computing the probability that the AoI exceeds a certain threshold $\Delta_{\text{th}}$, i.e., $\bar{F}_{AoI}(\Delta_{\text{th}})$, which is referred to as age violation probability. This is relevant for time-sensitive networking, where the age violation probability should not exceed a small value $\epsilon$, i.e.,

$$\bar{F}_{AoI}(\Delta_{\text{th}}) = \Pr[\Delta(t) \geq \Delta_{\text{th}}] \leq \epsilon. \tag{71}$$

**Example 13.** *Consider an M/M/1 last-come first-serve transmission with preemption. According to [147], the age survival function for this case is given by*

$$\bar{F}_{AoI}(t) = \begin{cases} \frac{\lambda}{\lambda-\mu}e^{-\mu t} - \frac{\mu}{\lambda-\mu}e^{-\lambda t}, & \lambda \neq \mu \\ (\lambda t + 1)e^{-\lambda t}, & \lambda = \mu \end{cases} \tag{72}$$

*which is depicted in Fig. 17 for different values of $\mu$ and $\lambda$. Obviously, the age survival function decays with the age, and more rapidly for higher values of $\lambda$ and/or $\mu$. For example, $\lambda = 3$ and $\mu = 3$, the age survival function is about $10^{-2}$ for $\Delta(t) > 2$ s. This means that the AoI exceeds 2 seconds for only 1% of the time. This is because the queue updates arrive and are served more rapidly resulting in fresher information at the receiver side. For the same setup, the age survival percentage becomes even lower ($\approx 0.5\%$) when increasing the service rate $\mu$ to 5. This is because the queue can deliver data faster, improving information freshness. Meanwhile, for $\lambda = 1$*

*and $\mu = 1$, the AoI exceeds 2 s for more than 50% of the time, since the amount of updates and service time are low, which results in less fresh data at the receiver.*

The age survival function is the right tail of the AoI for relatively large $\Delta_{th}$ and thus can be used to analyze/enforce ultra-reliable fresh update processes. Notably, the bounds, limiting forms, and EVT framework discussed in Section III may be exploited for accurately approximating the age survival function in scenarios where closed-form expressions are retractable or extremely complex for insightful analysis.

## VII. Analysis & Design of Large-Scale URLLC

Here, we overview some key mathematical and algorithmic tools for designing and/or assessing the performance of large-scale URLLC systems.

### A. Meta-Distribution

In general, the performance of wireless networks is highly influenced by their spatial configuration because phenomena like large- and small-scale fading, shadowing, and co-channel interference, are location-dependent. Hence, it is not surprising that the stochastic geometry, which allows modeling the network as a point process, constitutes a fundamental theoretical tool for the analysis and characterization of large-scale wireless systems such as cellular, *ad-hoc*, vehicular, and RF wireless power transfer [18], [80], [148]–[158].

Within the stochastic geometry framework, the most popular performance metric is the transmission success probability, which is defined as the probability that the corresponding SINR is greater than a given threshold $\gamma_{th}$, i.e., $p_s(\gamma_{th}) \triangleq \Pr[\text{SINR} \geq \gamma_{th}] = 1 - F_{\text{SINR}}(\gamma_{th})$. However, large-scale URLLC systems cannot directly benefit from $p_s(\gamma_{th})$, neither for system design nor analysis, since the computation of such metric involves averaging over space (and/or time).[18] Instead, a more refined metric has been proposed and leveraged in the last years: the *meta-distribution* of the SINR, which is defined as [80], [153]–[157]

$$p_m(\gamma_{th}, \xi) \triangleq \mathbb{P}^!\big[\Pr[\text{SINR} \geq \gamma_{th} \mid \Phi] \geq 1 - \xi\big], \qquad (73)$$

where $\Phi$ denotes the point process on $\mathbb{R}^d$ corresponding to the $d$−dimensional deployment of the network nodes. Moreover, $\mathbb{P}^!$ denotes the reduced Palm measure[19] of $\Phi$ given that there is an active transmitter (as in the case of bipolar networks) or receiver (as in cellular networks) at a prescribed location and the SINR is measured at the receiver. Meanwhile, $1 - \xi$, with $\xi \in [0, 1]$, denotes a target per-node success probability, where $\xi \ll 1$ in the context of URLLC. Note that $p_m(\gamma_{th}, \xi)$ denotes the fraction of users that attain an SINR of at least $\gamma_{th}$ in $(1 - \xi)$ fraction of the time, whereas $p_s(\gamma_{th})$ just characterizes the fraction of users that can communicate successfully. Observe that $p_s(\gamma_{th})$ can be obtained from $p_m(\gamma_{th}, \xi)$ as $p_s(\gamma_{th}) = \int_0^1 p_m(\gamma_{th}, \xi) d\xi$, thus, $p_m(\gamma_{th}, \xi)$ is indeed a broader performance metric.

Example 14 (Meta-Analysis of a Large-Scale Network). *Assume the setup described in Example 4, but where the interfering nodes are distributed according to a homogeneous Poisson point process $\Phi$ on $\mathbb{R}^2$ with density $\lambda$. Thus, the total number of interfering nodes is infinite, i.e., $K \to \infty$, but the number of interfering nodes in a given region $A \subset \mathbb{R}^2$ is finite and follows a Poisson distribution with mean $\lambda|A|$. power for simplicity, and assume no power control mechanism such that the average power (with respect to the Rayleigh small-scale fading) of the target receive signal and interfering signals are respectively given by $r_0^{-\alpha}$ and $V \triangleq \sum_{x \in \Phi} \|x\|^{-\alpha}$, where $\alpha > 2$. In this case, it can be shown that $\Upsilon = r_0^{-\alpha}/V$. Then, by using this in (25), which already addressed the channel fading randomness, while substituting it into (73), one obtains*

$$p_m(\gamma_{th}, \xi) \overset{\xi = 0}{=} \mathbb{P}^!\big(\exp(-\gamma_{th} r_0^\alpha V) \geq 1 - \xi\big)$$
$$= F_V\Big(-\frac{\ln(1-\xi)}{\gamma_{th} r_0^\alpha}\Big). \qquad (74)$$

*Since $V$, also referred to as interference without fading [148]), is obtained via transformations to RVs $\{\|x\|^{-\alpha}\}_{x \in \Phi}$, one may exploit latters' distributions, which were provided in [149] for Poisson networks, and/or a procedure exploiting the probability generating functional of a Poisson point process, to compute (74). Unfortunately, this has been proved to be a cumbersome task for general path-loss exponents, i.e., $\alpha > 2$. Fortunately, for some special cases, e.g., $\alpha \in \{3, 4, 5, 6, 8, 10, 12\}$, $f_V(v)$ has been derived in closed- or semi-closed form [148], [152]. For instance, for $\alpha = 4$, the PDF of $V$ is given by [148]*

$$f_V(v) \overset{\alpha = 4}{=} \frac{\pi \lambda}{2 v^{3/2}} \exp\Big(-\frac{\pi^3 \lambda^2}{4v}\Big). \qquad (75)$$

*By leveraging (75), one can get back to (74) to obtain*

$$p_m(\gamma_{th}, \xi) \overset{\xi = 0}{=} \int_0^{-\frac{\ln(1-\xi)}{\gamma_{th} r_0^\alpha}} f_V(v) dv$$
$$\overset{\alpha = 4}{=} \int_0^{-\frac{\ln(1-\xi)}{\gamma_{th} r_0^\alpha}} \frac{\pi \lambda}{2 v^{3/2}} \exp\Big(-\frac{\pi^3 \lambda^2}{4v}\Big) dv$$
$$\overset{(a)}{=} \frac{1}{\sqrt{\pi}} \Gamma\Big(\frac{1}{2}, \frac{\pi^3 \lambda^2}{4v}\Big)\Big|_{v=0}^{v=-\frac{\ln(1-\xi)}{\gamma_{th} r_0^\alpha}}$$
$$\overset{(b)}{=} \frac{1}{\sqrt{\pi}} \Gamma\Big(\frac{1}{2}, -\frac{\pi^3 \lambda^2 \gamma_{th} r_0^\alpha}{4 \ln(1-\xi)}\Big), \qquad (76)$$

*where (a) follows from solving the indefinite integral by exploiting [159, eq. (2.325.6)], while (b) comes from evaluating the extremes.*

The above expression is expected to be tight because of the same reasons given in Example 4. This is corroborated in Fig. 18, where $p_m(\gamma_{th}, \xi)$ is plotted against $1 - \xi$. In Fig. 18, we do not only plot $p_m(\gamma_{th}, \xi)$ but also the success probability $p_s(\gamma_{th})$. The latter metric reveals that the average success probability in the network is 90% and 99.85% for a density of $10^{-5}$ and $10^{-7}$ nodes/m$^2$, respectively, but says nothing about the reliability performance per node. Instead, $p_m(\gamma_{\text{th}}, \xi)$ reveals directly the percentage of nodes achieving a reliability of $1 - \xi$. Based on the results in Fig. 18, we can observe that only 44% of the nodes are operating with a success probability not inferior to 99% when $\lambda = 10^{-5}$ nodes/m², while that percentage of nodes increases up to 99.4% when $\lambda = 10^{-7}$ nodes/m².

---

[18]Recall that URLLC mandates a radical departure from expected utility-based approaches relying on averaged quantities [12].

[19]The Palm measure or Palm probability is the probability of an event given that the point process contains a point at some location. Similarly, the Palm distribution is the conditional point process distribution given that a point (the typical point) exists at a specific location [148].

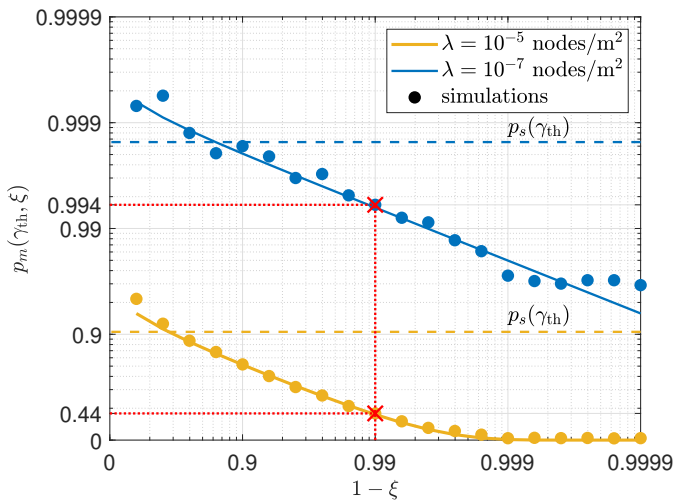

Fig. 18. Meta-distribution of the SIR as a function of $1 - \xi$ for $\lambda \in \{10^{-5}, 10^{-7}\}$ nodes/m². We set $\gamma_{\text{th}} = -10$ dB, $r_0 = 80$ m, and $\alpha = 4$.

TABLE XI
Main clustering methods and example algorithms/implementations

| Method | Description | Implementations |
|---|---|---|
| Partitioning | Partitioning the data by grouping similar items. The results vary according to a pre-defined number of clusters. | • $K$−Means<br>• $K$−Medoids<br>• $K$−Modes |
| Hierarchical | Decomposing the data hierarchically. The number of clusters is not specified in advance. | • Agglomerative<br>• Divisive |
| Others | Special methods that group the items according to their density or another abstract model. | • Density-based<br>• Model-based |

Note that $p_m(\gamma_{\text{th}}, \xi)$ can be leveraged to properly adjust the per-link reliability via rate control, which does not affect the SINR realizations, thus potentially guaranteeing network-wide reliability guarantees [157].

Unfortunately, MC methods for estimating $p_m(\gamma_{\text{th}}, \xi)$ are in general computationally expensive as they require separate averaging over fading and point realizations. Meanwhile, analytical closed-form characterizations are not often attainable, as in the case of Example 14 for any $\alpha > 2$. Therefore, efficient numerical and approximate analytical methods for obtaining $p_m(\gamma_{\text{th}}, \xi)$ take significant relevance here. The most prominent methods include moment-matching, the Gil-Pelaez theorem, and bounding approaches such as Markov inequalities and Paley-Zygmund bound (refer to Section III) [80], [154]. Interestingly, the beta distribution (by default defined in [0, 1]) and the Fourier-Jacobi expansion framework fit typical meta-distributions with high accuracy in most cases [154].

Notably, similar to the SINR or SIR meta-distribution, one can formulate the meta-distribution of other relevant figures, e.g., the energy in wireless-powered networks [155], the secrecy rate in the presence of eavesdroppers [153], and the number of transmission attempts to accomplish a success or reliability guarantee. In fact, outside the stochastic geometry framework, there is the so-called *meta-probability*, which refers to stochastic models describing the uncertainty in the system's probabilistic parameters arising, e.g., due to limited data and modeling assumptions. Therefore, meta-probability provides a high-level understanding of the probabilistic nature of the system and thus is crucial for URLLC design. Such a framework is leveraged in [4],

[6], [28], [29], [75] to impose robust guarantees on URLLC design/performance.

### B. Clustering

Large-scale URLLC design may demand large-scale optimization procedures. The recent few years have witnessed a growing trend towards exploiting ML tools for large-scale optimization since conventional convex/non-convex optimization approaches do not scale well with the number of variables and constraints. Specifically, decentralized ML approaches such as FL, multi-agent RL, and deep RL can be applied to reduce the state space in extremely high dimensional problems. Meanwhile, clustering is another powerful, yet simple, (unsupervised) ML tool to address large-scale optimization problems and constitutes the focus of our discussions in the following.

Clustering relies on building groups or hierarchies among *points/observations* based on their (dis-)similarities. The main clustering methods and example algorithms are summarized in Table XI and described briefly in the following.

*1) Partitioning methods:* The most popular partitioning method, K-Means (K-Modes for categorical data), iteratively relocates the cluster centers by computing the points' mean (mode). K-Means scales well on huge data sets, even though, it is very sensitive to outliers and fails to perform well on arbitrary shapes of data. These issues are mitigated by K-Medoids, which chooses actual data points as centers (medoids or exemplars) and can be used with arbitrary dissimilarity measures.

*2) Hierarchical methods:* There are two main types of hierarchical methods [160]: *i)* agglomerative, where each point starts in its own cluster, and pairs of clusters are merged as one moves up the hierarchy; and *ii)* divisive, where all points start in one cluster, and splits are performed recursively as one moves down the hierarchy. For clusters' merging/splitting, not only a distance metric but also a linkage criterion specifying the dissimilarity of clusters as a function of the pairwise distances of points in the clusters must be specified. Merges/splits are often determined in a greedy manner, and the results are presented in a dendrogram.

*3) Other methods:* In density-based clustering, the points that are highly dense are grouped together, which allows the

identification of low-density/extreme points. Meanwhile, model-based approaches involve applying a model to find the best cluster structures.

In the context of large-scale URLLC, clustering algorithms can be mainly leveraged for:

- Configuration of NOMA groups.
  NOMA promotes spectral and energy efficiency, which are both relevant performance figures in large-scale systems. A key challenge here lies in the optimal grouping of users exploiting the same orthogonal spectrum resource. It has been shown in [18], [161] that users with the most heterogeneous characteristics/requirements should be clustered together for optimum performance.

- Pilot allocation.
  Assigning orthogonal pilots to all users becomes inefficient, or even unaffordable, as the network grows. This issue can be addressed by resorting to clustering based on *i)* channel covariance matrices such that the pilots are reused among the users having sufficiently orthogonal channel subspaces [162], [163], and/or *ii)* traffic profiles such that devices with smaller activation probabilities (more stringent reliability requirements) can be grouped in clusters with smaller (greater) pilot pools [15].

- Data aggregation.
  In large-scale systems, devices may be clustered in local area networks, which then connect to the core networks through a gateway or data aggregator [151], [157], [164]–[168]. Instead of a traditional spatial clustering relying on geometric distances and/or energy-related measures [151], [157], [164]–[167], a more advisable approach in the context of large-scale URLLC lies in jointly considering traffic profiles (both in terms of traffic intensity and performance requirements), and network dynamics. Interestingly, the authors in [168] have initially proposed a priority-based spatial aggregation algorithm that gives priority to URLLC devices through power ramping in the physical random access channel.

- Coordinated multi-point.
  In coordinated multi-point scenarios, a group of network nodes (e.g., base stations) transmits/receives the same data to/from a device, thus inducing spatial diversity, which is key for supporting large-scale URLLC. Here, clustering plays a key role in selecting the set of network nodes for jointly serving each device. This can be implemented based on the instantaneous reference signal received power [169], SNR [170], distance [171] or even links LOS/non-LOS classification [20] from multiple network nodes.

Next, we illustrate how clustering can also be leveraged for access resource scheduling in large-scale systems.

EXAMPLE 15 (CLUSTERING-BASED ACCESS SCHEDULING).
*Consider a set $\mathcal{N} = \{1, 2, \cdots, N\}$ of devices with a given activation matrix $\mathbf{A} \in [0,1]^{N \times N}$, where $A_{i,j}$ denotes the probability that devices $i, j \in \mathcal{N}$ become simultaneously*

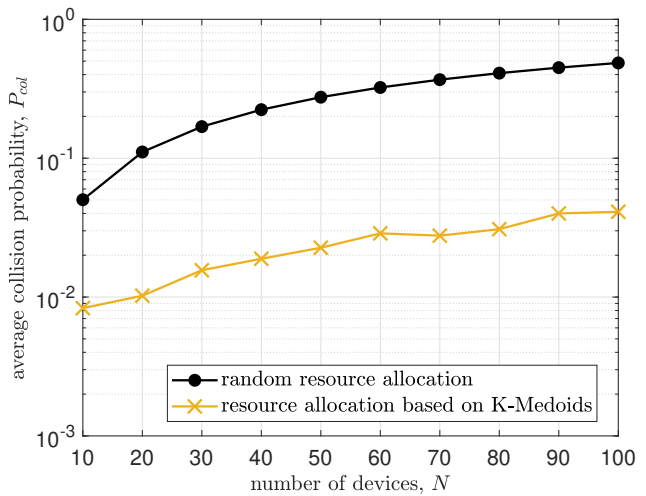

Fig. 19. Average collision probability as a function of the number of devices. We set $A_{i,j} = (i + j + |i - j|)^{-1}$, $\forall i, j \in \{1, 2, \cdots, N\}$ and $L = 8$.

*active, thus, $\mathbf{A}$ is symmetric, while $A_{i,j} \leq A_{i,i}$ and $A_{i,j} \leq A_{j,j}$. Such activation matrix can be learned over time [172], [173]. Moreover, assume that there is a set $\mathcal{L}$ of $L < N$ orthogonal spectrum resources either in time/frequency/code domain that can be used by the devices to access the wireless medium. Consider that a given spectrum resource $l \in \mathcal{L}$ is allocated to a sub-set of devices $\mathcal{N}_l \subset \mathcal{N}$, and assume $\cup_{l \in \mathcal{L}} \mathcal{N}_l = \mathcal{N}$ and $\mathcal{N}_{l_1} \cap \mathcal{N}_{l_2} = \phi$, $\forall l_1, l_2 \in \mathcal{L}$ and $l_1 \neq l_2$. A collision occurs if at least two devices attempt to access the medium by exploiting the same orthogonal resource, thus, the collision probability within the $l-$th spectrum resource is given by*

$$P_{col}^{(l)} = 1 - \prod_{\forall i, j \in \mathcal{N}_l, j > i} (1 - A_{i,j}), \quad (77)$$

*while*

$$P_{col} = \frac{1}{L} \sum_{l \in \mathcal{L}} P_{col}^{(l)} \quad (78)$$

*denotes the average collision probability in the network.*

*The optimum resource allocation comes from solving $\arg\min_{\mathcal{N}_l, \forall l \in \mathcal{L}} P_{col}$, which is an NP-hard problem due to its combinatorial structure. Interestingly, a sub-optimal but also scalable and simple approach relying on K-Medoids clustering can be designed by adopting $d_{i,j} = \mathbb{1}[i \neq j] A_{i,j}$ as dissimilarity measure between points (devices) $i, j \in \mathcal{N}$. In this way, two devices with small joint activation probability are likely to be clustered together. Fig. 19 illustrates the attainable performance gains of such an approach with respect to a traditional pure random allocation. Here, we would like to emphasize that the adopted K-Medoids approach is by no means optimized, and better results may be attainable by, e.g., more tightly coupling the dissimilarity measure with the problem's objective.*

### C. Compressed Sensing

Compressed sensing comprises a set of mathematical tools for fully or partially recovering sparse signals from a small set of measurements. The sparsity $f(\mathbf{s})$ of a given signal $\mathbf{s}$ is defined as the number of non-zero entries, i.e., $f(\mathbf{s}) = ||\mathbf{s}||_0$. Moreover, $\mathbf{s}$ is considered to be sparse if $f(\mathbf{s})$ is sufficiently small compared with the dimension of $\mathbf{s}$.

Assume $\mathbf{s} \in \mathbb{C}^n$ is sparse, i.e., $f(\mathbf{s}) \ll n$, and needs to be estimated from $m < n$ samples that are collected via a sensing matrix $\mathbf{H} \in \mathbb{C}^{m \times n}$ as

$$\mathbf{y} = \mathbf{Hs} + \mathbf{w}, \tag{79}$$

where $\mathbf{y} \in \mathbb{C}^m$ and $\mathbf{w} \in \mathbb{C}^m$ denote respectively the receive and noise signals. Then, compressed sensing problems are commonly formulated to find the sparsest signal such that the estimated measurement noise power is upper bounded by a pre-configured noise level $\eta$ as

$$\mathbf{s}^{\star} = \mathrm{argmin}_{\mathbf{s}} f(\mathbf{s}), \ \mathrm{s.t.} \ ||\mathbf{y} - \mathbf{Hs}||_2^2 \leq \eta. \tag{80}$$

Observe that the number of possible vectors $\mathbf{s}$ that satisfy $||\mathbf{y} - \mathbf{Hs}||_2^2 \leq \eta$ is potentially infinite since $\mathbf{y} = \mathbf{Hs}$ is an underdetermined system of equations. However, the sparsity-reduction objective function forces the solution to be unique, the sparsest one. Unfortunately, since $f : \mathbb{C}^n \rightarrow \{0, 1, \cdots, n\}$, there is no way but to rely on a combinatorial search to get the optimum solution of (80). The complexity of this approach increases exponentially in $n$, thus, the $\ell_0$−norm minimization approach is infeasible for most real-world applications [174].

Several techniques and heuristics have been suggested in the last few years to solve (80), including regularization, greedy, and message-passing approaches. Specifically,

- regulation techniques, e.g., the $\ell_1$−norm approximation of the $\ell_0$−norm [175], [176] and the alternative direction method of multipliers [177], rely on transforming (80) into a convex problem via regularized, often iterative, procedures;
- greedy algorithms, e.g., orthogonal matching pursuit [178], make a local optimal selection at each time with either the hope of ultimately finding the global optimum solution or just a sufficiently good local solution with low complexity and/or in a reasonable amount of time; and
- message-passing algorithms exploit factor graphs [179], thus, the *a posteriori* distribution of the signal to be reconstructed.

In all cases, properly choosing/designing the sensing matrix, such that it complies with the restricted isometric property, mutual coherence, and null space property [174], strengthens sparse signal recovery algorithms.

In wireless communications, compressed sensing is exploited for solving problems related to sparse estimation/detection and support identification e.g., [174], [180]–[185]. The latter relates to a partial recovering of $\mathbf{s}$, specifically, of the indexes of the non-zero elements of $\mathbf{s}$. Therefore, it can mimic the multi-user detection problem in massive MTC systems with grant-free and sporadic random access e.g., [174], [180]–[183], and also more recently has found applications in sparse coding mechanisms for URLLC, e.g., [184], [185]. Indeed, sparse coding, where the

---

**Algorithm 4:** AMP for MTC activity detection

**Input:** $\mathbf{Y}, \psi$
1 Initialization: $\hat{\mathbf{S}}^0 = \mathbf{0}$, $\mathbf{R}^0 = \mathbf{Y}$, $t = 0$
2 **repeat**
3     $\hat{\mathbf{s}}_n^{t+1} = \vartheta((\mathbf{R}^t)^H \boldsymbol{\varphi}_n + \hat{\mathbf{s}}_n^t), \ \forall n$
4     $\mathbf{R}^{t+1} = \mathbf{Y} - \boldsymbol{\Phi}(\hat{\mathbf{S}}^{t+1})^T + \frac{1}{\tau_p} \mathbf{R}^t \sum_{\forall n} \vartheta'((\mathbf{R}^t)^H \boldsymbol{\varphi}_n + \hat{\mathbf{s}}_n^t)$
5     $t = t + 1$
6 **until** *convergence or maximum number of iterations*;
7 $\forall n$: set $\hat{\alpha}_n = 1$ if $||\hat{\mathbf{s}}_n^t||_2 \geq \psi$, $\hat{\alpha}_n = 0$ otherwise

**Output:** $\{\hat{\alpha}_n\}$

---

encoding and decoding stages are jointly modeled as a compressed sensing problem, is emerging as a potential enabler for URLLC due to favorable reliability-latency tradeoffs.

We next discuss an example linking both the multi-user detection and support identification problems. A simple approximate message-passing (AMP) solving algorithm relying on iterative thresholding is also provided. Readers are encouraged to refer to [174] for additional examples of how the compressed sensing framework can be applied to wireless communications-related problems.

EXAMPLE 16 (SPARSE MTC ACTIVITY DETECTION [180]).
*Consider the uplink communication between a coordinator with $M$ antennas and $N$ single antenna devices, out of which only a few are simultaneously active. Quasi-static fading is assumed, and the small-scale channel between the $n$−th device and the coordinator is denoted as $\mathbf{h}_n$. The coherence interval comprises $\tau$ samples, out of which $\tau_p$ are used for user identification (and potentially channel estimation). The coordinator is aware of the pilot sequence associated with each device, whereas $\mathbf{h}_n$ coefficients, which change independently between consecutive coherence intervals, are unknown.*

*The signal $\mathbf{Y} \in \mathbb{C}^{\tau_p \times M}$ received at the coordinator in a given user identification phase can be written as*

$$\mathbf{Y} = \sum_{n=1}^{N} \sqrt{\tau_p \bar{\gamma}} \alpha_n \boldsymbol{\varphi}_n \mathbf{h}_n^T + \mathbf{W}$$
$$= \sqrt{\tau_p \bar{\gamma}} \boldsymbol{\Phi} \mathbf{S}^T + \mathbf{W} \tag{81}$$

*where $\alpha_n$ denotes the device activity indicator for device $n$, $\bar{\gamma}$ is the average SNR of the signal received at the coordinator from each active device, $\sqrt{\tau_p \bar{\gamma}} \boldsymbol{\varphi}_n \in \mathbb{C}^{\tau_p}$ with $||\boldsymbol{\varphi}||_2^2 = 1$ is the pilot sequence of the $n$−th device, and $\mathbf{W}$ is the power-normalized AWGN samples. Observe that the signals arrive at the coordinator with the same average SNR, which can be motivated by the use of statistical inverse power control at the devices. Moreover, $\boldsymbol{\Phi} = [\boldsymbol{\varphi}_1, \boldsymbol{\varphi}_2, \ldots, \boldsymbol{\varphi}_N] \in \mathbb{C}^{\tau_p \times N}$ constitutes the pilot matrix, while $\mathbf{S} = [\mathbf{s}_1, \mathbf{s}_2, \cdots, \mathbf{s}_N] \in \mathbb{C}^{M \times N}$, where $\mathbf{s}_n \triangleq \alpha_n \mathbf{h}_n$, is an effective channel matrix. Here, $\mathbf{S}$ has a sparse structure as the rows corresponding to inactive users are zero. Therefore, the activity detection problem reduces to finding the non-zero rows of $\mathbf{S}$.*

*Algorithm 4 iteratively solves the above problem. Note that $\vartheta(\cdot)$ is a denoising function, which for the multi-*

*user detection problem can be defined as in [180, eq. (11)]. Meanwhile, $\vartheta(\cdot)'$ is the first order derivative of $\vartheta(\cdot)$ and $\mathbf{R}^t$ is the signal residual at iteration $t$. In step 1, all the devices are assumed inactive, while the estimate $\hat{\mathbf{S}}^t = [\hat{\mathbf{s}}_1^t, \hat{\mathbf{s}}_2^t, \cdots, \hat{\mathbf{s}}_N^t]$ and the residual signals are then iteratively updated in steps 3 and 4, respectively. In the case of the latter, a crucial term containing $\vartheta(\cdot)'$, called the Onsager term, is included as it has been shown to substantially improve the performance of the algorithm. Observe that the estimated activity indicators are returned as the output of the algorithm after hard-thresholding. Specifically, the $\ell_2-$norm of each $\hat{\mathbf{s}}_n^t$ is compared with a given threshold to decide whether the $n-th$ device is active.*[20]

Unfortunately, AMP algorithms such as that illustrated in Algorithm 4 face several inconveniences. On the one hand, prior distributions of the device activity patterns and channels are usually unknown. On the other hand, the design of the denoising function $\vartheta(\cdot)$ becomes tedious, if not impossible, in most practical cases of known prior distributions. Indeed, typical prior assumptions that are difficult to handle include correlated activation patterns and/or other than zero-mean Gaussian channels. To alleviate these issues, expectation propagation algorithms [186], [187] have emerged as a more flexible (less model-dependent) approach. Moreover, it should be noted that although the basic expectation propagation algorithms are in general more computationally complex than the AMP, several state-of-the-art complexity-reduction techniques (e.g., see [188]–[191]) may be implemented to make them even more appealing in practice. Alternatively, data-driven compressed sensing may be exploited, which has proven useful for lower complexity and latency in [192]. Therein, the authors present variations of neural networks for the compressed sensing task together with prospective performance results.

Notice that configuring the hyperparameters of the algorithm might also be quite challenging, especially for URLLC systems with stringent performance requirements. In the case of Algorithm 4, the main issue is related to properly configuring the decision threshold $\psi$. Notice that a relatively large (small) $\psi$ reduces the probability of misdetection (false-alarm), but at the cost of triggering many false-alarm (misdetection) events. In fact, this is not only an issue of AMP implementations, but of most state-of-the-art procedures, as those based on regulation techniques, greedy, and other message-passing algorithms.

In general, prior procedures aiming at estimating certain raw features of $\mathbf{s}$ can potentially mitigate the previous issue and assist the posterior estimation based on compressed sensing. For instance, the authors in [193] proposed coordinated pilot transmission mechanisms that can be run prior to the multi-user detection phase, as that described in Example 16, to detect the signal sparsity level (number of active devices), i.e, $f(\mathbf{s})$, in each coherence interval. With such information, one can directly modify the step 7 in Algorithm 4 such that only the $f(\mathbf{s})$ devices with the greatest associated equivalent channel powers are declared as active, thus, completely avoiding the tuning of a decision threshold. The potential performance gains of such an approach are illustrated in Fig. 20 in terms of activity detection error rate. Note that the regularized multi-user detection algorithms proposed in [194]–[197] directly exploit this information as well, although no work prior to [193] had specified how such information could be obtained. In any case, a completely deterministic/accurate prior on $f(\mathbf{s})$ may not be available in practice, and one may need to consider the probability mass function of $\widehat{f(\mathbf{s})}$ for the multi-user detection phase design of large-scale grant-free URLLC systems. As an alternative, the authors of [183] propose a dynamic adaptive compressive sensing multi-user detection for uplink URLLC networks without any a priori knowledge of the user activity sparsity level. Specifically, they leverage the temporal correlation between active user sets in adjacent time slots to estimate the true user activity sparsity level iteratively.

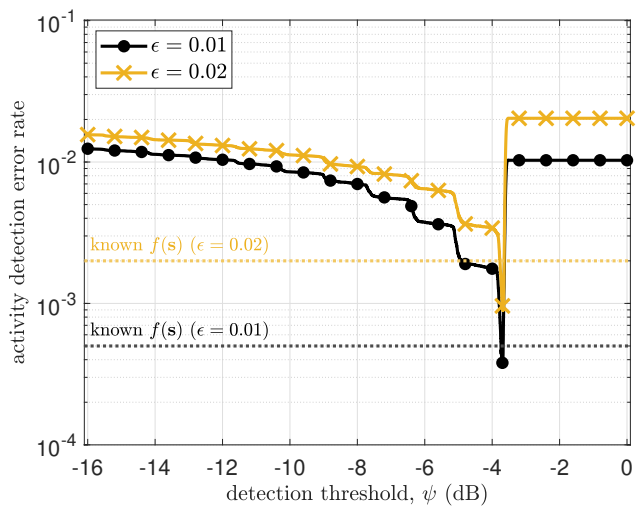

Fig. 20. Average error rate as a function of the detection threshold. We have assumed Rayleigh fading channels, i.i.d activation of the devices with $\Pr[\alpha_n = 1] = \epsilon$ and $\Pr[\alpha_n = 0] = 1 - \epsilon$, and Bernoulli pilots as in [180]. Moreover, we set $\tau_p = 48$, $M = 64$, $N = 200$, and $\bar{\gamma} = 20$ dB.

Finally, compressed sensing algorithms are highly sensitive to noise, and thus perform poorly in low SNR conditions. This motivates massive MIMO implementations, with the corresponding SNR gains and channel hardening [180]–[182]. Interestingly, one may exploit diversity (a fundamental URLLC enabler), e.g., in frequency or space, for channel hardening. This has not been explored so far in the context of multi-user detection with compressed sensing. The reason may be that diversity and sparsity are often conceptually opposed. In any case, diversity-sparsity trade-offs are worth investigating for supporting/enabling URLLC.

## D. Mean Field Game Theory

Mean-field (MF) game theory is an attractive tool for describing and optimizing networks with an asymptotically

---

[20] The channel estimates could be easily obtained as well as they match the entries of $\mathbf{S}$ associated with the non-zero activity indicators. Nevertheless, they could be still refined in a last step by leveraging traditional MMSE channel estimators but considering only the group of detected devices.

large number of inter-device interactions [198].

Consider a very large set $\mathscr{S}$ of nodes, and let $s \in \mathscr{S}$ to represent a rational node/agent with an associated set of states $\mathscr{X}_s \subset \mathbb{R}^n$ and set of actions $\mathscr{A}_s$. Then, one can formulate a strategic game $\mathscr{G} = (\mathscr{S}, \mathscr{A}, \mathscr{X}, \{u_s\})$, where $\mathscr{A} = \mathscr{A}_1 \times \cdots \times \mathscr{A}_{|\mathscr{S}|}$ and $\mathscr{X} = \mathscr{X}_1 \times \cdots \times \mathscr{X}_{|\mathscr{S}|}$ denote the overall action and state profiles, respectively, while $u_s$ is the average utility function of $s$. While traditional game theory approaches consider the individual interactions among agents, MF game theory takes advantage of the massiveness of the network assuming that $|\mathscr{S}| \to \infty$ to approximate the solution.

The common underlying assumptions are:

- at any given time instance $t \in [0, T]$, individual agents' states $\mathbf{x}_s(t) \in \mathscr{X}_s$ and actions $\mathbf{a}_s(t) \in \mathscr{A}_s$ have a negligible impact on the game;
- the optimization criterion is invariant to the permutation of agents' indices;
- agents are indistinguishable, i.e., they share a common state and action space. This allows reducing the game to a generic agent with states $\tilde{\mathbf{x}}(t) \in \tilde{\mathscr{X}}$ and action $\tilde{\mathbf{a}}(t) \in \tilde{\mathscr{A}}$ playing against the MF distribution.

Let the state $\tilde{\mathbf{x}}(t)$ to evolve as $d\tilde{\mathbf{x}}(t) = \boldsymbol{\mu}(t) dt + \Lambda d\mathbf{w}(t)$, where $\boldsymbol{\mu} : [0, T] \to \mathbb{R}^n$ is a deterministic function, $\mathbf{w} : [0, T] \to \mathbb{R}^n$ is a random process, and $\Lambda \in \mathbb{R}^{n \times n}$ is a constant matrix. Given a state transition from time $t \in [0, T]$ to $T : \tilde{\mathbf{x}}(t) \to \tilde{\mathbf{x}}(T)$, the utility function for the generic agent is given by

$$u\big(t, \tilde{\mathbf{x}}(t), \tilde{\mathbf{a}}(t)\big) = \mathbb{E}_{\tilde{\mathbf{x}}}\left[\int_t^T f\big(\tau, \tilde{\mathbf{x}}(\tau), \tilde{\mathbf{a}}(\tau)\big) d\tau + c\big(T, \tilde{\mathbf{x}}(T)\big)\right], \quad (82)$$

where $f : [0, T] \times \tilde{\mathscr{X}} \times \tilde{\mathscr{A}} \to \mathbb{R}$ is the instantaneous payoff, and $c : T \times \tilde{\mathscr{X}} \to \mathbb{R}$ is a terminal cost associated with reaching state $\tilde{\mathbf{x}}(T)$. The MF distribution, i.e.,

$$\rho\big(t, \tilde{\mathbb{S}}(t)\big) = \lim_{|\mathscr{S}| \to \infty} \frac{1}{|\mathscr{S}|} \sum_{s=1}^{|\mathscr{S}|} \mathbb{I}\{\mathbf{x}_s(t) = \tilde{\mathbf{x}}(t)\}, \quad (83)$$

which approximates the collective behavior of the whole population, is an implicit argument of $f$.

To solve the game, agents depart from a local estimate of (83) using other agents' initial states, and then run an iterative procedure. The latter consists in alternating between solving the optimal action from the Hamilton-Jacobi-Bellman equation, i.e.,

$$\frac{\partial u\big(t, \tilde{\mathbf{x}}(t), \tilde{\mathbf{a}}(t)\big)}{\partial t} + \max_{\tilde{\mathbf{a}}(t)}\left[\boldsymbol{\mu}^T(t) \frac{\partial u\big(t, \tilde{\mathbf{x}}(t), \tilde{\mathbf{a}}(t)\big)}{\partial \tilde{\mathbf{x}}}\right.$$
$$\left. + f\big(t, \tilde{\mathbf{x}}(t), \tilde{\mathbf{a}}(t)\big) + \frac{1}{2}\mathbf{tr}\left(\Lambda^2 \frac{\partial^2 u\big(t, \tilde{\mathbf{x}}(t), \tilde{\mathbf{a}}(t)\big)}{\partial \tilde{\mathbf{x}}^2}\right)\right] = 0, \quad (84)$$

for a given MF distribution, and an improved estimate of the MF distribution from the Fokker-Planck-Kolmogorov equation, i.e.,

$$\frac{\partial \rho\big(t, \tilde{\mathbf{x}}(t)\big)}{\partial t} + \frac{\partial}{\partial \tilde{\mathbf{x}}}\left[\frac{\partial u\big(t, \tilde{\mathbf{x}}(t)\big)}{\partial \tilde{\mathbf{x}}} \rho\big(t, \tilde{\mathbf{x}}(t)\big)\right] - \mathbf{tr}\left(\frac{\Lambda^2}{2} \frac{\partial^2 \rho\big(t, \tilde{\mathbf{x}}(t)\big)}{\partial \tilde{\mathbf{x}}^2}\right) = 0, \quad (85)$$

given the action obtained from (84). Notice that the solution of an MF game, known as MF equilibrium, is guaranteed to converge to the Nash equilibrium in the asymptotic case

$|\mathscr{S}| \to \infty$. However, since practical network deployments have a large but finite number of nodes, an MF game converges to an $\epsilon$-approximate Nash equilibrium [199] in most of these settings. That is, some players may benefit when deviating from the equilibrium, thus violating the Nash equilibrium conditions, but the expected payoff gain will not be superior to $\epsilon$.

In MF games, no information is exchanged among agents; thus, this approach is effective in reducing the signaling overhead of the network. This is appealing for meeting the ultra-low latency requirements in massive/dense URLLC systems. Take an ultra-dense network [48], where the cells may experience high levels of interference due to their proximity, as an example. This demands effective interference management techniques, which may induce a severe signaling overhead if the cells treat the interfering links individually. Instead, the problem can be recast as an MF game, which allows modeling the mutual interference experienced by all cells as an average interference over the state distribution (83) [200]. Therefore, each cell can locally determine a transmission policy without coordinating with its interfering cells.

With respect to the ultra-high reliability requirements in massive/scalable URLLC systems, one may need to incorporate tail/risk statistics (refer to Section III) in the instantaneous payoff function. Other interesting applications are those related to studying neural networks and multi-agent RL, and simplifying rare events simulations (refer to Section IV). Regarding the latter, note that modeling how each agent interacts with its peers may require significant computational effort if traditional MC methods or agent-based models are used. Fortunately, with MF game theory we can significantly reduce the complexity by abstracting the massive interactions in the evolution of one agent.

The three main underlying assumptions of the MF games framework enumerated above constitute also critical limitations. Hence, future research directions may consider an MF game with partial observability of the agents' state-action space [201] and/or multi-population MF games [202] in which different groups of agents have not only a different state-action space but also conflicting targets, hence mimicking the current QoS heterogeneity in wireless networks. Another limitation of MF games is that solving (84)- (85) becomes computationally expensive for high-dimensional states due to the course of dimensionality. This limits the applications of classical MF game approaches for real-time applications and networks where devices have limited computational capabilities. In view of this, there is a growing interest in using DL or RL to solve MF games. Indeed, neural networks are used in [203] to approximate the optimal solutions of equations (84)- (85) at the cost of violating the conditions for convergence of the MF game framework. In another approach, an FL-based MF game approach was proposed in [204] to allow the devices to periodically exchange their model parameters so that convergence is guaranteed.

## VIII. Conclusions & Future directions

URLLC is about taming the tail of reliability and latency under the uncertainty arising from the stochastic nature of wireless communications. Moreover, most URLLC use cases are very different from conventional human-type communications characterized by traffic with smaller payloads and sporadic arrivals, large device density, and heterogeneous service demands. Thus, designing and/or analyzing URLLC systems mandates adopting a vastly different set of statistical tools and methodologies. Aiming to fill the corresponding gap in existing surveys and overview papers on URLLC, this article presented a unified framework for designing and/or analyzing URLLC. Specifically, several relevant statistical tools and methodologies were introduced, and their applications to URLLC were highlighted together with thorough discussions on how these tools can be used interdependently towards designing/analyzing URLLC systems. As a unique feature of the tutorial-like nature of this article, each of the introduced tools and methodologies was further elaborated through concrete numerical examples and selected applications. This article is targeted toward graduate students, early-stage researchers, and professionals interested in getting a comprehensive overview of the statistical tools and methodologies relevant to URLLC, and will help foster more research in URLLC and its evolution towards next-generation wireless systems.

In what follows, we present some potential future research directions for the discussed tools and methodologies.

*1) From URLLC to dependable communications:* Current 5G approaches for meeting URLLC requirements based on tweaking the system design are not scalable nor efficient. This is especially evident for use cases related to emerging sophisticated cyber-physical systems, which rely on embedded, decentralized, real-time computations and interactions, where physical and software components are deeply intertwined [205]. The design and analysis of such systems mandate a departure from link-specific URLLC to holistic dependability. For this, there is a series of associated challenges, which are mainly related to i) the scalable modeling of the different kinds of adverse events, e.g., errors, faults, failures (refer to [206]), and their cross-correlation, ii) the extrapolation of dependability concepts and tools as those overviewed in Section II, which are originally conceived for systems engineering, to wireless communications engineering in different domains/layers, and iii) the development of data-driven dependable mechanisms. Moreover, notice that a thorough joint optimization of the communication, control, computing, and sensing processes is fundamental for resource-efficient truly-dependable systems. The target should be flexible and adaptable URLLC networks that allow customization without significantly sacrificing manageability to efficiently support diverse applications. Communication and control co-design, quantum computing and communication, and joint communication and sensing, together with data-driven optimization/design approaches, are instrumental for this and may continue receiving dedicated research in the coming years. All in all, the notion of dependability is expected to play a more dominant role in beyond 5G wireless systems to provide a richer performance measure of critical systems [8], [9].

The term xURLLC, or next generation URLLC, refers to URLLC evolution towards 6G [207]. It may be noted that *reliability* and *latency* were the two main performance indicators in 5G URLLC. In the 6G era, xURLLC will encompass other important performance indicators, such as the mean time between failures (refer to Section II-A), the distribution of failures, jitter, security [208], AoI [209], etc., to reflect the needs of the various vertical sectors that it will serve. Examples of prospective enablers of xURLLC include novel cell-free massive MIMO architecture to support low latency communication [210], semantic aided wireless communication framework [211], ML-aided channel access schemes [212], discrete signaling based co-existence schemes with other services [213], and quantum computation protocols to ensure user privacy and data security in distributed networks [214].

Notice that we do not suggest that a separate version of URLLC is implemented for each control application in 5G/6G. Instead, our argument is to design a flexible and adaptable URLLC network that can support various control applications efficiently. The idea is to develop a framework that provides a set of common functionalities and mechanisms while allowing customization and optimization for different application requirements. ORAN and data-driven approaches may be handy here. We believe that a carefully designed, versatile architecture that enables co-design without sacrificing the efficiency and manageability of the underlying network infrastructure can be practical and lead to significant advancements.

*2) Statistical bounds, inequalities, approximations, and risk measures on small sample sizes:* The bounds and approximations discussed in Section III inherently rely on a statistical pre-processing of data samples. For instance, applying Markov's inequality (26) mandates computing the expected value of a certain random process and the GPD fitting of the conditional CDF (33) is data-driven. Therefore, the confidence in such approaches increases with the sample size. However, the number of samples corresponding to a certain relevant random process may be strictly limited in many URLLC use cases, which calls for novel probabilistic bounds/approximations/risk-metrics with accompanied confidence guarantees. As commented in Example 5, they may be designed using the same tools and approaches outlined in Section III but enriched with additional procedures considering the data finitude, e.g., exploiting the central limit theorem. Finally, further research is needed to wisely exploit risk-related metrics, other than the traditional value-at-risk and expected shortfall, in the design/analysis of URLLC systems.

*3) Ultra-efficient sampling for rare event simulations:* The methods discussed in Section IV rely on statistical models, including some underlying prior knowledge, for proper sampling. Therefore, efficient approximations/upper bounds and optimization of target distributions are needed to reduce computational complexity and/or increase ac-curacy. In addition, many URLLC use cases require efficient sampling to assess novel probabilistic metrics with a significant number of samples from the probability space where the URLLC event exists. Therefore, further research is needed to properly link efficient sampling algorithms and URLLC research. One direction may be using Hamiltonian Monte Carlo methods, particularly for use cases with large dimensions, as it is a non-Markovian-based method able to explore the probability space efficiently by avoiding the dependencies of the Markov chain [215], [216]. Similarly, variational Bayes methods [95], [96], poorly explored in the URLLC literature, deserve more attention.

*4) Fixed to variable length coding:* The FBL channel codes considered in Section V are fixed-length codes where each codeword has the same number of bits. Variable-length codes, on the other hand, encode message symbols to codewords with a variable number of bits. This has the advantage that the coding rate can be adjusted to the channel conditions leading to enhanced resource efficiency, albeit requiring a feedback channel [217]. The involvement of the feedback loop incurs extra latency and an additional error source (as illustrated through the example in Section V-B). A recent study has shown that noise in the feedback link can cause a significant increase in the minimum average latency when applying variable-length codes to URLLC use cases [218]. This raises the question *whether variable-length coding schemes are suitable for URLLC,* or alternately *how to design variable-length codes for URLLC applications with FBL transmissions?*

*5) FBL queuing and dependable AoI:* Notably, SNC may help identify and define new delay-bounded QoS metrics incorporating the channel's and the arrival and service processes' stochasticity. However, the relationship between SNC and FBL theory/coding still requires further investigation, especially the characterization of the tail distribution and the delay bounds in extreme URLLC [141]. Meanwhile, open challenges in the context of information freshness include the characterization of AoI tail distributions for different network models, part of which was characterized in [147]. Moreover, future research should focus on critical decision-making in inter-disciplinary problems such as communication control co-design, which demands timely and reliable fresh information. In such cases, the link between AoI and dependability metrics, such as mean time to first failure, deserves particular attention.

*6) Strengthening the large-scale URLLC design/analysis:* In Section VII-B, we only referred to hard-clustering, where a point can only belong to exactly one cluster. However, soft (or fuzzy) clustering [219], [220], where a point can potentially belong to multiple clusters, may be more appealing for assessing the inherent complexity/heterogeneity of wireless networks, especially URLLC. Meanwhile, tools and methodologies related to meta-distribution, compressed sensing, and MF games rely heavily on underlying models and assumptions. Since ultimately *"all models are wrong"* (quote from George Box), say imperfect, they must be investigated considering modeling imperfections/uncertainties. Finally, regarding Example 16 in Section VII-C and the discussions that followed it, we would like to emphasize that multi-user detection in grant-free URLLC systems can be assisted by a preliminary sparsity level estimation phase as discussed in [193]. However, there are numerous associated research directions yet to explore, including *i)* sparsity level estimators exploiting prior traffic history knowledge and/or optimized for MIMO operation, and *ii)* a joint optimization of the sparsity level estimation and multi-user detection phases since the total number of symbols for these tasks might be very limited and accuracy of the sparsity level estimation is intrinsically imperfect. Data-driven sparse detection is another prominent research direction, where further studies could be directed to learning confidence.

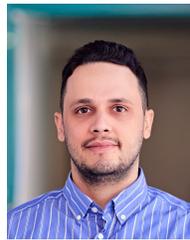

**Onel L. A. López** was born in Sancti-Spíritus, Cuba, in 1989. He received the B.Sc. (1st class honors, 2013), M.Sc. (2017), and D.Sc. (with distinction, 2020) degree in Electrical Engineering from the Central University of Las Villas (Cuba), the Federal University of Paraná (Brazil) and the University of Oulu (Finland), respectively. From 2013-2015 he served as a specialist in telematics at the Cuban telecommunications company (ETECSA). He is a collaborator to the 2016 Research Award given by the Cuban Academy of Sciences, a co-recipient of the 2019 IEEE European Conference on Networks and Communications (EuCNC) Best Student Paper Award, the recipient of the 2020 best doctoral thesis award granted by Academic Engineers and Architects in Finland TEK and Tekniska Föreningen i Finland TFiF in 2021, and the recipient of the 2022 Young Researcher Award in the field of technology in Finland. He authored the book entitled "Wireless RF Energy Transfer in the Massive IoT Era: towards sustainable zero-energy networks", Wiley, Dec 2021. He currently holds an Assistant Professorship (tenure track) in sustainable wireless communications engineering at the Centre for Wireless Communications (CWC), Oulu, Finland. His research interests include wireless connectivity, sustainable and dependable IoT, energy harvesting, wireless RF energy transfer, machine-type communications, and cellular-enabled positioning systems.

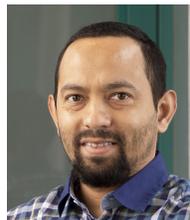

**Nurul H. Mahmood** was born in Bangladesh. He received the M.Sc. degree in Mobile Communications from Aalborg University, Denmark in 2007, and the Ph.D. degree in wireless communications from the Norwegian University of Science and Technology (NTNU), Norway in 2012. He is currently a Senior Research Fellow and Adjunct Professor in Critical Machine Type Communications at CWC, University of Oulu, Finland. He is the coordinator of wireless connectivity research in the Finnish 6G Flagship program and the University of Oulu's project manager for Hexa-X-II - EU's flagship 6G research project. He was an Assoc. Professor with the Department of Electronics Systems at Aalborg University, Denmark prior to joining the University of Oulu in December 2018. Nurul has an h-index of 21 and has published one edited book, 26 journal papers, and over 50 conference papers. His research interests include resilient communications for wireless industrial networks and resource optimization with a focus on URLLC and MTC.

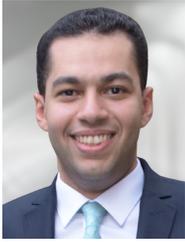

**Mohammad Shehab** is a postdoctoral Researcher at the University of Oulu, Finland. He obtained his B.Sc from Alexandria University, Egypt in 2011. He worked as a teaching assistant at Alexandria University and the Arab Academy from 2012-2015. He obtained two M.Sc degrees from the Arab Academy (2014) and the University of Oulu (2017). He obtained his doctoral degree in 2022 from Oulu with a focus on energy-efficient QoS provisioning in IoT. He won the best student paper award in ISWCS 2017, and the Nokia Foundation award consecutively for 2018 and 2019. His current research directions include machine learning, UAVs, and proactive resource allocation. His work so far has resulted in 13 conference papers, 8 journal papers, and 1 patent.

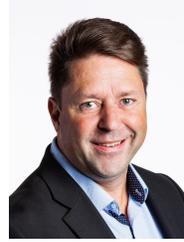

**Matti Latva-aho** received the M.Sc., Lic.Tech. and Dr. Tech (Hons.) degrees in Electrical Engineering from the University of Oulu, Finland in 1992, 1996, and 1998, respectively. From 1992 to 1993, he was a Research Engineer at Nokia Mobile Phones, Oulu, Finland after which he joined the Centre for Wireless Communications (CWC) at the University of Oulu. Prof. Latva-aho was Director of CWC during the years 1998-2006 and Head of the Department for Communication Engineering until August 2014. Currently, he is a professor at the University of Oulu on wireless communications and Director for the National 6G Flagship Programme. He is also a Global Research Fellow at Tokyo University. His research interests are related to mobile broadband communication systems and currently, his group focuses on 6G systems research. Prof. Latva-aho has published over 500 conference or journal papers in the field of wireless communications. He received Nokia Foundation Award in 2015 for his achievements in mobile communications research.

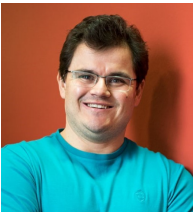

**Hirley Alves** received the B.Sc. and M.Sc. degrees from the Federal University of Technology-Paraná (UTFPR), Brazil, in 2010 and 2011, respectively in electrical engineering, and the dual D.Sc. Degree from the University of Oulu and UTFPR, in 2015. In 2017, he was an Adjunct Professor in machine-type wireless communications with the Centre for Wireless Communications (CWC), University of Oulu, Oulu, Finland. Currently, he is an Associate Professor and the Head of the Machine-type Wireless Communications Group. He is actively working on massive connectivity and ultra-reliable low latency communications for future wireless networks, 5GB and 6G, and full-duplex communications. He coordinates the MTC activities for the 6G Flagship Program under the Massive Wireless Automation Thematic Area. He is a co-recipient of the 2017 IEEE International Symposium on Wireless Communications and Systems (ISWCS) Best Student Paper Award and the 2019 IEEE European Conference on Networks and Communications (EuCNC) Best Student Paper Award co-recipient of the 2016 Research Award from the Cuban Academy of Sciences. He has been the organizer, chair, TPC, and tutorial lecturer for several renowned international conferences. He is the General Chair of the ISWCS'2019 and the General Co-Chair of the 1st 6G Summit, Levi 2019, and ISWCS 2021.

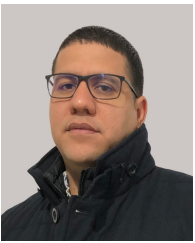

**Osmel Martínez Rosabal** was born in Camagüey, Cuba, in 1989. He received a B.Sc. (Hons., 2013) degree and M.Sc. (Hons., 2021) degrees from the Central University of Las Villas (Cuba) and the University of Oulu (Finland), respectively, both in Electrical Engineering. From 2013-2018 he served as a specialist in telematics at the Cuban telecommunications company (ETECSA). He is currently pursuing his Ph.D. in Wireless Communications Engineering with a focus on sustainable wireless energy transfer systems, at the Centre for Wireless Communications (CWC), University of Oulu, Finland.

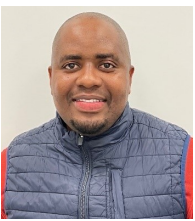

**Leatile Marata** received the BEng-Electronics and Telecommunications (2013) and the MEng-Telecommunications (2017) from the Central University of Las Villas and the Botswana International University of Science and Technology (BIUST), respectively. He is currently pursuing his doctoral studies with the Centre for Wireless Communications (CWC), University of Oulu. His main research interests are signal processing for wireless communications, Bayesian inference, sparse signal recovery, and compressed sensing.